\documentclass[11pt]{article}
\pdfoutput=1
\usepackage[margin=1.0in]{geometry}
\usepackage{amsmath,amssymb,graphicx}
\usepackage{hyperref}
\usepackage{slashed}
\usepackage{wasysym}
\usepackage{hhline}

\usepackage[utf8]{inputenc}

\usepackage[greek,english]{babel}

\usepackage[caption=false]{subfig}
\usepackage{booktabs}

\usepackage[font=small,labelfont=bf]{caption}
\usepackage{cite}
\usepackage{enumerate}

\newcommand{\be}{\begin{equation}}
\newcommand{\ee}{\end{equation}}
\newcommand{\beq}{\begin{equation}}
\newcommand{\eeq}{\end{equation}}

\newcommand{\beqn}{\begin{eqnarray}}
\newcommand{\eeqn}{\end{eqnarray}}
\newcommand{\bea}{\begin{eqnarray}}
\newcommand{\eea}{\end{eqnarray}}

\usepackage[table]{xcolor}
\usepackage{bm}

\definecolor{gray}{cmyk}{0,0,0,0.05}
\newcolumntype{a}{>{\columncolor{gray}} l}

\numberwithin{equation}{section}

\usepackage{soul}

\begin{document}


\begin{center}


~
\vskip 1cm

{\LARGE \bf 
Detecting long-lived multi-charged particles \\[0.2cm] in neutrino mass models
with MoEDAL
}

\vskip 1.cm

{\large 
Martin Hirsch$^{(a)}$,
Rafa\l{} Mase\l{}ek$^{(b)}$ and
Kazuki Sakurai$^{(b)}$
}

\vskip 0.7cm

$^{(a)}${\em
Instituto de F\'{\i}sica Corpuscular (CSIC-Universitat de Val\`{e}ncia), \\
 C/ Catedr\'atico Jos\'e Beltr\'an 2, E-46980 Paterna (Val\`{e}ncia), Spain }

\vskip 0.3cm

$^{(b)}${\em
Institute of Theoretical Physics, Faculty of Physics,\\University of Warsaw, ul.~Pasteura 5, PL-02-093 Warsaw, Poland\\[0.1cm]
}

\end{center}

\vskip 0.5cm
\begin{abstract}
A certain class of neutrino mass models predicts long-lived particles
whose electric charge is four or three times larger than that of
protons.  Such particles, if they are light enough, may be produced at
the LHC and detected.  We investigate the possibility of observing
those long-lived multi-charged particles with the MoEDAL detector,
which is sensitive to long-lived particles with low velocities
($\beta$) and a large electric charge ($Z$) with $\Theta \equiv
\beta/Z \lesssim 0.15$.  We demonstrate that multi-charged scalar
particles with a large $Z$ give three-fold advantage for MoEDAL;
reduction of $\Theta$ due to strong interactions with the detector,
and enhancement of the photon-fusion process, which not only increases
the production cross-section but also lowers the average production
velocity, reducing $\Theta$ further.  To demonstrate the performance
of MoEDAL on multi-charged long-lived particles, two concrete neutrino
mass models are studied. In the first model, the new physics sector is
non-coloured and contains long-lived particles with electric charges
2, 3 and 4.  
{
A model-independent study finds MoEDAL
can expect more than 1 signal event at the HL-LHC ($L = 300$ fb$^{-1}$)
if these particles are lighter than 600, 1100 and 1430 GeV, respectively.  
These compare with the current ATLAS limits 
650, 780 and 920 GeV for $L = 36$ fb$^{-1}$.}
The second model has a coloured new physics sector, which
possesses long-lived particles with electric charges 4/3, 7/3 and
10/3. 
{
The corresponding MoEDAL's mass reaches at the HL-LHC 
are 1400, 1650 and 1800 GeV, respectively,
which compare with the current CMS limits 
1450, 1480 and 1510 GeV for $L = 36$ fb$^{-1}$.}
{
In a model-specific study
we explore the parameter space of neutrino mass generation models and identify the
regions that can be probed with MoEDAL at the end of Run-3 and
the High-Luminosity LHC.}

\end{abstract}


\vskip 1cm

\section{Introduction}
\label{sec:intro}

Finding new particles or at least setting limits on their existence is
one of the main objectives of the experiments at the Large Hadron
Collider (LHC). for Run 1 and 2, the focus was mainly on searches in
the missing energy channel, motivated by models involving Dark Matter
candidates, such as supersymmetry.  Recently, a lot of effort has been
put into the analyses for more unorthodox signatures, which include,
for example, multi-jet \cite{Aad:2020nyj, Sirunyan:2018xwt} (-lepton
\cite{Sirunyan:2019bgz}), mono-jet
\cite{Aad:2021egl,Sirunyan:2017hci}, displaced vertices
\cite{Aad:2019xav, Sirunyan:2020cao} and disappearing track
\cite{Aaboud:2017mpt, Sirunyan:2020pjd} signatures.

Signatures of long-lived particles are one of such exotic signatures
that attracted attention recently.\footnote{For a broad review of LHC
  searches for long-lived particles, see \cite{Alimena:2019zri}.}  The
search strategy varies depending on the lifetime of the particle.  For
intermediate lifetimes, ${\cal O}(1)\,{\rm cm} \lesssim c \tau
\lesssim {\cal O}(10)\,{\rm cm}$, searches for displaced vertices and
disappearing tracks are effective.  If the particles are electrically
charged and have very long lifetimes, $c \tau \gtrsim {\cal
  O}(1)\,{\rm m}$, ATLAS and CMS detectors may register them as a
slow-moving particle in their muon systems or a particle with an
anomalously high ionising power in their electromagnetic calorimeters.
Searches of this kind are often called heavy stable charged
particle (HSCP) searches.

It has been pointed out recently that the MoEDAL (Monopole and Exotics
Detector at the LHC) detector may also be capable of searching for
charged meta-stable particles ($c \tau \gtrsim {\cal O}(1)$\,m) and
give independent (and sometimes complementary) constraints on them
from ATLAS and CMS \cite{Acharya:2020uwc, Felea:2020cvf}.  MoEDAL is
largely a passive detector that has been designed primarily to look
for magnetic monopoles \cite{Acharya:2014nyr, Pinfold:2009oia, Pinfold:2019nqj,
  Acharya:2019vtb}. It is located around the interaction point in the
VErtex LOcator (VELO) cavern of the LHCb experiment. The principal
component 
\footnote{MoEDAL consists also of paramagnetic trapping volumes (MMTs)
  able to capture long-lived highly ionising particles for later
  examination, and of array of pixel devices (TimePix) used to monitor
  highly ionising background in the cavern. Since these detectors are
  of no importance for our analysis, we do not discuss them any
  further.}  of the MoEDAL detector is a large array (circa 120 $m^2$) of
nuclear track detectors (NTDs) made of multiple CR39 and Makrofol
plastic sheets stacked together.  Magnetic monopoles are expected to
have a large magnetic charge required by the Dirac quantization
condition, $Q_m = 2 \pi n/Q_e$ ($n \in \mathrm Z$), and therefore be
highly ionising.  Such particles, if they travel through an NTD panel,
would leave a microscopic damage along their trajectories, which can
be revealed by dismantling NTD panels and putting plastic sheets in an
etching solution. The signature of highly ionising particle is the
presence of double cone shaped etch-pits, with sizes in the range 20
to 50 $\mu m$, collinear in all sheets of plastic within a single NTD
module. Combining information from multiple etch-pits allows to
reconstruct particle trajectories and infer their electric charge with
resolution better than $0.05e$, with $e$ being elementary charge.
During Run-2, NTD panels were exposed for a year, after which they
were disassembled and transported to an external laboratory, where the 
plastic sheets were etched and scanned using optical microscopes. For
the Run-3 data taking period, an automated system controlled by artificial
intelligence is being prepared to accelerate analyses.

Not only magnetic monopoles but any electrically charged particles
with high ionisation power can leave a similar signature in an NTD.  The
condition for electrically charged particles to be detected by MoEDAL
is given by $\Theta \equiv \beta/Z < 0.15$, where $\beta$ is the
velocity of the particle and $Z \equiv Q/|e|$ is the electric charge
measured in the proton charge units. It imposes an effective cut-off
on fast-moving particles, allowing to highly suppress SM background.
In order to further suppress the background to a negligible level, NTD
panels are located at distances $\sim 2$\,m from the interaction
point.  Therefore, those anomalously heavy ionising particles must
also be long-lived at least with $c \tau \gtrsim {\cal O}(1)$\,m to
reach the NTD panels.  Detection of singly charged particles at MoEDAL
has recently been studied~\cite{Felea:2020cvf}.  The study focused on
supersymmetric staus and considered a gluino cascade decay chain
involving two long-lived (LL) particles, $\tilde \chi_1^0$ and $\tilde
\tau^{\pm}$;\, $pp \to \tilde g \tilde g$, $\tilde g \to jj [\tilde
  \chi^0_1]_{\rm LL}$, $[\tilde \chi^0_1]_{\rm LL} \to \tau^{\pm}
    [\tilde \tau^{\mp}]_{\rm LL}$.  
It has been demonstrated that MoEDAL can probe regions of the
parameter space which have not been excluded by the HSCP searches
currently performed by ATLAS and CMS.  In Ref.~\cite{Acharya:2020uwc}
the study has been extended to a  more comprehensive list of
supersymmetric particles, as well as for doubly charged scalars and
fermions.  In this paper we augment these studies by investigating 
MoEDAL's performance for multiply charged long-lived particles with $Z
\gtrsim 2$.  We also demonstrate the importance of the previously
neglected photon-fusion production channel for scalar particles and
update the earlier results obtained in Ref.~\cite{Acharya:2020uwc}.

Searches for long-lived multi-charged particles are interesting, since
such particles often appear in phenomenological models. For example,
an electroweak triplet scalar field with hypercharge $Y=1$ is
introduced in the type-II seesaw model for neutrino masses. 
Another example is a doubly charged Higgsino, which
can be found in supersymmetric left-right symmetric models.

The smallness of the observed neutrino masses has motivated a variety
of models proposed in the literature. In particular, radiative
neutrino mass models have a long tradition
\cite{Zee:1980ai,Cheng:1980qt,Zee:1985id,Babu:1988ki}, since they
provide automatically a suppression for neutrino masses, for a review
see \cite{Cai:2017jrq}.  Systematic classifications for different
radiative models have been published for 1-loop \cite{Bonnet:2012kz},
2-loop \cite{Sierra:2014rxa} and even 3-loop \cite{Cepedello:2018rfh}
models. For our current paper, we will consider a special class of
1-loop models, first discussed in \cite{Arbelaez:2020xcg}. 1-loop
neutrino mass models often add a discrete symmetry to the SM gauge
group. The puprose of this is twofold. First, it allows to remove
unwanted tree-level contributions and, second, such models contain a
candidate for the Dark Matter in the universe. The proto-type model
for this class is the scotogenic model \cite{Ma:2006km}.  The models
presented in \cite{Arbelaez:2020xcg} are orthogonal to this ansatz in
the sense that they are automatically the leading contribution to the
neutrino mass, without the addition of a new ad-hoc symmetry. The
resulting models cannot contain SM singlet fermions \footnote{Since
  this would result in a seesaw type-I contribution to the neutrino
  mass}, instead they contain multiply charged particles, both
fermionic fields and scalars. Due to the small neutrino mass {\em and}
the high electric charge these new fields are typically very
long-lived particles.  In this paper we consider two concrete models
of this class and derive MoEDAL's sensitivities to the long-lived
particles predicted in them.  We provide model-independent MoEDAL
detection reach for types of particles anticipated in the selected
class of models, and we also interpret our numerical results in the
concrete examples to identify viable parameter regions that can be
explored by MoEDAL at the LHC Run-3 and the High Luminosity LHC
(HL-LHC), under the condition that the given parameter set can fit the
observed data for the neutrino masses.

The rest of the paper is organised as follows.  In Section
\ref{sec:models} we briefly review the two radiative neutrino mass
models studied in this paper and discuss the lifetimes of
multi-charged particles in the models.  We then describe our analysis
method for the estimation of the expected signal events at MoEDAL in
Section \ref{sec:analysis}.  In Section \ref{sec:model-1}
(\ref{sec:model-2}) we present our numerical results for the first
(second) neutrino mass models.  The first half of each of these two
sections focuses on the model-independent sensitivities for the
multi-charged long-lived particles with $Z \geq 2$ treating their
lifetimes as a free parameter.  We then show, in the second half of
the sections, the model-dependent results for Model-1 and Model-2, and
identify the parameter regions that can fit the neutrino mass data and
also will be explored by MoEDAL at the LHC Run-3 and HL-LHC.  
We conclude our study in Section
\ref{sec:disc}. 

\section{Neutrino mass models and multi-charged long-lived particles}
\label{sec:models}

In this section we briefly review two variants of radiative neutrino
mass models with similar features. These models have been introduced
and discussed for the first time in Ref.~\cite{Arbelaez:2020xcg}.  In
this class of models, the Standard  Model (SM) is extended with two
scalar fields, $S_1$ and $S_3$, which are singlet and triplet
representations of $SU(2)_L$, respectively, and three pairs of
vector-like fermions $(F_i, \bar F_i)$ $(i = 1,2,3)$ in $SU(2)_L$ in
doublet representations. In the first model (Model-1), all beyond the
Standard Model (BSM) fields are colour singlets, while the new fields
are in colour (anti-)triplet representations in the second model
(Model-2).

\begin{table}[t!]
\centering
\renewcommand{\arraystretch}{1.3}
\begin{tabular}{c|c|c|c|c}
           & $S_1$ & $S_3$ & $F_i$ &  $\bar F_i$  \\
           \hline
~Spin~     & 0 & 0 & $\frac{1}{2}$ & $\frac{1}{2}$ \\
           \hline
~$SU(2)_L$~ & ${\bf 1}$ & ${\bf 3}$ & ${\bf 2}$ & ${\bf 2}$ \\ 
           \hline
~$U(1)_Y$~ & 2 & 3 & $\frac{5}{2}$ & $-\frac{5}{2}$ \\
\hline
~Lepton number~ & $-2$ & $-4$ & $-3$ & $3$ \\
\end{tabular}
\caption{The BSM fields and their quantum numbers in Model-1.  The
  index $i$ ($=1,2,3$) distinguishes the three copies of $(F, \bar F)$
  fields. Since lepton number is broken in the model, the lepton 
number assignment in this table should be understood as being valid 
in the limit $\lambda_5 \to 0$, see text.}
\label{tab:charge}
\end{table}

The quantum numbers of the BSM fields in Model-1 are summarised in Table
\ref{tab:charge}.  With these charge assignments the BSM Lagrangian is
given by
\beqn
{\cal L}_{\rm BSM} ~=~ {\cal L}_{\rm kin}
&-& \left[ (h_{ee})_{ij} e^c_i e^c_j S_1^\dagger 
+ (h_F)_{ij} L_i F_j S_1^\dagger 
+ (h_{\bar F})_{ij} L_i \bar F_j S_3 \,+\, {\rm h.c.} \right]
\nonumber \\
&-& \left[ \lambda_5 H H S_1 S_3^\dagger \,+\, {\rm h.c.} \right]
\nonumber \\
&+& \lambda_2 |H|^2 |S_1|^2 
+ \lambda_{3a} |H|^2 |S_3|^2 + \lambda_{3b} |H S_3|^2
+ \lambda_4 |S_1|^4
\nonumber \\
&+& \lambda_{6a} |S_3^\dagger S_3|^2 + \lambda_{6b} |S_3 S_3|^2
+ \lambda_7 |S_1|^2 |S_3|^2 \,.
\label{eq:L1}
\eeqn
Here the mass terms of the BSM fields ($m_{S_1}^2 |S_1|^2 +
m_{S_3}^2 |S_3|^2 + m_{F_i} \bar F_i F_i$) are included in ${\cal
  L}_{\rm kin}$.  It should be stressed that the $\lambda_5$ term
breaks the lepton number symmetry and is therefore indispensable for
neutrino mass generation. In the limit of $\lambda_5\to 0$ all BSM
fields have definite lepton numbers, as quoted in table
\ref{tab:charge}.  Note that, in each operator the BSM fields appear
even times, except for the $h_{ee}$ term.  This means one can
introduce a BSM parity by assigning an odd (even) charge to the BSM
(SM) fields.  When the BSM parity is exact, i.e. $(h_{ee})_{ij} = 0$,
the lightest BSM particle cannot decay.  The $(h_{ee})_{ij}$ coupling
therefore controls the decay lifetime of the lightest BSM particle. We
must stress that the limit $(h_{ee})_{ij} = 0$ is not allowed
phenomenologically, since it would lead to a charged stable relic
excluded by cosmology.

Since a non-vanishing $\lambda_5$ breaks lepton number it may be taken
to be small and the theory can remain technically natural in the sense
of t'Hooft.  Similarly, if either $(h_F)_{ij}$ or $(h_{\bar F})_{ij}$
were absent, lepton number would be conserved and if both are absent
simultaneously the BSM fermion number symmetry, $F_i \to e^{i \theta}
F_i$, $\bar F_i \to e^{-i \theta} \bar F_i$, becomes exact.  Therefore
a configuration in which both $(h_F)_{ij}$ and $(h_{\bar F})_{ij}$ are
small is radiatively stable, i.e.~technically natural.

\begin{figure}[!t]
\centering
\includegraphics[width=0.38\textwidth]{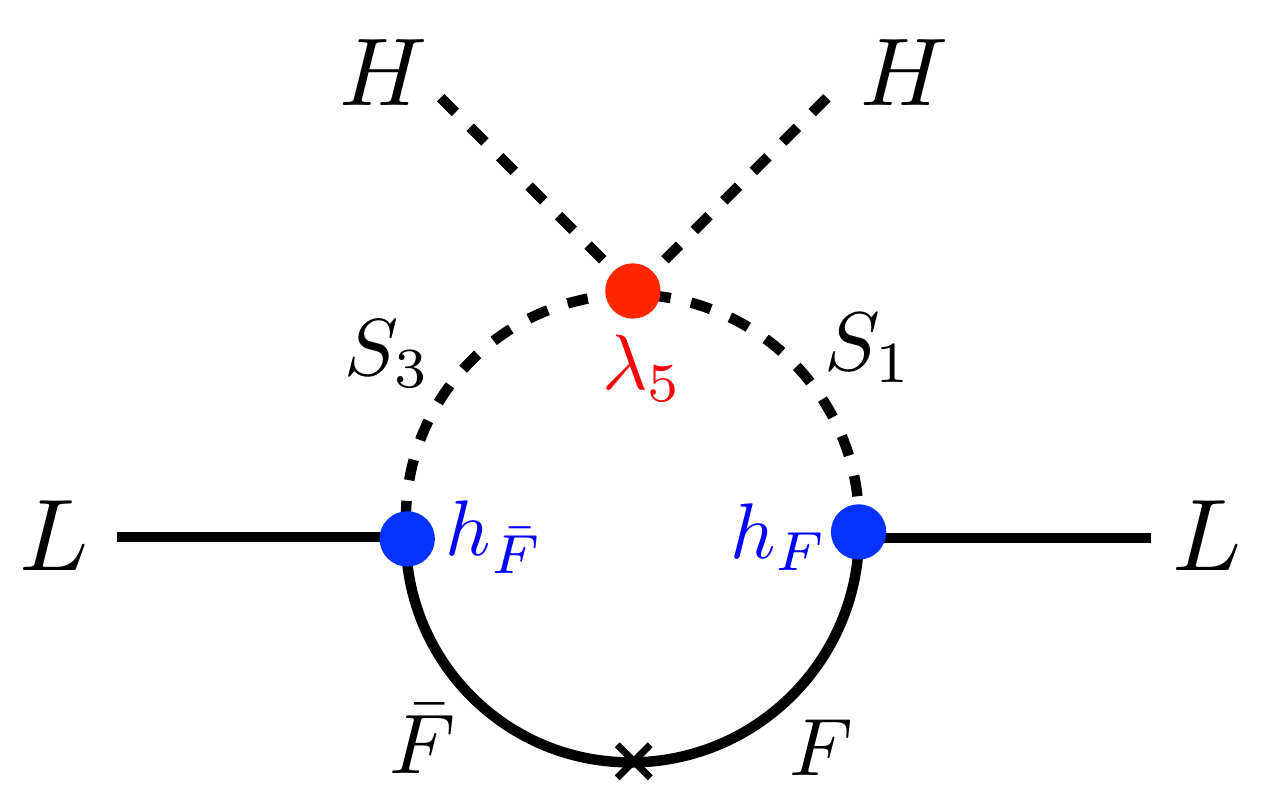}
\caption{The one-loop diagram generating the neutrino masses.}
\label{fig:massgen}
\end{figure}

In this class of models a dimension-5 operator for neutrino masses, 
roughly of order 
\beq
\frac{\lambda_5 N_c}{32 \pi^2 \Lambda^2} \left[ 
   h_{\bar F}^T m_F h_F + h_F^T m_F h_{\bar F} \right]_{ij}
\cdot
L_i H L_j H \,,
\label{eq:dim5}
\eeq 
is radiatively generated by integrating out the BSM fields through the
diagram in Fig.~\ref{fig:massgen}, where $N_c$ is the number of
colours of the particles in the loop and $\Lambda$ is the mass scale
of the BSM particles, with $\Lambda \sim \max ( m_{F_i}, m_{S_1},
m_{S_3})$.  Assuming all heavy masses are approximately equal to each
other and $h_F$ and $h_{\bar F}$ are diagonal, the neutrino masses are
roughly given by 
\beq m_\nu ~\sim~ 0.05 \cdot N_c \cdot \left(
\frac{\lambda_5}{10^{-6}} \right) \left( \frac{h_F h_{\bar
    F}}{10^{-4}} \right) \left( \frac{1\,{\rm TeV}}{\Lambda}
\right)\,{\rm eV}\,.
\label{eq:mnu}
\eeq
Note that in the numerical parts of this work, we do not use 
Eq. (\ref{eq:mnu}). Instead, we fit neutrino masses and 
angles, taking the full 1-loop calculation as presented in 
\cite{Arbelaez:2020xcg}, based on the general formulas given in 
\cite{Cordero-Carrion:2019qtu}.

The BSM fields have large $SU(2)_L$ and $U(1)_Y$ charges, which
results in the presence of multiply charged particles in the mass
eigenstates.  In particular, the $S_3$ field includes doubly, triply
and quadruply charged particles, denoted by $S^{+2}, S^{+3}$ and
$S^{+4}$, respectively.  The $S^{+2}$ mixes with the $S_1$ field.
The mass matrix for doubly charged scalars is given by 
\beq
\begin{pmatrix}
m_{S_1}^2 - \frac{1}{2} \lambda_2 v^2 & \frac{1}{2} \lambda_5 v^2 \\
\frac{1}{2} \lambda_5 v^2 & m_{S_3}^2 - \frac{1}{2} \lambda_{3a} v^2 \\
\end{pmatrix}
\label{eq:mmat}
\eeq   
Those multi-charged particles may be long-lived, if
\beq
m_{S_3} \ll m_{S_1}, m_F\,,
\label{eq:hierarchy}
\eeq 
Throughout this paper we assume this mass hierarchy. To simplify the
parameter space searches, we will also assume that $\lambda_i v^2 \ll
m^2_{S_3}$, for $i=2,(3a),(3b),5$, i.e. except for $\lambda_5$ the
exact values of the quartic couplings do not matter numerically. This
assumption results in the members of the multiplet $S_3$ being nearly
degenerate in mass. With the assumption $m_{S_3} \ll m_{S_1}$, the
lighter and heavier mass eigenstates of doubly charged particles are
$S^{+2} \simeq S_3$ and $S_1^{+2} \simeq S_1$, respectively. The mass
eigenvalues can be approximated as
\beqn
m^2_{S_1^{+2}} &=& m_{S_1}^2 - \frac{1}{2} \lambda_2 v^2 \,,
\\ 
m^2_{S^{+2}} &=& m_{S_3}^2 - \frac{1}{2} \lambda_{3a} v^2 \,.
\label{eq:m+2}
\eeqn
The mass degeneracy between $S^{+3}$ and $S^{+4}$ is also lifted due
to the electroweak symmetry breaking.  The physical masses are given
by
\beqn
m^2_{S^{+3}} &=& m_{S_3}^2 - \frac{1}{2} 
    ( \lambda_{3a} + \frac{1}{2} \lambda_{3b} ) v^2 \,,
\label{eq:m+3}
\\ 
m^2_{S^{+4}} &=& m_{S_3}^2 - \frac{1}{2} 
 \left(\lambda_{3a} + \lambda_{3b} \right) v^2 \,,
\label{eq:m+4}
\eeqn

\begin{figure}[!t]
\centering
\includegraphics[width=0.8\textwidth]{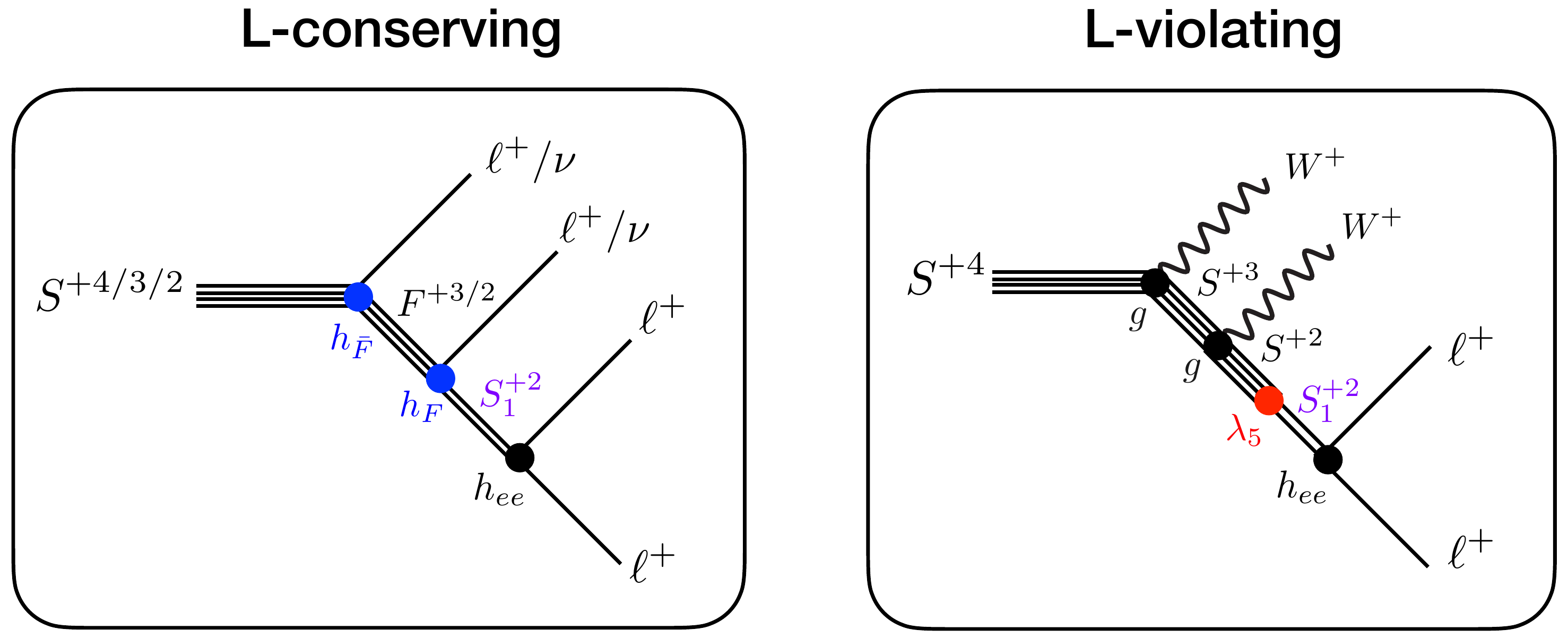}
\caption{Diagrams for possible decays for $S^{+4}$, $S^{+3}$ and $S^{+2}$. The
same particle decays into both $L=-4$ (left) and $L=-2$ (right) final
states, thus its lepton number is not fixed, once either $\lambda_5$
or $h_F\times h_{\bar F}$ is non-zero. For simplicity, we assign the
lepton number violation to $\lambda_5$ in the following. Observing
both decay modes would be an experimental demonstration of lepton number
violation.}
\label{fig:dec}
\end{figure}

With the mass assumption Eq.~\eqref{eq:hierarchy}, $S^{+4},S^{+3}$ and
$S^{+2}$ are lighter than other BSM particles.  These particles carry
lepton number $L=-4$ (in the unphysical limit $\lambda_5=0$) and decay
directly into the Standard Model particles.\footnote{The decays among
  triplet states, e.g.~$S^{+4} \to \ell^+ \nu S^{+3}$ via an off-shell
  $W$-boson, are chirally suppressed and proportional to the mass
  squared of the final state lepton, and therefore subdominant. } Due
to the large electric charge, the multiplicity of the decay final
states tend{s} to be large, incurring strong phase-space suppression
factors.

The main decay modes for $S^{+4}$ are 4-body modes; $S^{+4} \to 4
\ell^+$ and $S^{+4} \to W^+ W^+ \ell \ell$, as shown in
Fig.~\ref{fig:dec}. The approximate formulae for the partial decay
rates, valid for $v^2 \ll m_{S_3} \ll m_{S_1}, m_F$, are given 
roughly by \cite{Arbelaez:2020xcg}
\beqn
\Gamma(S^{+4} \to \ell_\alpha^+ \ell_\beta^+ \ell_\gamma^+ \ell_\delta^+)
&\sim&
\frac{|(h_{ee})_{\alpha \beta}|^2}{192 \cdot (4 \pi)^5}
\cdot
\left| \left\langle \frac{h^{\gamma \delta}_{F \bar F}}{m_F} \right\rangle \right|^2
\cdot
\frac{m^7_{S_3}}{m^4_{S_1}} \,,
\label{eq:s4_4l}
\\
\Gamma(S^{+4} \to W^+ W^+ \ell_\alpha^+ \ell_\beta^+ )
&\sim&
\frac{|g_2^2 (h_{ee})_{\alpha \beta}|^2}{48 \cdot (4 \pi)^5}
\cdot
\left( \frac{\lambda_5 v^2}{m^2_{S_1}} \right)^2 
\cdot
\frac{m^5_{S_3}}{m^4_{W}} \,,
\label{eq:s4_2w2l}
\eeqn
 where \beq \left| \left\langle \frac{h^{\gamma \delta}_{F \bar
    F}}{m_F} \right\rangle \right| ~\equiv~ \left| \sum_k
\frac{h_F^{\gamma k} h_{\bar F}^{\delta k} + h_F^{\delta k} h_{\bar
    F}^{\gamma k} }{m_{F_k}} \right|
\eeq 
is an effective reduced coupling introduced in \cite{Arbelaez:2020xcg}.  
It is worth noting that the decay mode $\Gamma(S^{+4} \to 4 \ell^+)$
is proportional to $h_F h_{\bar F}$, while the decay $\Gamma(S^{+4}
\to W^+ W^+ \ell^+ \ell^+)$ is proportional to
$\lambda_5$. Neutrino masses require that the product of these
parameters is small, see Eq. (\ref{eq:mnu}). We stress again, that we
do not use approximate Eqs. (\ref{eq:s4_4l})-(\ref{eq:s4_2w2l}) in our
numerical work.  Instead, the widths are calculated numerically at
each parameter point using {\tt MadGraph 5}
\cite{Alwall:2007st,Alwall:2011uj,Alwall:2014hca}.  A simultaneous
observation of both final states would be an experimental proof of
lepton number violation.

The two competing main decay modes of $S^{+3}$ are $S^{+3} \to \ell^+
\ell^+ \ell^+ \nu$ and $S^{+3} \to W^+ \ell^+ \ell^+$.  As for
$S^{+4}$, the former is $L$-conserving and the latter $L$-violating.
The partial decay rates are approximately given by
\beqn
\Gamma(S^{+3} \to \ell_\alpha^+ \ell_\beta^+ \ell_\gamma^+ \nu_\delta^+)
&\sim&
\frac{|(h_{ee})_{\alpha \beta}|^2}{192 \cdot (4 \pi)^5}
\cdot
\left| \left\langle \frac{h^{\gamma \delta}_{F \bar F}}{m_F} \right\rangle \right|^2
\cdot
\frac{m^7_{S_3}}{m^4_{S_1}} \,,
\label{eq:s3_3l}
\\
\Gamma(S^{+3} \to W^+ \ell_\alpha^+ \ell_\beta^+ )
&\sim&
\frac{|g_2 (h_{ee})_{\alpha \beta}|^2}{16 \cdot (4 \pi)^3}
\cdot
\left( \frac{\lambda_5 v^2}{m^2_{S_1}} \right)^2 
\cdot
\frac{m^3_{S_3}}{ m^2_{W}} \,.
\label{eq:s3_w2l}
\eeqn
Notice that the approximate formula for the decay rate
Eq.~\eqref{eq:s3_3l} is identical to Eq.~\eqref{eq:s4_4l} for the
$S^{+4}$ decay.  This is because these two decays are related by 
$SU(2)_L$ symmetry.

Finally, the partial decay rates for the two main decay modes of $S^{+2}$, 
$S^{+2} \to \nu \nu \ell^+ \ell^+$ and $S^{+2} \to \ell^+ \ell^+$ 
are given by
\beqn
\Gamma(S^{+2} \to \ell_\alpha^+ \ell_\beta^+ \nu_\gamma^+ \nu_\delta^+)
&\sim&
\frac{|(h_{ee})_{\alpha \beta}|^2}{192 \cdot (4 \pi)^5}
\cdot
\left| \left\langle \frac{h^{\gamma \delta}_{F \bar F}}{m_F} \right\rangle \right|^2
\cdot
\frac{m^7_{S_3}}{m^4_{S_1}} \,,
\label{eq:s2_2l2n}
\\
\Gamma(S^{+2} \to\ell_\alpha^+ \ell_\beta^+ )
&\sim&
\frac{|(h_{ee})_{\alpha \beta}|^2}{4 \pi} 
\cdot
\left( \frac{\lambda_5 v^2}{m^2_{S_1}} \right)^2 
\cdot
m_{S_3}\,.
\label{eq:s2_2l}
\eeqn
The approximate formula for 4-body decay mode is again
the same as Eqs.~\eqref{eq:s4_4l} and \eqref{eq:s3_3l} due to the
$SU(2)_L$ symmetry.  This 4-body decay is subdominant unless
$\lambda_5$ is extremely small.

The second variant of this type of radiative neutrino mass models
(Model-2), that we will study in the numerical parts later on, can
be obtained simply from Model-1 by giving colour (anti-)triplet charges
to the BSM fields in Model-1, with some related changes in hyper-charge 
assignments.  We distinguish the fields and particles
in Model-2 with tilde from those in Model-1.  Namely, we write Model-2
fields as $\tilde S_{1}, \tilde S_{3}, \tilde F_i, \tilde {\bar F}_i$.
In order to break the BSM parity and let the lightest BSM particle
decay, the $h_{ee}$ operator in Model-1 should be replaced by 
\beq
(h_{ee})_{ij} e^c_i e^c_j S_1^\dagger ~\to~ (h_{ed})_{ij} d^c_i e^c_j
\tilde S_1^\dagger \,.
\label{eq:hed}
\eeq

This implies that the difference of hypercharges between $\tilde S_1$
and $S_1$ must be $-2/3$.  In order to keep all the other operators in
Eq.~\eqref{eq:L1}, the hyper-charge of the other BSM fields must be
modified also by this amount.  
Also, assuming the operators in Eq.~\eqref{eq:hed} do not break the lepton number 
(we assume it is broken by the non-vanishing $\lambda_5$ coupling),
we must assign $L = -1$ for $\tilde S_1$,
and the lepton number assignment for all other fields must be modified accordingly.
In summary, the BSM fields of Model-2
can be obtained by the following replacements from Model-1: \beqn
\begin{array}{rcl}
\mbox{\bf Model-1} & ~ & \mbox{\bf Model-2} \\
S_1 ({\bf 1}, {\bf 1}, 2)_{-2} & \to & \tilde S_1 (\bar {\bf 3}, {\bf 1}, 4/3)_{-1} \\
S_3 ({\bf 1}, {\bf 3}, 3)_{-4} & \to & \tilde S_3 (\bar {\bf 3}, {\bf 3}, 7/3)_{-3} \\
F ({\bf 1}, {\bf 2}, 5/2)_{-3} & \to & \tilde F ({\bf 3}, {\bf 2}, 11/6)_{-2} \\
\bar F ({\bf 1}, {\bf 2}, -5/2)_{3} & \to & \tilde {\bar F} (\bar {\bf 3}, {\bf 2}, -11/6)_{2} \,,
\end{array}
\label{eq:replace}
\eeqn
where the three numbers in the brackets represent the representations
of $SU(3)_c \times SU(2)_L \times U(1)_Y$
and the subscripts denote our assignment of the lepton number.

With this charge assignment, the Lagrangian of Model-2 is the same as
that for Model-1 in Eq.~\eqref{eq:L1} with the replacement in
Eqs.~\eqref{eq:hed} and \eqref{eq:replace}.  The mechanism and formula
for neutrino mass generation, Eqs.~\eqref{eq:dim5} and \eqref{eq:mnu},
are unchanged (with $N_c = 3$).  Due to the shift of hypercharge, the
electric charge of Model-2 fields are also shifted from those of
Model-1 by $-2/3$.  For example, the $SU(2)$ triplet field, $\tilde
S_3$, includes the three states with electric charges $Z = 4/3$, $7/3$
and $10/3$, denoted by $\tilde S^{+4/3}$, $\tilde S^{+7/3}$ and
$\tilde S^{+10/3}$, respectively.  The $\tilde S^{+4/3}$ mixes with
the singlet field $\tilde S_1$ and give two physical states, as in
Model-1 for $S^{+2}$ and $S_1$.  The formula for the physical
masses, Eqs.~\eqref{eq:m+2}, \eqref{eq:m+3} and \eqref{eq:m+4}, are
unchanged with the obvious replacement; $S^{+2} \to \tilde S^{+4/3}$,
$S^{+3} \to \tilde S^{+7/3}$ and $S^{+4} \to \tilde S^{+10/3}$.  The
formulae for the partial decay rates of $\tilde S^{+10/3}$, $\tilde
S^{+7/3}$, $\tilde S^{+4/3}$ are unchanged from Eq.~\eqref{eq:s4_4l}
to Eq.~\eqref{eq:s2_2l} with the replacement $\ell^+_\alpha \to \bar
d_\alpha$ in the final state and $(h_{ee})_{\alpha \beta} \to
(h_{ed})_{\alpha \beta}$.

\section{The estimation of signal events}
\label{sec:analysis}

In our numerical analysis we first implement the model described in
the previous section in {\tt SARAH} \cite{Staub:2013tta,
  Staub:2012pb}.  The output is plugged into {\tt SPheno}
\cite{Porod:2003um,Porod:2011nf} to calculate the mass spectra, mixing
matrices and scalar two-body and fermionic three-body decays.
Finally, these information are fed into {\tt MadGraph 5}
\cite{Alwall:2007st,Alwall:2011uj,Alwall:2014hca}, which is used for
event generation as well as numerical evaluation for the decay rates
and the production cross-sections.  For the cross section calculation
we use the {\texttt {LUXqed17}$\_$plus$\_$PDF4LHC15$\_$nnlo$\_$100}. This PDF, based
on Manohar et al. \cite{Manohar:2016nzj,Manohar:2017eqh}, combines QCD
partons from {\texttt {PDF4LHC15}} \cite{Butterworth:2015oua} with an
improved calculation of the photon density in the proton.

In order to estimate the expected {number of} signal events in the MoEDAL
detector, we closely follow the procedure described in detail in
\cite{Acharya:2014nyr,Acharya:2020uwc}.  First, Monte Carlo events for
the pair production of exotic particles at the 13 TeV LHC are
generated using {\tt MadGraph 5}.  Having the velocity vectors of
charged particles in hands, we check whether or not there is a
MoEDAL's NTD panel in the direction of a charged particle.  We
implement this by a function $\omega({\bf p}_i)$ of the particle $i$'s
momentum vector ${\bf p}_i$.  Namely, $\omega({\bf p}_i) = 1$ if there
is an NTD panel in the direction of ${\bf p}_i$ and 0 otherwise.  Here
we assume MoEDAL's Run-3 configuration where all NTD panels are facing
to the interaction point at right angle.  In this case, given that
a charged particle hits an NTD panel, the detection efficiency can be
modeled by the step function \beq \Theta \left( 0.15 \cdot Z_i -
\beta_i \right), \eeq where $Z_i$ and $\beta_i$ are the electric
charge and the magnitude of the velocity of the particle $i$ ($i =
1,2$), respectively, and $\Theta(x) = 1$ for 
$x \geq {  0}$ and $0$
otherwise.  Finally, the probability that a charged particle with
lifetime $\tau_i$ survives until it reaches an NTD panel is given by
\beq 
P_{\rm surv}({\bf p}_i) ~=~ \exp \left( - \frac{L_{\rm
    NTD}({\bf p}_i) }{ \beta_i \gamma_i c \tau_i } \right), 
\eeq 
where $L_{\rm NTD}({\bf p}_i)$ is the distance from the interaction point to
the NTD panel in the direction of the momentum vector ${\bf p}_i$ of
the charged particle $i$.
In summary, the number of expected signal events with the integrated 
luminosity ${\cal L}$ is given by
\beq\label{eq:N}
N_{\rm sig} ~=~ \sigma \cdot {\cal L} \cdot \epsilon,
\eeq
with
\beq\label{eq:prob}
\epsilon ~=~ \left\langle \,
\sum_{i=1,2}
\omega({\bf p}_i) \cdot P_{\rm surv}( {\bf p}_i ) \cdot \Theta \left( 0.15 \cdot Z_i - \beta_i  \right)
\right\rangle_{\rm MC} \,,
\eeq
where $\sigma$ is the production cross-section of the exotic particles
and $\langle \cdots \rangle_{\rm MC}$ represents the Monte Carlo average.

\section{Numerical analysis for colour singlet models}
\label{sec:model-1}

As described in the previous section, exotic particles in Model-1 are
colour singlets.  Among them the scalar $SU(2)$-triplet, $S_3$, may be
long-lived when the other exotic states (a scalar $SU(2)$-singlet,
$S_1$, and vector-like $SU(2)$-triplet fermions, $F$ and $\bar F$) are
heavier than $S_3$ and the model parameters are fitted to explain the
neutrino masses.  In this section, we first investigate the collider
properties of the three mass eigenstates of $S_3$ and derive the model
independent exclusion reach in the first subsection.  We then
interpret the results for Model-1 and identify the parameter region
that can be probed by MoEDAL at the Run-3 of LHC.

\subsection{Results for colour-singlet multi-charged particles}

The $S_3$ and its conjugate state have three almost mass degenerate 
eigenstates denoted by $S^{4 \pm}$, $S^{3 \pm}$ and $S^{2 \pm}$ with
electric charges $\pm4$, $\pm3$ and $\pm2$, respectively.  All
these particles can be long-lived and potentially contribute to the
signal at MoEDAL depending on the model parameters.
There are three types of production mechanisms;
(i) Drell-Yan pair production process exchanging a $Z$ or $\gamma$
(ii) photon fusion pair production process
(iii) associated productions, $S^{4\pm} S^{3\mp}$ and $S^{3\pm} S^{2\mp}$, 
via a $W^\pm$ exchange.

\begin{figure}[!t]
\centering
\includegraphics[width=0.6\textwidth]{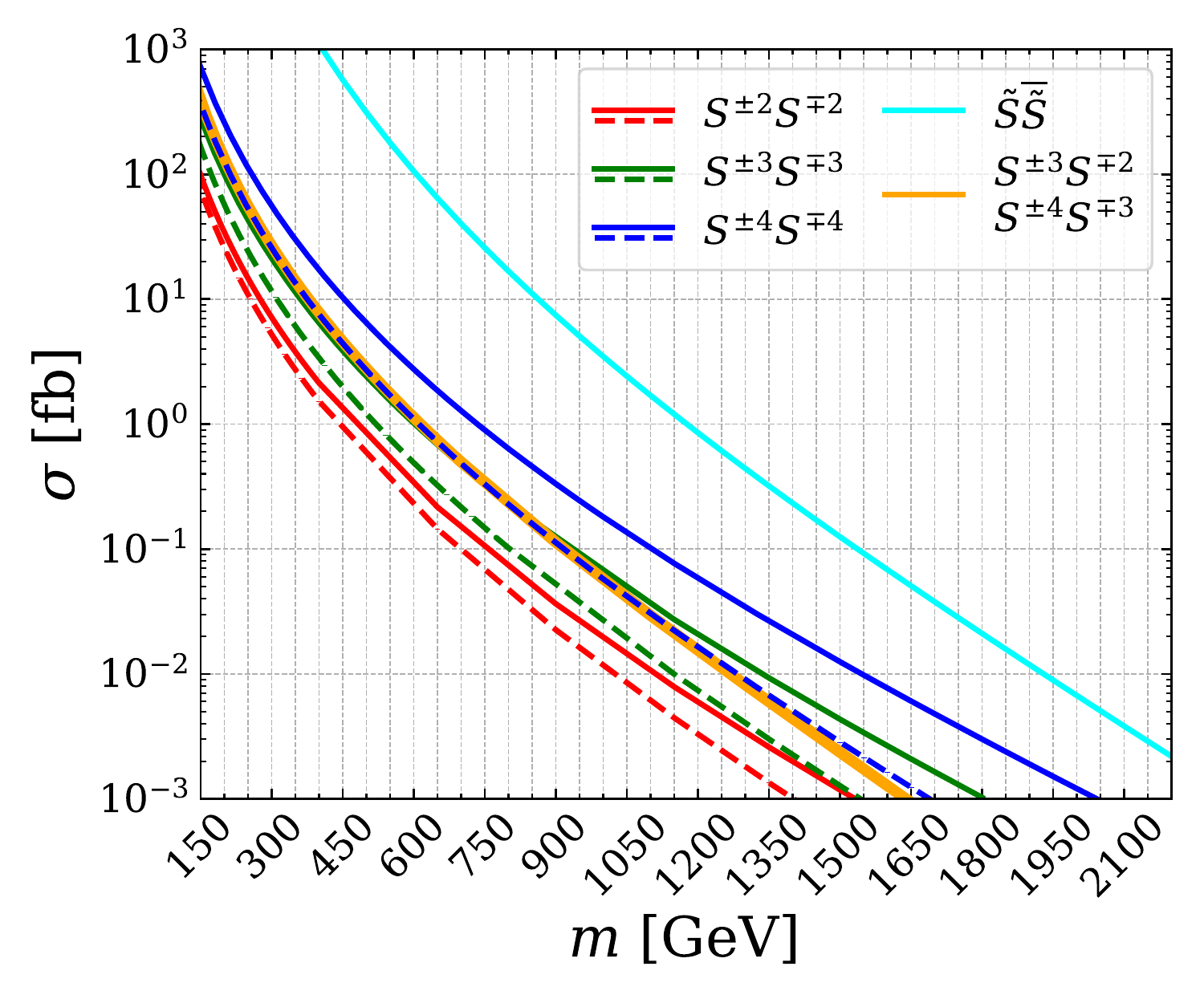}
\caption{The leading-order cross-sections of various production modes
  in Model-1 and -2.  The dashed curves correspond to the
  cross-sections {\it without} the photon fusion process.  The orange
  curve represents cross-sections for associated productions, $pp \to
  S^{\pm 3} S^{\mp 2}$ or $S^{\pm 4} S^{\mp 3}$ (their cross-sections
  are the same), mediated by a $s$-channel $W$-boson.  The cyan curve
  shows the QCD cross-sections of the single pair production mode in
  Model-2, i.e.~$pp \to \tilde S^{+4/3} \tilde S^{-4/3}$, $\tilde
  S^{+7/3} \tilde S^{-7/3}$ and $\tilde S^{+10/3} \tilde S^{-10/3}$.}
\label{fig:xsec}
\end{figure}

We show in Fig.~\ref{fig:xsec} the pair production cross-sections for
$S^{\pm4}$ (blue), $S^{\pm3}$ (green) and $S^{\pm 2}$ (red) by the
solid curves, which include both contributions from the Drell-Yan
($s$-channel $Z/\gamma$) and photon fusion processes.  To see the
impact of the photon fusion process, we also show the cross-sections
{\it without} the photon fusion by the dashed curves.  We see that the
photon fusion contribution is very important for scalar particles with
higher electric charge multiplicity, $|Q| = Z e$.  For example, at the
mass around 300 GeV, the photon fusion contribution enhances the
cross-section by $\sim 30$\,\% for $S^{\pm2}$, while for $S^{\pm4}$ it
makes up more than $50$\,\% of the total.  This is because the
contribution of the $s$-channel photon exchange diagram is
proportional to $|Q|^2$, while the cross-section of the photon fusion
process is proportional to $|Q|^4$.  Another reason why the photon
fusion is important is as follows.  The Drell-Yan $s$-channel process
exchanging a spin-1 gauge boson generally suffers from a $p$-wave
suppression in the scalar particle production, since the $s$-wave
production mode is forbidden by the angular momentum conservation.
This also explains another feature we see in the plot that the photon
fusion becomes relatively more enhanced in the higher mass region,
because at higher masses having large velocities costs more energy and
the cross-section gets suppressed by the parton distribution function.
The associated production, $pp \to S^{\pm4} S^{\mp3}$ (or $S^{\pm3}
S^{\mp2}$, since these modes have the same leading-order
cross-section), mediated by the $s$-channel $W$-boson, is also shown
by the orange curve.

The size of this cross-section is as large as the $S^{\mp3}$ pair
production cross-section and give non-negligible effects when studying
the implications for the concrete model in the next subsection.
Finally we show the cross-section for the pair productions of coloured
particles in Model-2 by the cyan curve, which we will discuss in
detail in the next section.

\begin{figure}[!t]
\centering
\includegraphics[width=0.6\textwidth]{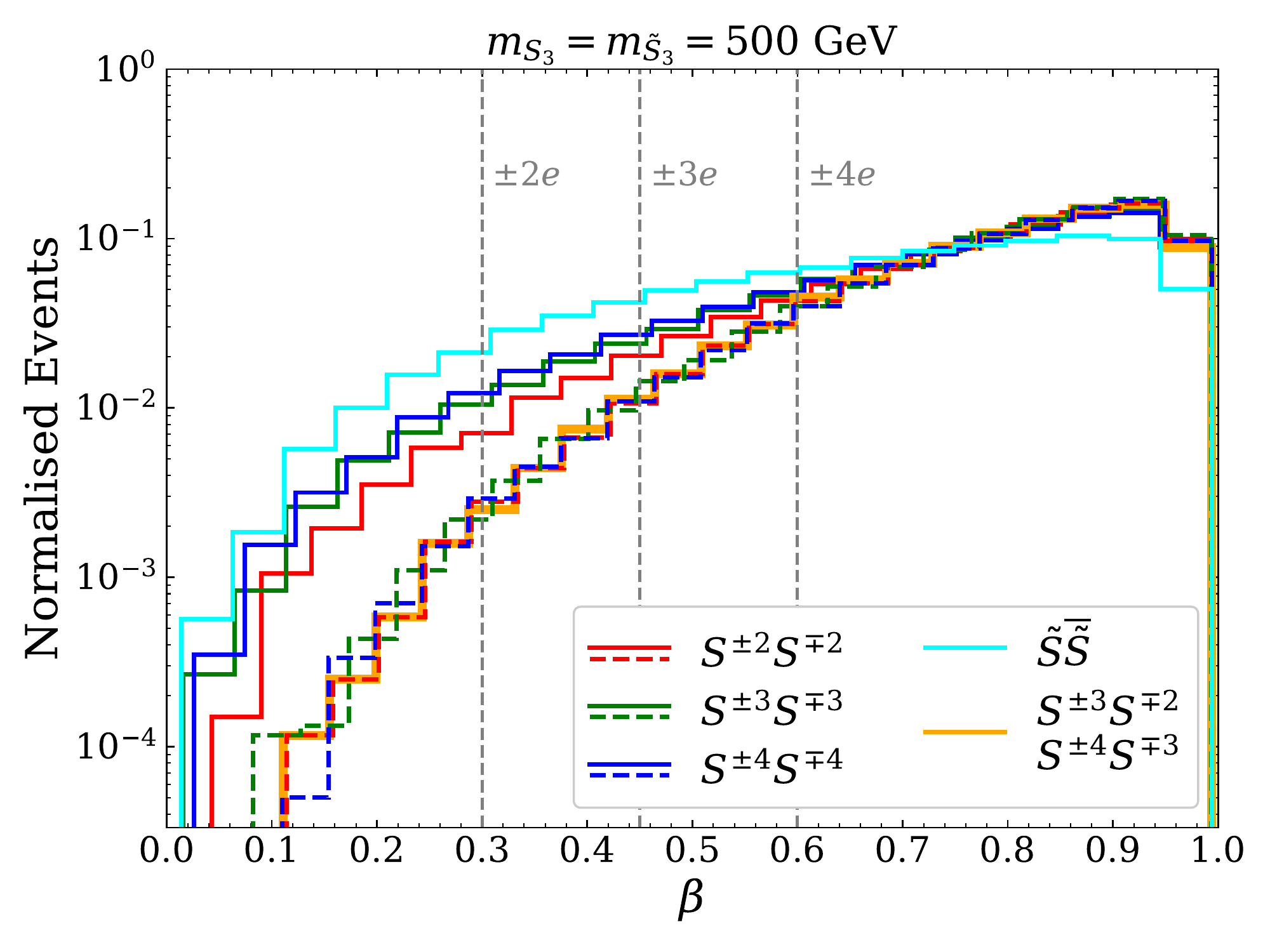}
\caption{Velocity distributions of the produced multi-charged
  particles in various production modes in Model-1 and -2.  
 {The masses are set to be $500$ GeV for all particles.}
  The dashed
  curves correspond to the velocity distributions {\it without} the
  contribution from the photon fusion process.  The orange curve
  include the contributions from two associated production modes, $pp
  \to S^{+3} S^{-2}$ and $S^{+4} S^{-3}$, mediated by a $s$-channel
  $W$-boson.  The cyan curve shows the velocity distribution by the
  QCD production in Model-2, i.e.~$pp \to \tilde S^{+4/3} \tilde
  S^{-4/3}$, $\tilde S^{+7/3} \tilde S^{-7/3}$ and $\tilde S^{+10/3}
  \tilde S^{-10/3}$. 
  }
\label{fig:vel}
\end{figure}

In Fig.~\ref{fig:vel} we show the velocity distributions for 
charged scalar production, {where the masses are taken to be $500$ GeV for all particles.}  
The same colour scheme is used as in
Fig.~\ref{fig:xsec}.  As in the previous plot, the solid histograms
are for the production {\it with} the photon fusion process and the
dashed histograms represent to the production {\it without} it.

As can be seen, the particle velocities are lower in general when the
photon fusion is included.  This is because the $s$-wave mode is
forbidden in the Drell-Yan processes and the produced particles are
required to have non-zero velocities.  This also explains why the
particles with larger $|Q|$ have lower velocities on average because
the photon fusion is relatively more enhanced for larger $|Q|$
compared to the Drell-Yan process.

As explained earlier, the MoEDAL detector is sensitive to particles
that have a large electric charge and small velocities with $\beta <
0.15 \cdot Z$.  The vertical dashed lines in Fig.~\ref{fig:vel} show
the threshold velocities for the particles with $Z (= |Q/e|) = 2$, 3
and 4.  As can be seen, the detection sensitivity is significantly
larger for a particle with larger $Z$ because (a) the threshold
velocity increases and (b) the average production velocity decreases.

\begin{figure}[t!]
\centering
      \includegraphics[width=0.35\textwidth]{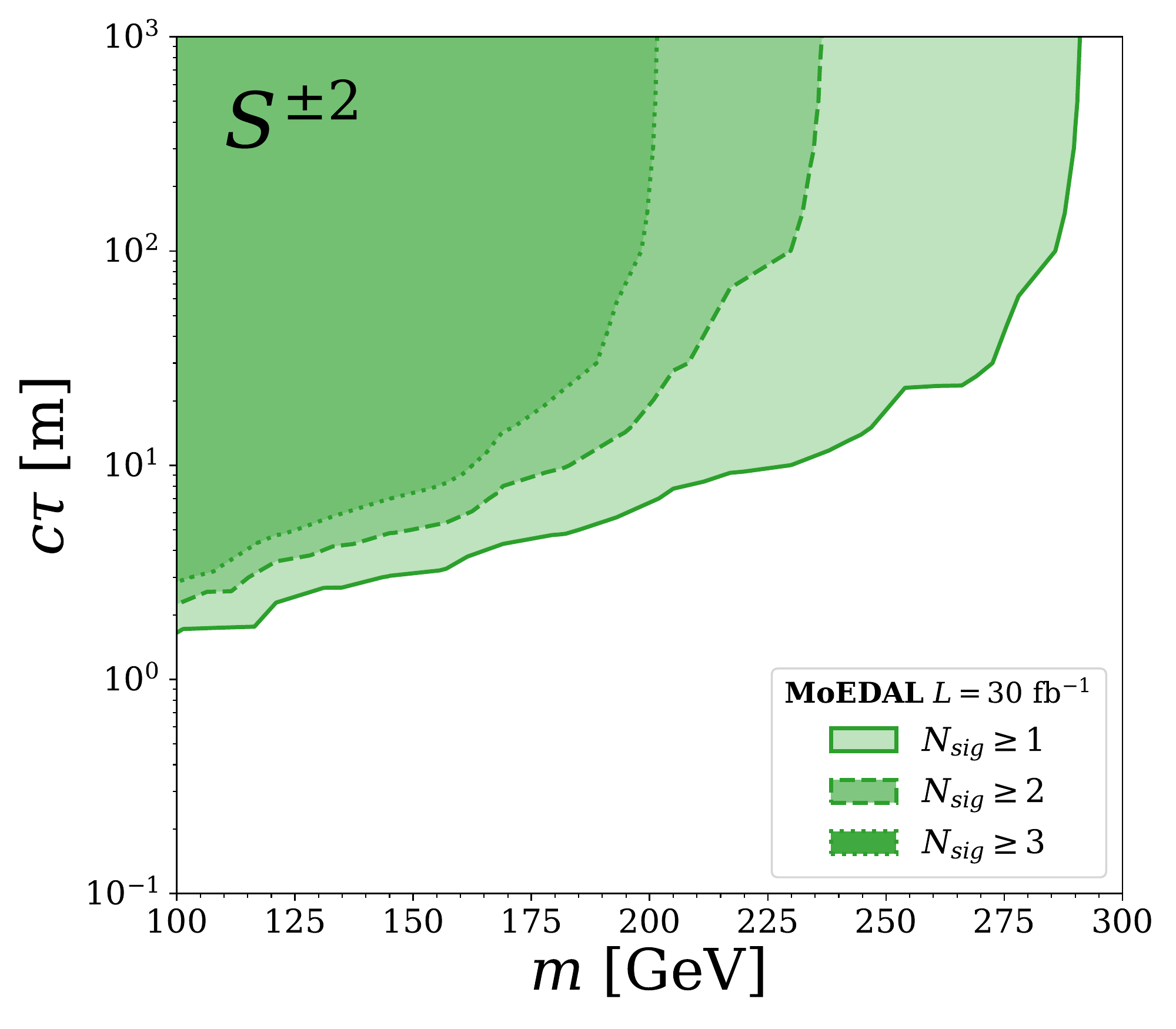} \hspace{5mm}
      \includegraphics[width=0.35\textwidth]{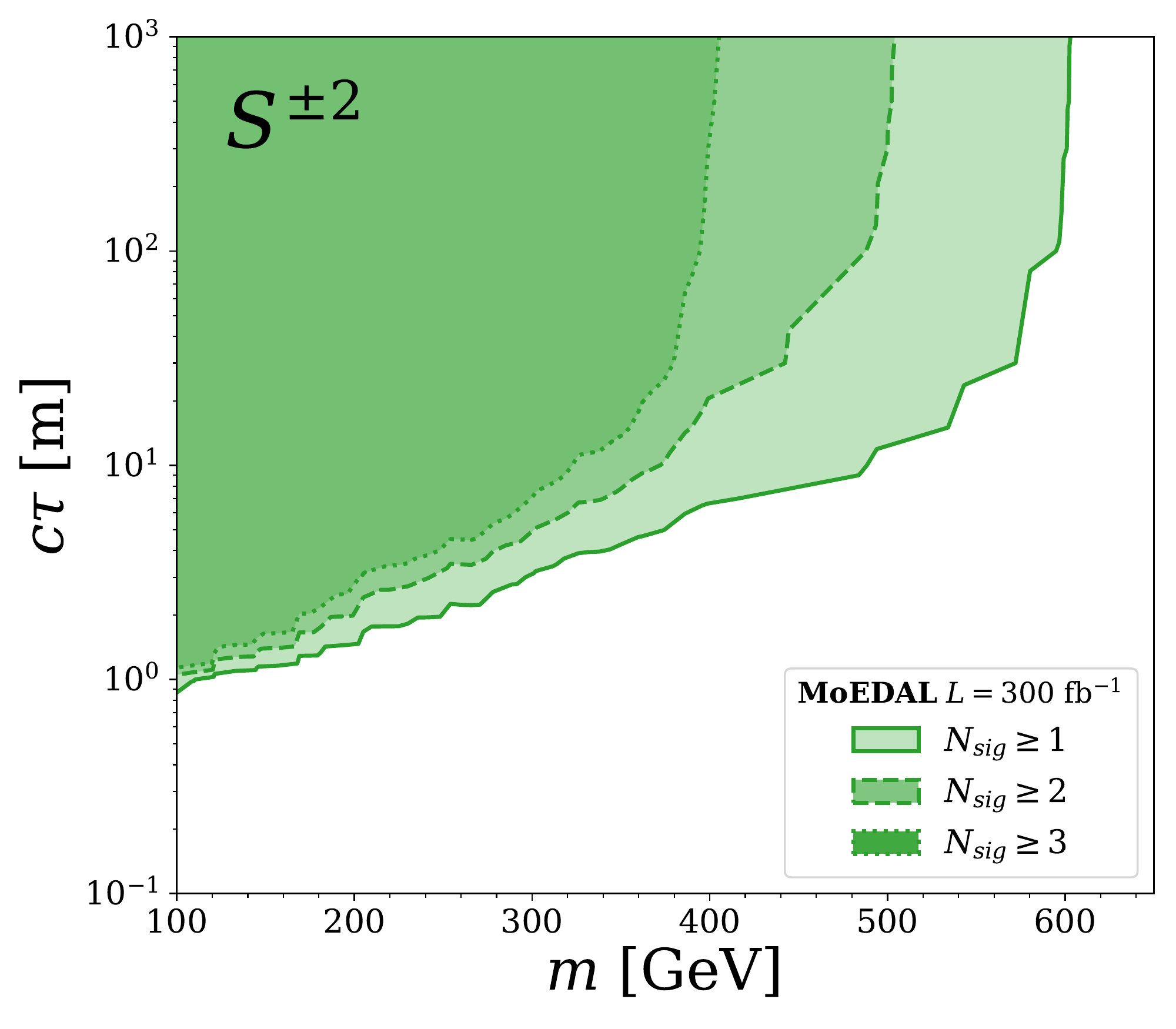}
      \includegraphics[width=0.35\textwidth]{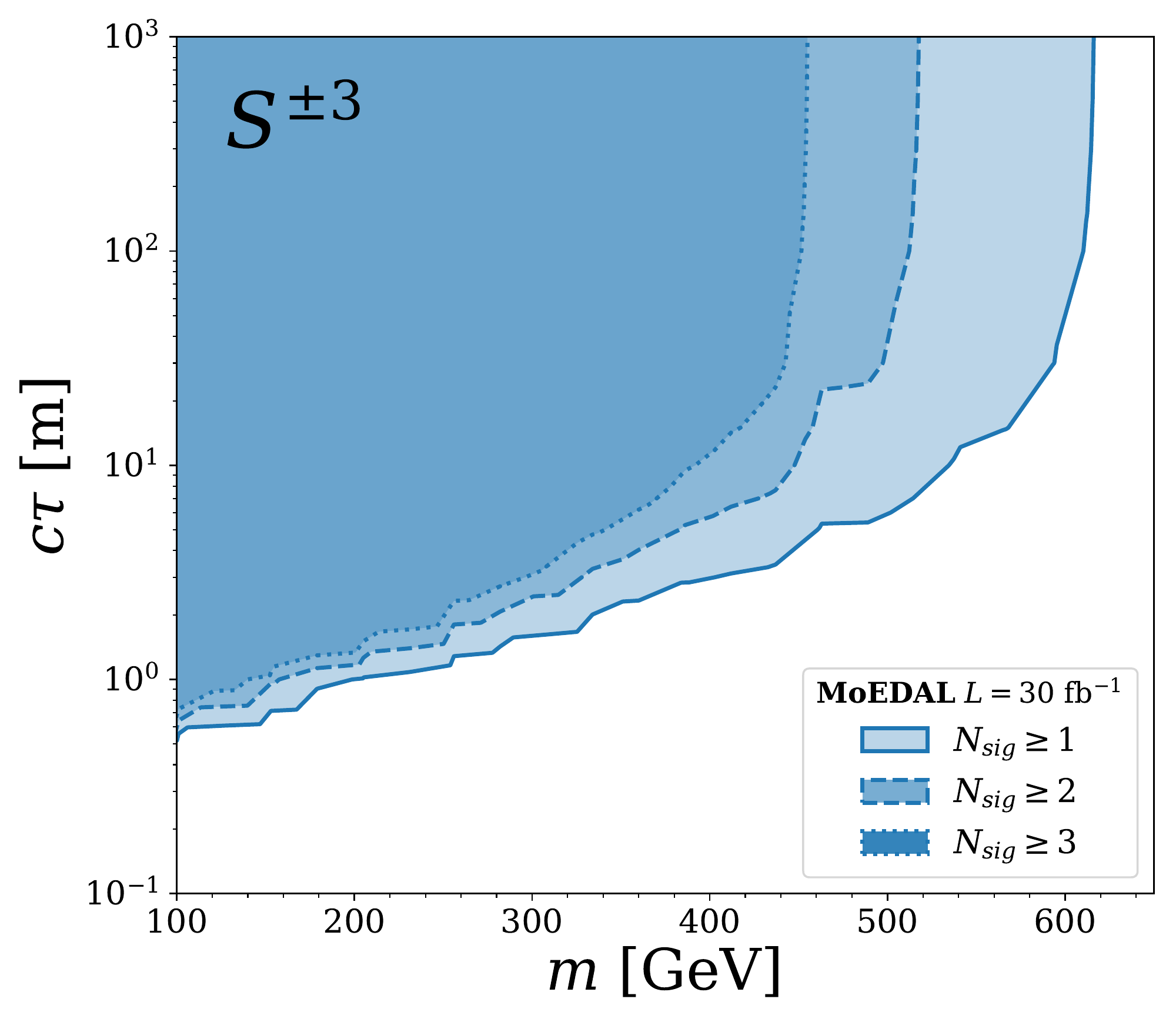}  \hspace{5mm}
      \includegraphics[width=0.35\textwidth]{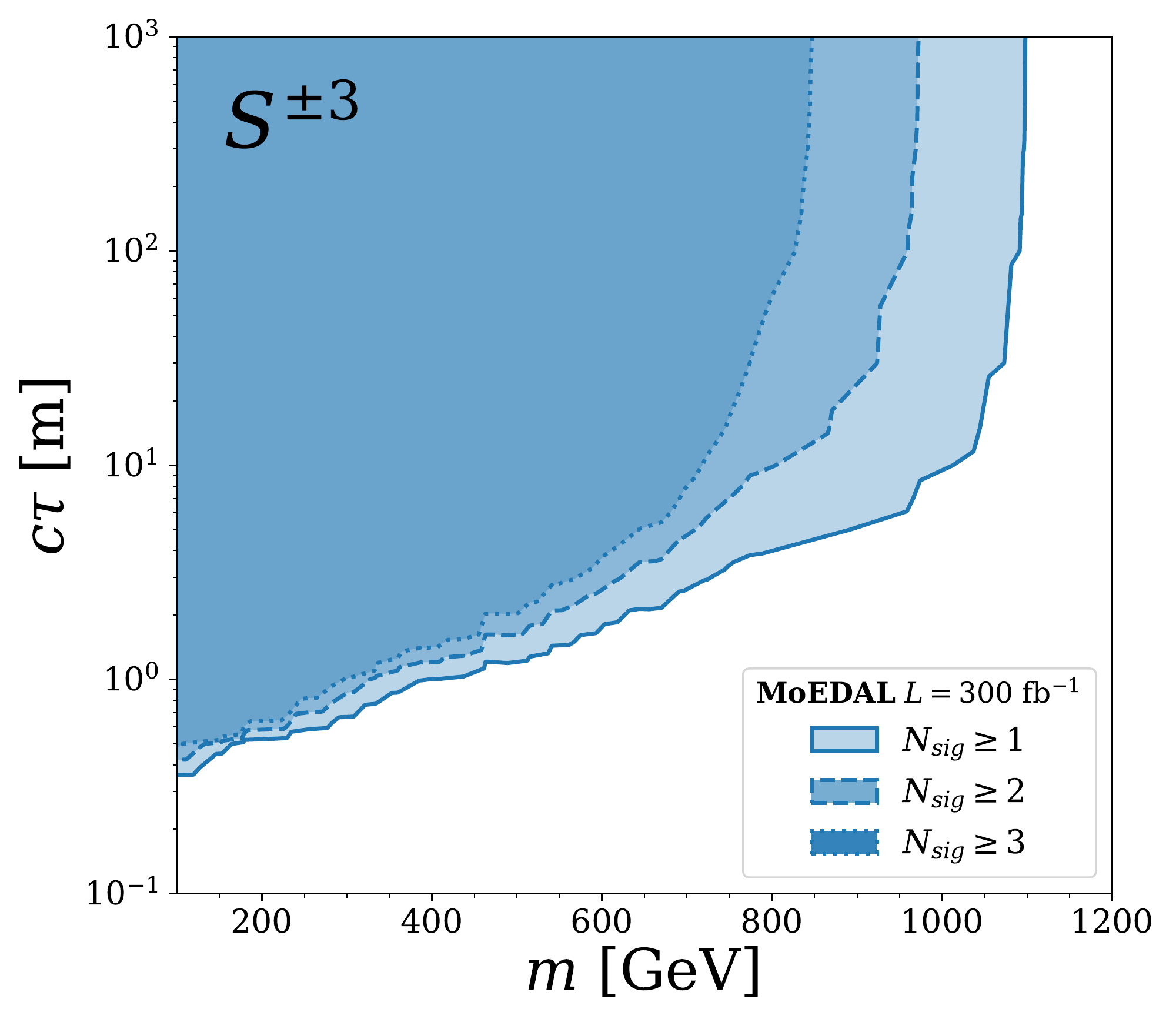}
      \includegraphics[width=0.35\textwidth]{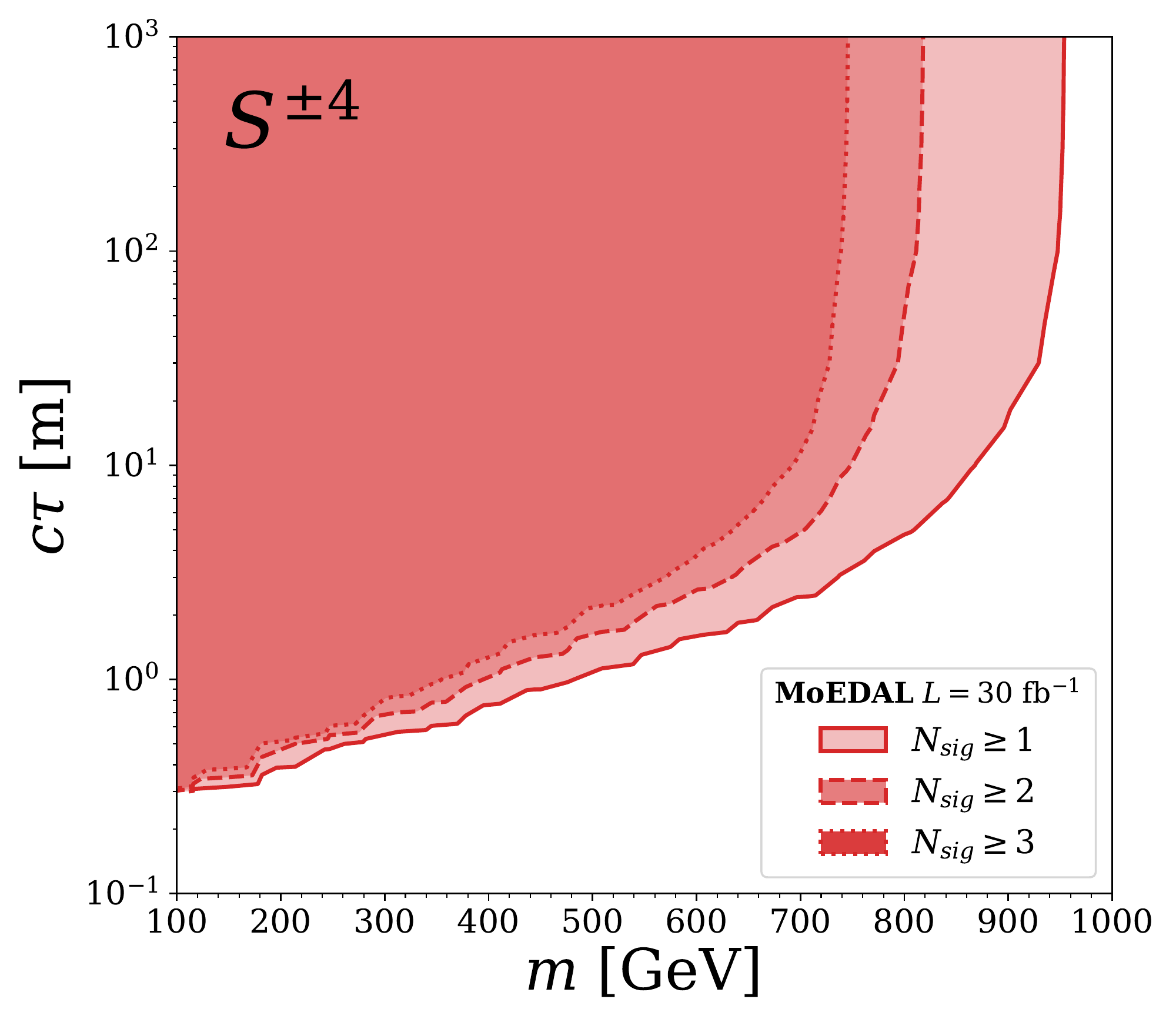}  \hspace{5mm}
      \includegraphics[width=0.35\textwidth]{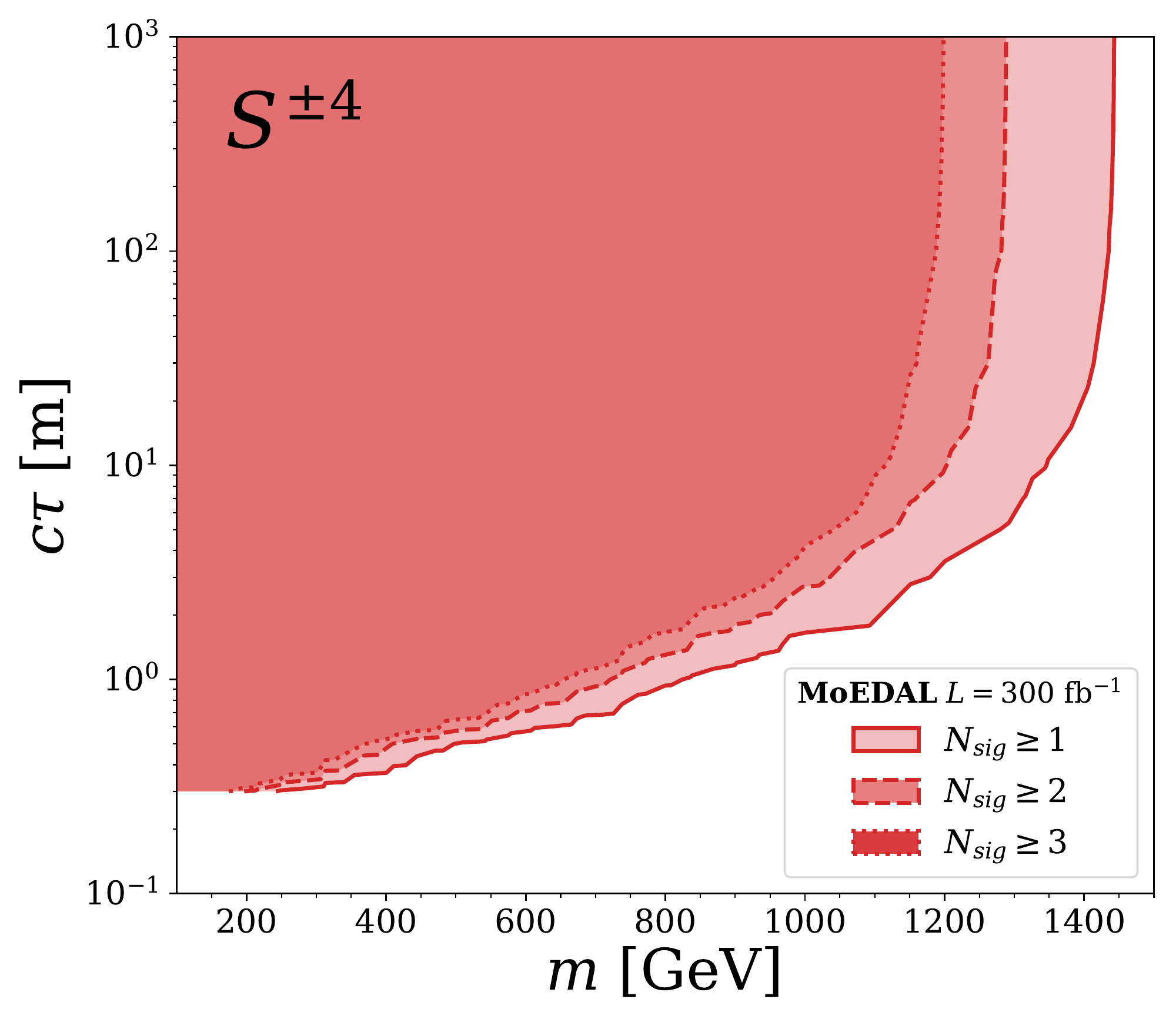}
      \includegraphics[width=0.35\textwidth]{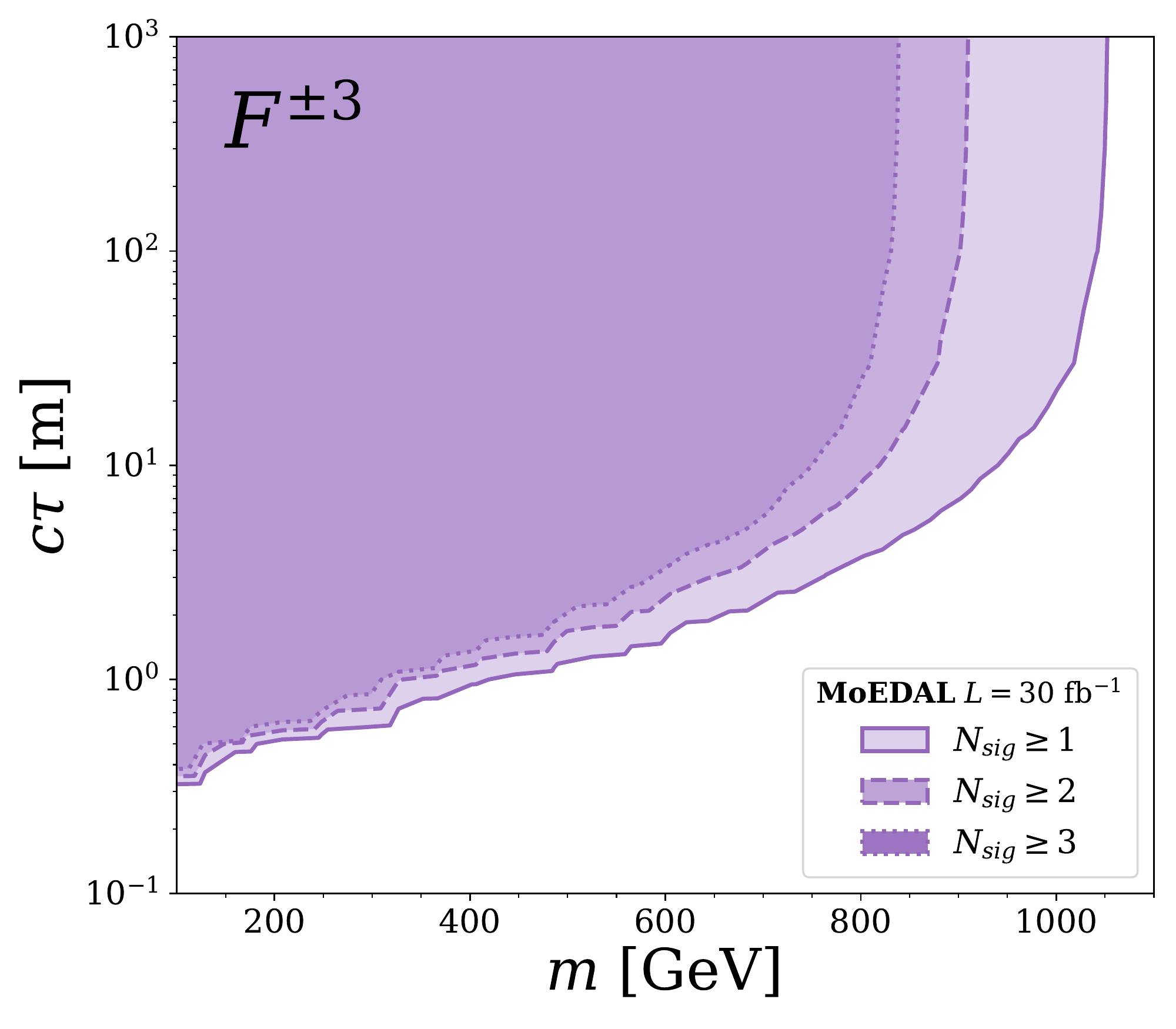}  \hspace{5mm}
      \includegraphics[width=0.35\textwidth]{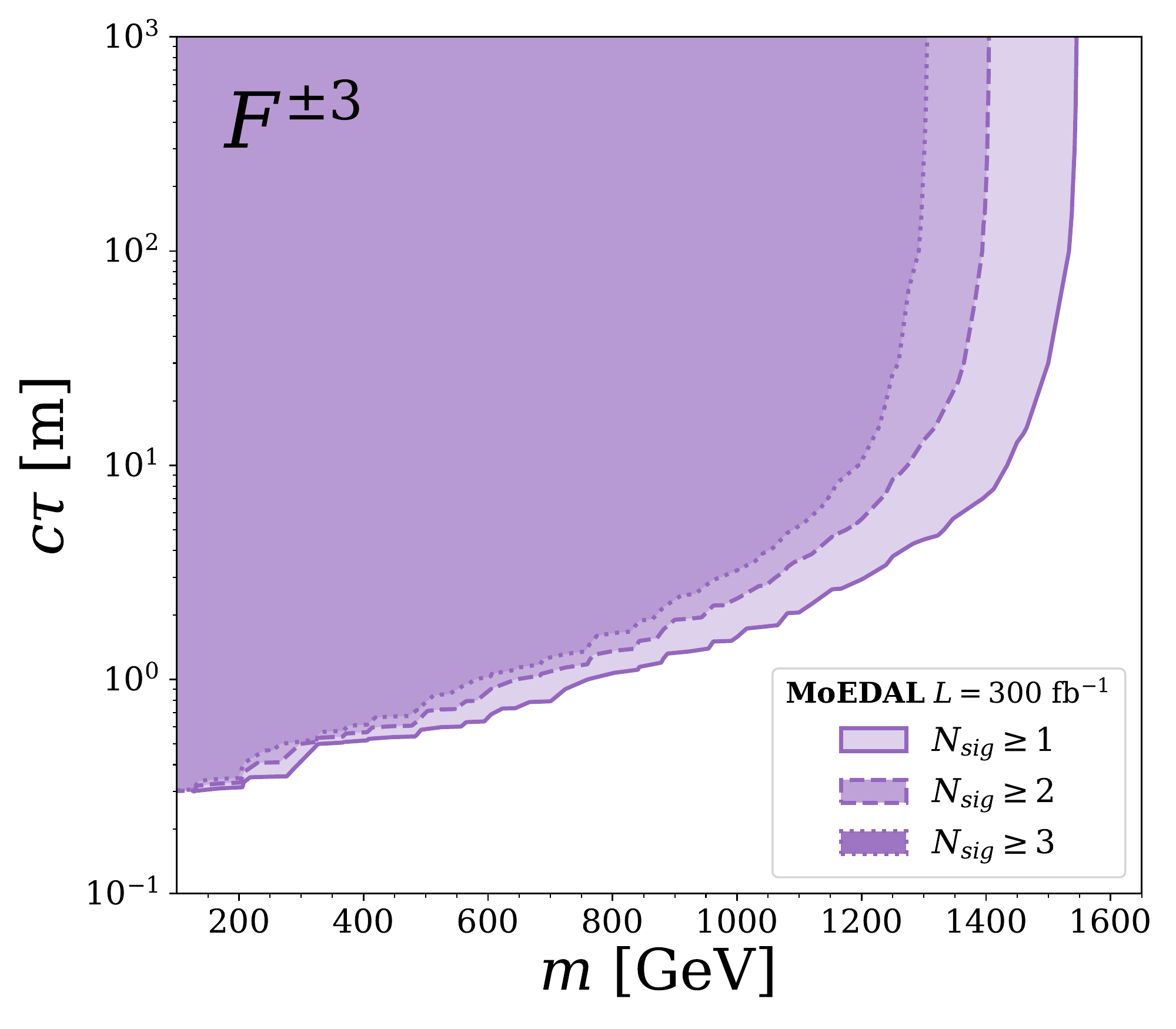}  
\caption{Model-independent detection reach at MoEDAL in the ($m$, $c
  \tau$) parameter plane for individual particles of Model-1.  The
  solid dashed and dotted contours correspond to $N_{\rm sig} = 1$, 2
  and 3, respectively.  In the left (right) panel the integrated
  luminosity of 30 (300) fb$^{-1}$ is assumed.}
\label{fig:lim_1}
\end{figure}
%
%

Fig.~\ref{fig:lim_1} shows the region in the ($m$ vs $c \tau$)
parameter plane in which the expected number of events detected by
MoEDAL, $N_{\rm sig}$, exceeds 1 (solid), 2 (dashed) and 3(dotted)
with the integrated luminosity of 30\,fb$^{-1}$ (left) and
300\,fb$^{-1}$ (right).  These luminosities are expected to be
delivered at the end of Run-3 and High-Luminosity LHC run,
respectively.  In this part of the study we only consider the
particle-antiparticle pair production of a single species, e.g.~$pp
\to S^{+4} S^{-4}$.  In this case the expected number of signal events
depends only on the mass and lifetime of the particle, which we treat
as free parameters.  The results are therefore model-independent.

The top panels in Fig.~\ref{fig:lim_1} show the expected sensitivities
(green regions) for the doubly charged scalar particle $S^{\pm2}$.
Since the expected background at MoEDAL with assumed luminosities is
much less than 1, $N_{\rm sig} = 3$ approximately corresponds to the
$95$\,\% CL exclusion when observing no signal event.  As can be seen,
the mass reach for $N_{\rm sig} = 1$ (3) with $c \tau \gtrsim 100$\,m
amounts to $m_{S^{\pm2}} \simeq 290$ (190) GeV in Run-3 (30 fb$^{-1}$)
This can be compared to the corresponding mass reach $m_{S^{\pm2}}
\simeq 160$ GeV for $N_{\rm sig} = 1$ obtained in the previous study
\cite{Acharya:2020uwc} where the photon fusion process was not taken
into account.  For HL-LHC with 300 fb$^{-1}$, the mass reach is
extended to $m_{S^{\pm2}} \simeq 600$ (400) for $N_{\rm sig} = 1$ (3).

The panels in the second (blue) and third (red) rows of
Fig.~\ref{fig:lim_1} show the expected sensitivities for triply
($S^{\pm3}$) and quadruply ($S^{\pm4}$) charged scalar particles,
respectively.  As can be seen, the mass reaches for $c \tau \gtrsim
100$\,m are significantly extended compared to that of $S^{\pm2}$.  In
the long lifetime limit, MoEDAL with the Run-3 luminosity can probe
$S^{\pm3}$ up to $m_{S^{\pm3}} \simeq 610$ (430) GeV, while for
$S^{\pm4}$ the reach is further extended up to $m_{S^{\pm4}} \simeq
960$ (700) GeV with $N_{\rm sig} = 1$ (3).  At the HL-LHC, the mass
reach for $S^{\pm3}$ is improved to be $1100$ (850) GeV, whilst the
mass reach for $S^{\pm4}$ is found to be 1430 (1200) GeV with $N_{\rm
  sig} = 1$ (3)

Our first neutrino mass model (Model-1) has a triply charged
vector-like fermion pair $F^{\pm3}$.  Although those fermions are not
expected to be long-lived, we show the model-independent MoEDAL
sensitivity for $F^{\pm3}$ for completeness.\footnote{ The MoEDAL
  sensitivity for doubly charged fermions has been studied in
  \cite{Acharya:2020uwc}.  } Since Dirac fermions have twice more
degrees of freedom than complex scalars and their Drell-Yan process is
not $p$-wave suppressed, their production cross-section is
significantly larger than the scalar particles with the same charge.
This leads to greater sensitivity than the scalar case.  With the
integrated luminosity of 30 fb$^{-1}$, we see that MoEDAL can probe
$F^{\pm3}$ up to $m_{F^{\pm3}} \simeq 1030$ (800) GeV for $N_{\rm
  sig}=1$ (3).  For 300 fb$^{-1}$ we see the mass reach improves up to
1550 (1300) GeV for $N_{\rm sig}=1$ (3).

{
One can notice in Fig. \ref{fig:lim_1} that for very high values of decay length $c\tau \geq 100$ m the detection reach for MoEDAL practically does not change. This is a consequence of the experimental setup and applies to all types of particles. In order to be detected at MoEDAL, a particle needs to deposit energy in all layers of the NTD panel, which means it cannot decay before reaching the detector, nor within it. When particle's decay length becomes large, the probability of reaching MoEDAL detector asymptotically approaches unity, and particle can be thought of being quasi-stable. Further increase in $c\tau$ provides little benefit for detection reach, which is now 
determined mostly by the production cross-section, luminosity and particles' velocity distribution, as can be seen from Eq. \eqref{eq:N} and Eq. \eqref{eq:prob}.
}

\begin{table}[t!]
\centering
\renewcommand{\arraystretch}{1.2}
\begin{tabular}{c||c|c||c|c}
           & current HSCP bound & HSCP (Run-3) & MoEDAL (Run-3) & MoEDAL (HL-LHC) \\
           & 36 fb$^{-1}$ \cite{Aaboud:2018kbe} & 300 fb$^{-1}$ \cite{Jager:2018ecz} & 30 fb$^{-1}$ & 300 fb$^{-1}$ \\        
\hhline{=||=|=||=|=}
$S^{\pm2}$ & ((650))  & -- & 190 (290) & 400 (600)   \\
\hline
$S^{\pm3}$ & ((780))  & -- & 430 (610) & 850 (1100)   \\
\hline
$S^{\pm4}$ & ((920))  & -- & 700 (960) & 1200 (1430)   \\
\hline
$F^{\pm3}$ & 1130   & 1500  & 800 (1030) & 1300 (1550)   \\
\end{tabular}
\caption{Summary for the model-independent mass reaches (in GeV) of
  the multi-charged particles in Model-1 by MoEDAL (2nd and 3rd
  columns).  The first column shows the current mass bounds from the
  ATLAS analysis \cite{Aaboud:2018kbe} with 36 fb$^{-1}$.  The numbers
  in the double-brackets correspond to our naive estimate of the mass
  bounds (see the text).  The second column represents the projected
  mass reach in Run-3 (300 fb$^{-1}$) obtained in
  \cite{Jager:2018ecz}.  The numbers outside (inside) the brackets in
  the third and fourth columns represent MoEDAL's mass reaches with
  $N_{\rm sig} \geq 3$ (1) assuming $L = 30$ (Run-3) and 300 (HL-LHC)
  fb$^{-1}$, respectively.  }
\label{tab:sum}
\end{table}

In Table \ref{tab:sum} we summarise the model-independent mass reaches
(in GeV) for the multi-charged particles in Model-1 obtained in this
subsection and compare these results with the current bounds and
future projections from the heavy stable charged particle (HSCP)
searches by ATLAS and CMS (if available).  The current mass bounds are
taken (estimated) from the ATLAS analysis \cite{Aaboud:2018kbe} with
the 36 fb$^{-1}$ data.  In the ATLAS analysis \cite{Aaboud:2018kbe},
the result was interpreted only for fermionic particles.  We estimated
the bounds on the scalar particles, $S^{\pm2}$, $S^{\pm3}$ and
$S^{\pm4}$, by naively imposing the cross-section bounds for fermionic
particles on the cross-sections of our scalar particles in Model-1,
assuming that the ATLAS' signal efficiency is not very sensitive to
the spins.  Numbers in double-brackets in the first column are
obtained in this way and should be taken with a grain of salt.  To our
knowledge the Run-3 projection is available only for fermionic
particles \cite{Jager:2018ecz}.  In \cite{Jager:2018ecz}, the
projected mass reach for $F^{\pm3}$ for Run-3 LHC was estimated at 1500
GeV.  The numbers outside (inside) the brackets in the third and
fourth columns represent MoEDAL's mass reaches with $N_{\rm sig} \geq
3$ (1), assuming $L = 30$ (Run-3) and 300 (HL-LHC) fb$^{-1}$,
respectively.

As can be seen in Table \ref{tab:sum} we do not have very good hope to
observe Model-1 particles at MoEDAL in Run-3 LHC, since the majority
of the region in which MoEDAL is sensitive has already been excluded
(except for a small ($\sim 40$ GeV) range for $S^{\pm 4}$).  The
situation is brighter at HL-LHC.  There, MoEDAL can explore $S^{\pm
  3}$, $S^{\pm 4}$ and $F^{\pm 3}$ in the regions that are allowed by
the current constraints from the HSCP searches.

\subsection{Interpretation of the results for Model-1}

In this subsection, we investigate the MoEDAL sensitivity for the
neutrino mass model with the non-coloured BSM sector described in the
previous section.  In the model-independent study presented in the
previous subsection, we treated the lifetimes of various particles as
free and independent parameters.  On the other hand, in this
subsection the lifetimes are calculated as functions of model
parameters which fit the experimental data about the neutrino masses
and mixing angles.

The Lagrangian of Model-1 is given in Eq.~\eqref{eq:L1}.  Among many
dimensionless parameters, phenomenologically important ones are
$(h_{ee})_{ij}$, $\lambda_5$ and the product $h_F h_{\bar F}$.  In the
following analysis, we assume only the 1-1 component of
$(h_{ee})_{ij}$ is non-zero for simplicity, but the extension to other
cases are trivial and will be discussed later.  The $h_{ee}$
($=(h_{ee})_{11}$) and $\lambda_5$ couplings control the $L$-violating
decays; $S^{+4} \to W^+ W^+ \ell^+ \ell^+$, $S^{+3} \to W^+ \ell^+
\ell^+$ and $S^{+2} \to \ell^+ \ell^+$, while the decay rate for the
$L$-conserving ones; $S^{+4} \to 4 \ell^+$, $S^{+3} \to \nu \ell^+
\ell^+ \ell^+$ and $S^{+2} \to \nu \nu \ell^+ \ell^+$, are determined
by $h_{ee}$ and $h_F h_{\bar F}$ (see Fig.~\ref{fig:dec}).  The
$L$-violating decays depend on the mass parameters $m_{S_3}$ and
$m_{S_1}$, while $L$-conserving ones depend also on $m_{F_i}$
($i=1,2,3$).  The above decay modes are the dominant ones and the
multi-charged particles can be long-lived when $m_{S_3} < m_{S_1},
m_{F_i}$.  We therefore assume this mass hierarchy and fix 
$m_{F_1} = 3$ TeV throughout this section for
simplicity.

The dimensionless couplings and mass parameters relevant to the decays
are also important for determination of the neutrino masses.  The
approximate formula for the neutrino masses is given in
Eq.~\eqref{eq:mnu}.  In the following numerical analysis, we perform a
parameter fit for the neutrino data and choose $h_F$ and $h_{\bar F}$
for the given $\lambda_5$ and mass parameters of the model, which will
be used in the calculation of the lifetimes and the expected signal
events in the MoEDAL detector.

\begin{figure}[t!]
\centering
      \includegraphics[width=0.46\textwidth]{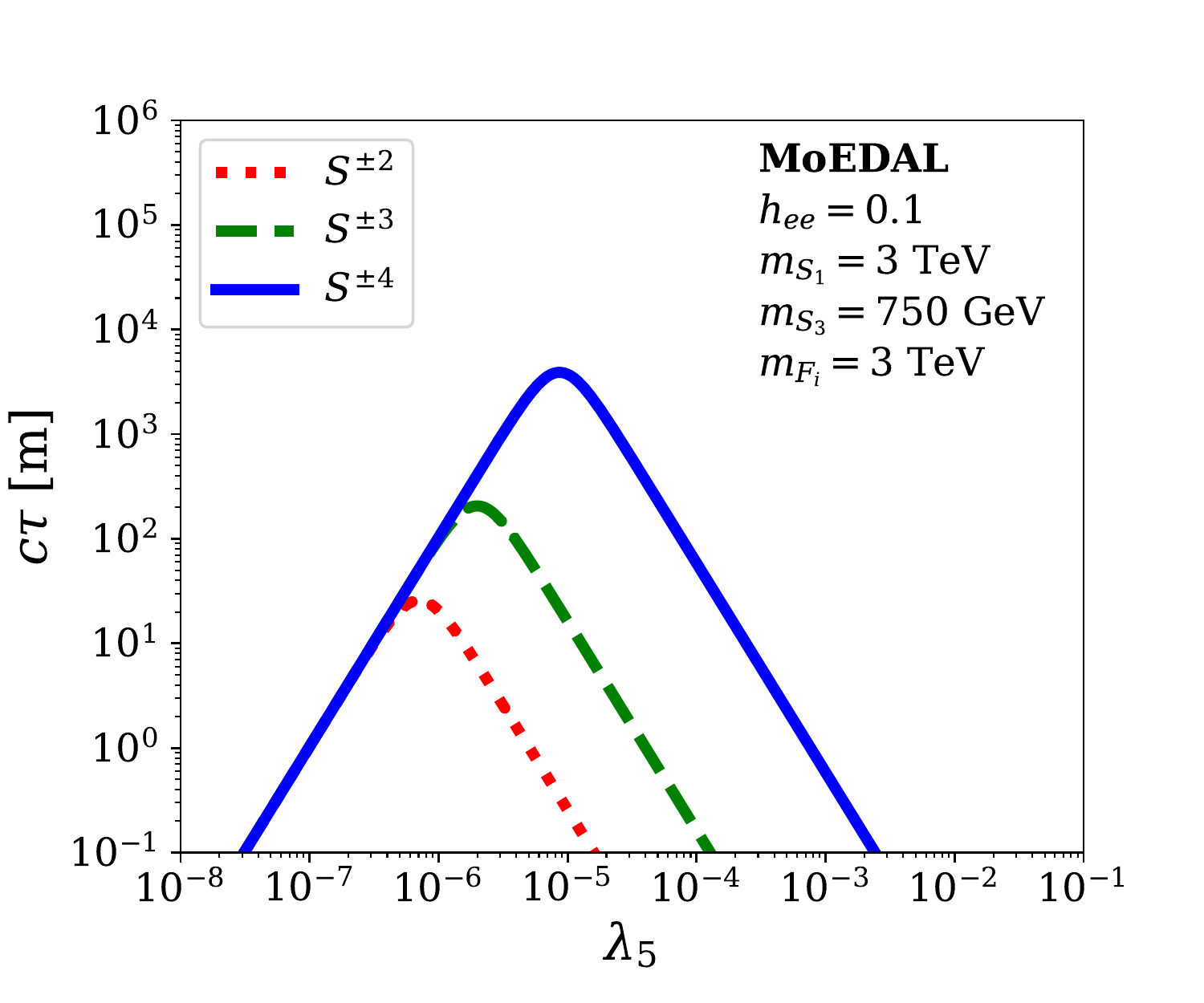}
      \includegraphics[width=0.46\textwidth]{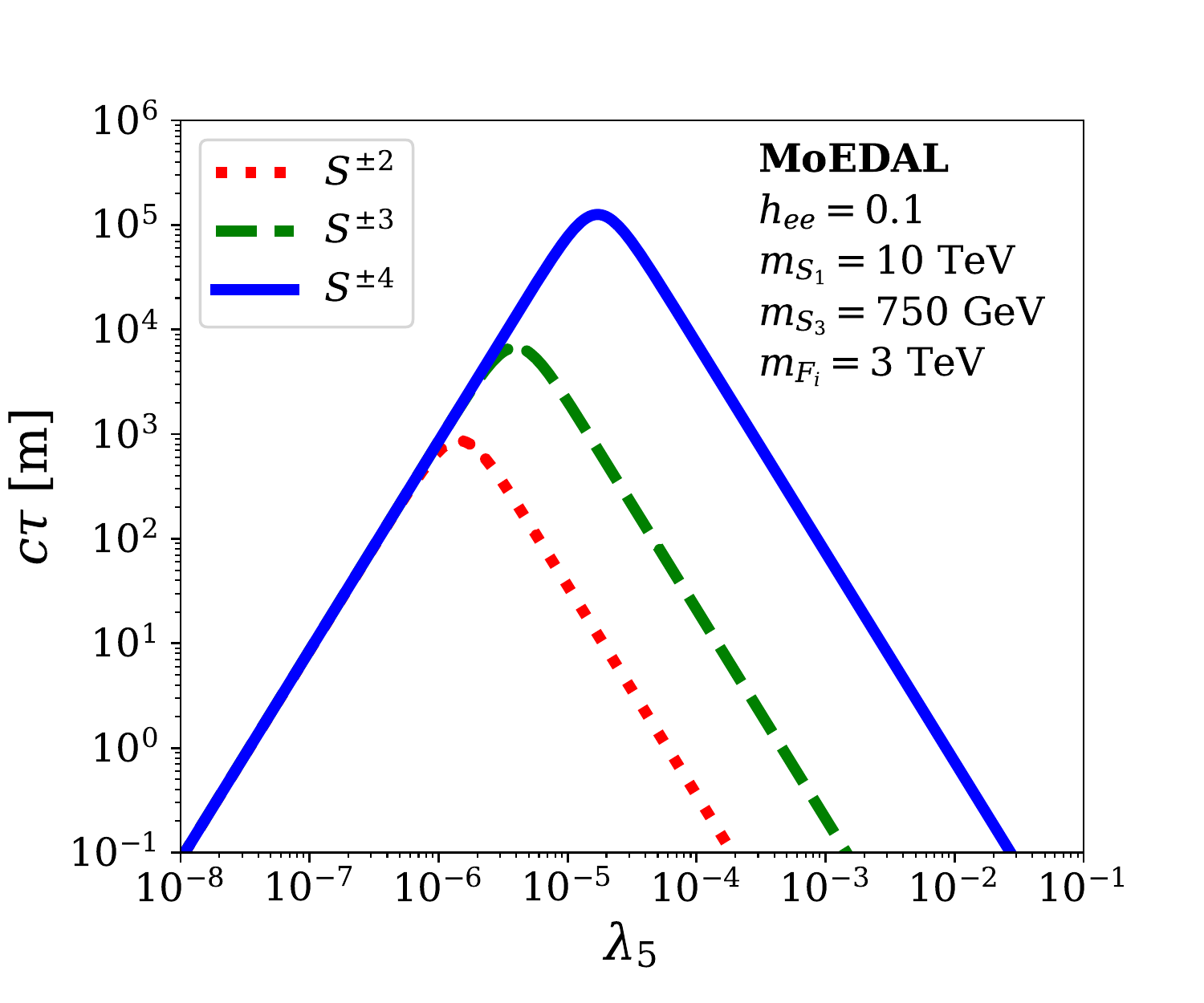}
\caption{Lifetimes of multi-charged particles, $S^{\pm4}$ (blue solid)
  $S^{\pm3}$ (green dashed) and $S^{\pm2}$ (red dotted), in Model-1 as
  functions of $\lambda_5$, where $h_F$ and $h_{\bar F}$ are fitted to
  the neutrino data.  The $m_{S_1}$ is taken to be 3 and 10 TeV in the
  left and right plots, respectively.  The other parameters are fixed
  as $h_{ee} = 0.1$, $m_{S_3} = 750$ GeV and $m_{F_i} = 3$ TeV ($i =
  1,2,3$).  }
\label{fig:lifetime}
\end{figure}

In Fig.~\ref{fig:lifetime} we show the lifetimes of $S^{\pm4}$ (blue
solid), $S^{\pm3}$ (green dashed) and $S^{\pm2}$ (red dotted) as a
function of $\lambda_5$, where $h_F$ and $h_{\bar F}$ are chosen to
fit the neutrino masses and mixing angles.  We take $m_{S_1} = 3$ TeV
in the left panel and 10 TeV in the right.  The other masses are fixed
as $m_{S_3} = 750$ GeV and $m_{F_i} = 3$ TeV.  The lifetimes are
calculated taking $(h_{ee})_{11} = 0.1$ for simplicity, while the
lifetimes for other values of $(h_{ee})_{ij}$ can be obtained by
simple rescaling since the lifetimes are proportional to $(\sum_{ij}
|(h_{ee})_{ij}|^2)^{-1}$.  In both plots, the black horizontal lines
indicates $c \tau = 1$\,m, which is the typical lifetime required for
detection at the MoEDAL.

On can see from the plots that in the region with $\lambda_5 \gtrsim
10^{-4}$, the lifetimes increase as $\lambda_5$ decreases.  This is
because in this region the $L$-violating diagrams, which is
proportional to $\lambda_5$ at the amplitude level, are dominant.  On
the other hand, in the region with $\lambda_5 \lesssim 10^{-6}$, the
response is opposite and the lifetimes decrease as $\lambda_5$
decreases.  This is due to the fact that in this region the dominant
decays are the $L$-conserving modes, proportional to $h_F$ and
$h_{\bar F}$, while the product $h_F h_{\bar F}$ is inversely related
to $\lambda_5$ through the neutrino mass fit.  In the region where the
$L$-conserving processes are dominant, the lifetimes of $S^{\pm 4}$,
$S^{\pm 3}$ and $S^{\pm 2}$ coincide.  This is anticipated since these
decays are related by $SU(2)_L$ and share the same formulae of decay
rates neglecting small effects of electroweak symmetry breaking, as
discussed in section \ref{sec:models}.  The lifetimes of multi-charged
particles are peaked at particular values of $\lambda_5$, at which the
sizes of $L$-conserving and $L$-violating decays become comparable.
This happens around $\lambda_5 \sim 10^{-5}$, $3\times 10^{-6}$ and
$10^{-6}$ for $m_{S^{\pm 4}}$, $m_{S^{\pm 3}}$ and $m_{S^{\pm 2}}$,
respectively, with a slight dependence of $m_{S_1}$ as can be seen
from the two plots in Fig.~\ref{fig:lifetime}.

\begin{figure}[t!]
\centering
      \includegraphics[width=0.4\textwidth]{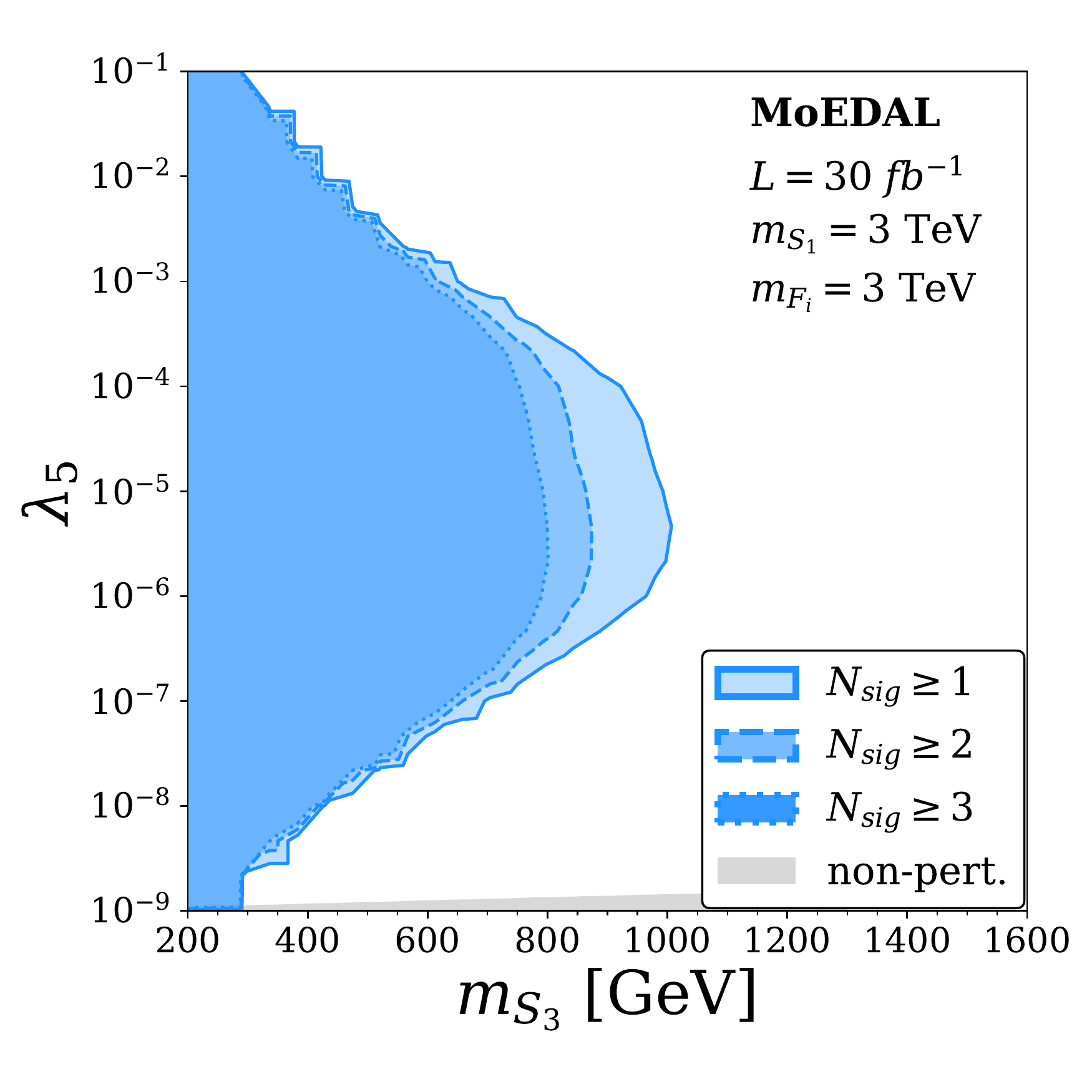} \hspace{5mm}
     \includegraphics[width=0.4\textwidth]{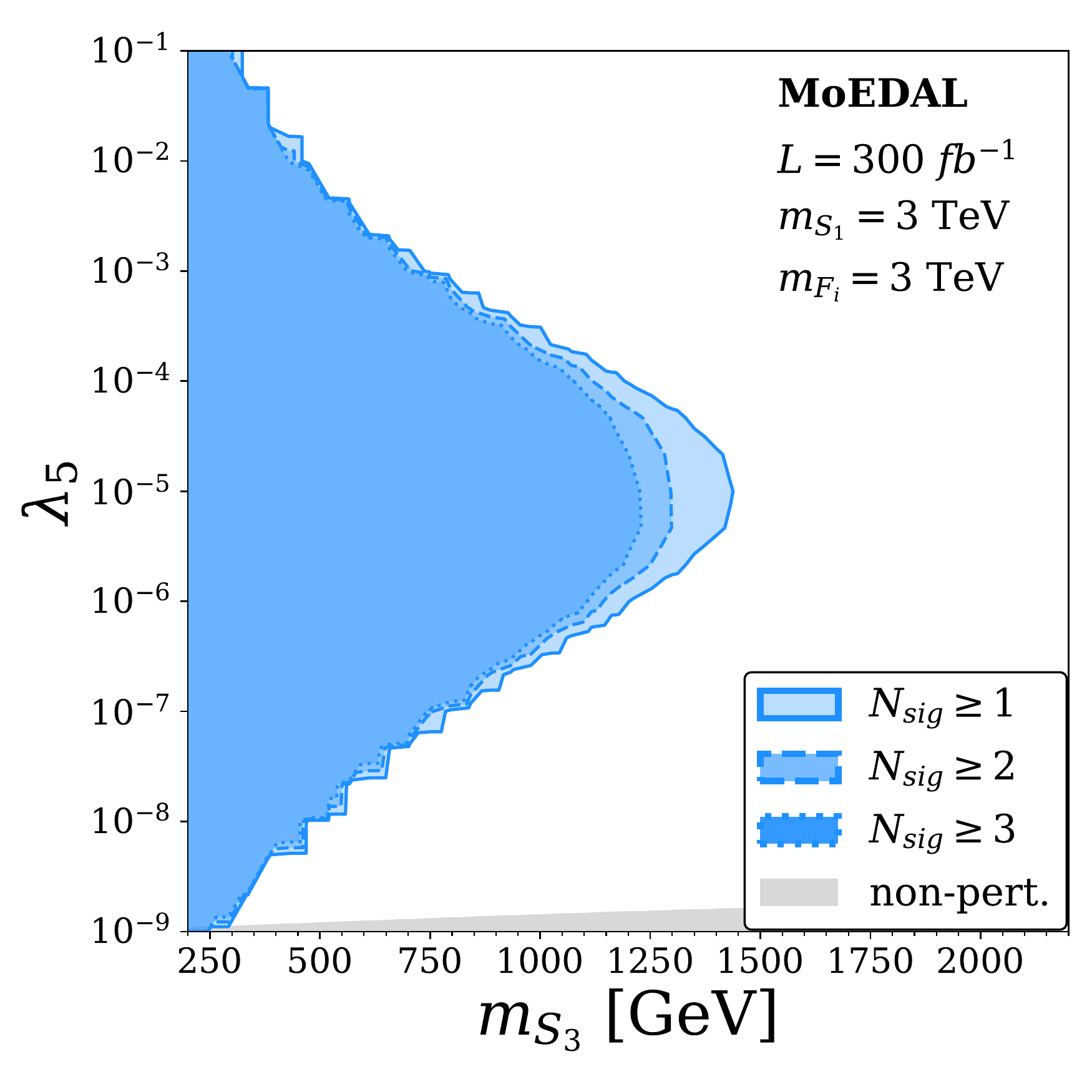}
     \includegraphics[width=0.4\textwidth]{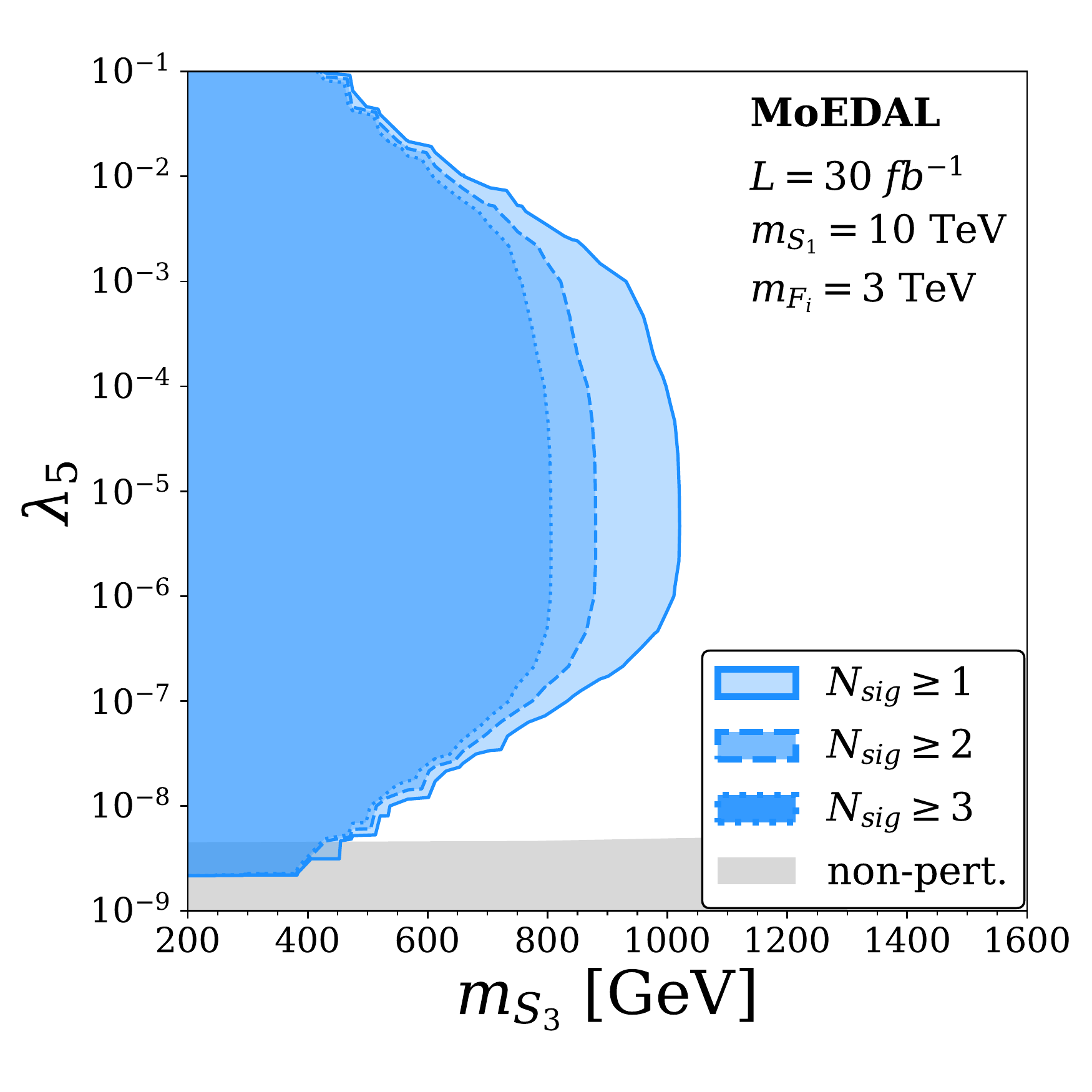} \hspace{5mm}
     \includegraphics[width=0.4\textwidth]{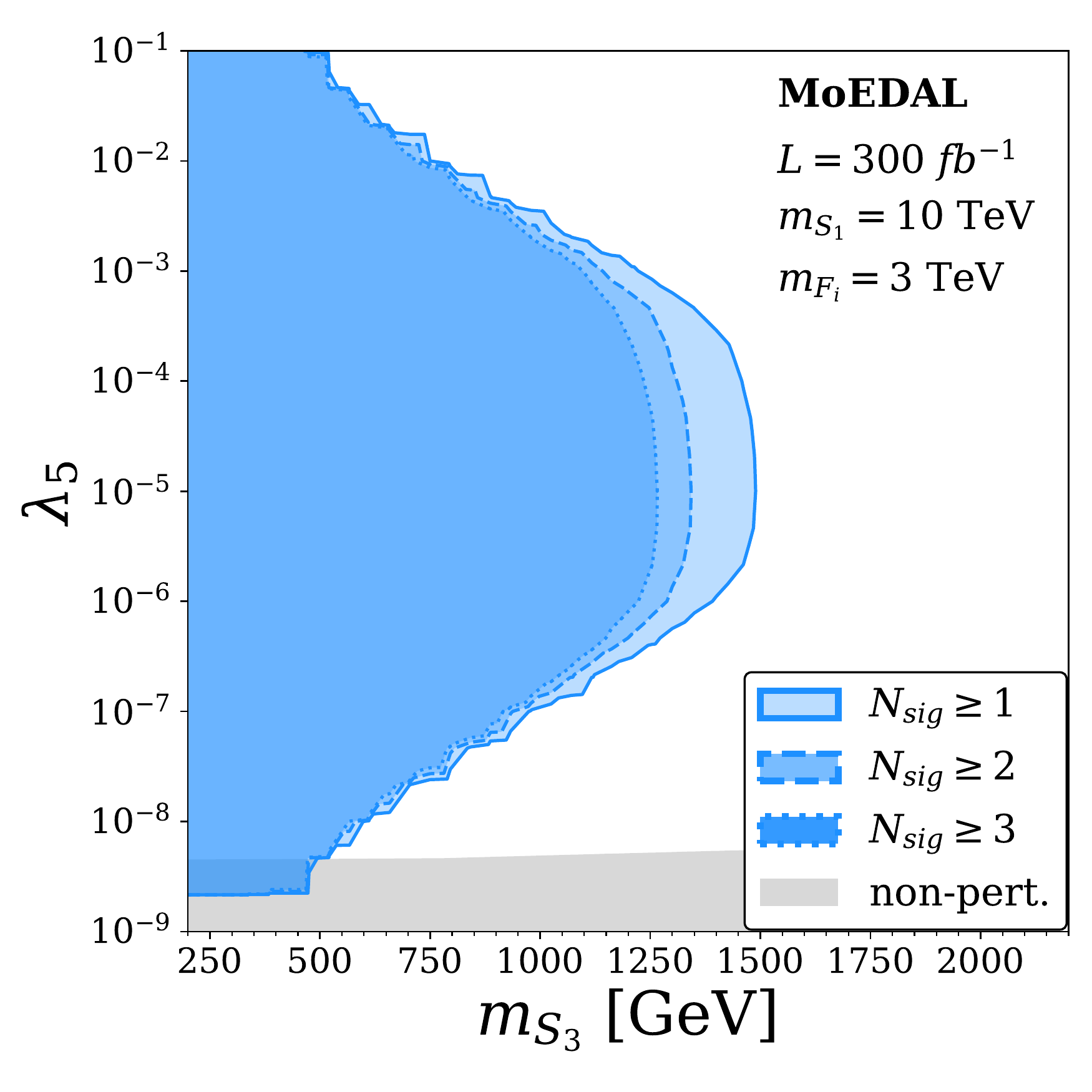}
\caption{MoEDAL's detection sensitivity for Model-1 in the ($m_{S_3}$,
  $\lambda_5$) plane.  In the regions inside the solid, dashed and
  dotted contours, MoEDAL expects to observe more than 1, 2 and 3
  events, respectively.  The region with $N_{\rm sig} \geq 3$ will be
  excluded at 95\% CL if MoEDAL does not observe any signal event.
  The left and right panels correspond to $L = 30$ and 300 fb$^{-1}$,
  respectively.  We take $m_{F_i} = 3$ TeV ($i = 1,2,3$) for all plots
  and $m_{S_1} = 3$ (10) TeV in the top (bottom) panels.  The $h_{F}$
  and $h_{\bar F}$ are fitted to the neutrino data.  In the grey
  region one of these couplings tend to be non-perturbative; ${\rm
    max}_{ij} \left( |(h_{F})_{ij}|, |(h_{\bar F})_{ij}| \right) \geq
  2$.  }
\label{fig:nsig}
\end{figure}

In Fig.~\ref{fig:nsig} we show the MoEDAL sensitivity for Model-1 in
the ($m_{S_3}$, $\lambda_5$) parameter plane.  The two plots on the
left assume an integrated luminosity of 30 fb$^{-1}$ (Run-3), while
in the right plots 300 fb$^{-1}$ (HL-LHC) is assumed.  Here we take
all production modes into account including associated productions $pp
\to S^{\pm 4} S^{\mp 3}$ and $S^{\pm 3} S^{\mp 2}$.  We set $m_{F_i} =
3$ TeV for all plots, while $m_{S_1}$ is taken to be 3 (10) TeV for
the top (bottom) plots.  In the region shaded with grey, around
$\lambda_5 \sim 10^{-9}$, ${\rm max}_{ij} \left( |(h_{F})_{ij}|,
|(h_{\bar F})_{ij}| \right) \geq 2$, so the model tends to be
non-perturbative.

As we see in the top right plot with $m_{S_1} = 3$ TeV and $L = 30$
fb$^{-1}$, MoEDAL can probe the model up to $m_{S_3} = 1000$ (800) GeV
with $N_{\rm sig} = 1$ (3) for $\lambda_5 \sim 10^{-5}$, at which the
lifetime of $S^{\pm 4}$ is maximized.  The reach of $m_{S_3}$ is
degraded to $\sim 600$ GeV for both $N_{\rm sig} = 1$ and 3 if
$\lambda_5$ is taken around $10^{-3}$ or $10^{-7}$.  For a larger
luminosity $L = 300$ fb$^{-1}$, the sensitivity is improved up to
$\sim 1450$ (1250) GeV for $N_{\rm sig} = 1$ (3) at $\lambda_5 \sim
10^{-5}$.  For different values of $\lambda_5$ the sensitivities are
degraded.  For example, MoEDAL can probe the model only up to $m_{S_3}
\sim 750$ GeV if $\lambda_5 \sim 10^{-3}$ or $10^{-7}$.
Moving to the bottom plots with $m_{S_1} = 10$ TeV, we see that the
contours of $N_{\rm sig} = 1$ and 3 become more flat and the highest
mass reaches become more stable for a variation of $\lambda_5$.  This
is because the decay rates get extra suppression due to large
$m_{S_1}$, and $c \tau$ comes down to the $\sim 10$\,m range only for
$\lambda_5 \lesssim 10^{-7}$ or $\gtrsim 10^{-3}$ as can be seen in
the right plot of Fig.~\ref{fig:lifetime}.  In the range $10^{-6}
\gtrsim \lambda_5 \gtrsim 10^{-4}$, MoEDAL can probe the model up to
$m_{S_3} \sim 1000$ (800) GeV for $N_{\rm sig} = 1$ (3) with $L=30$
fb$^{-1}$ (Run-3).  For the larger luminosity $L=300$ fb$^{-1}$
(HL-LHC), the reach is extended to 1500 (1250) GeV for $N_{\rm sig} =
1$ (3).  In all plots the maximum reaches for $m_{S_3}$ roughly
correspond to the maximum reaches for $m_{S^{\pm 4}}$ in the previous
subsection.

\begin{figure}[t!]
\centering
    \includegraphics[width=0.4\textwidth]{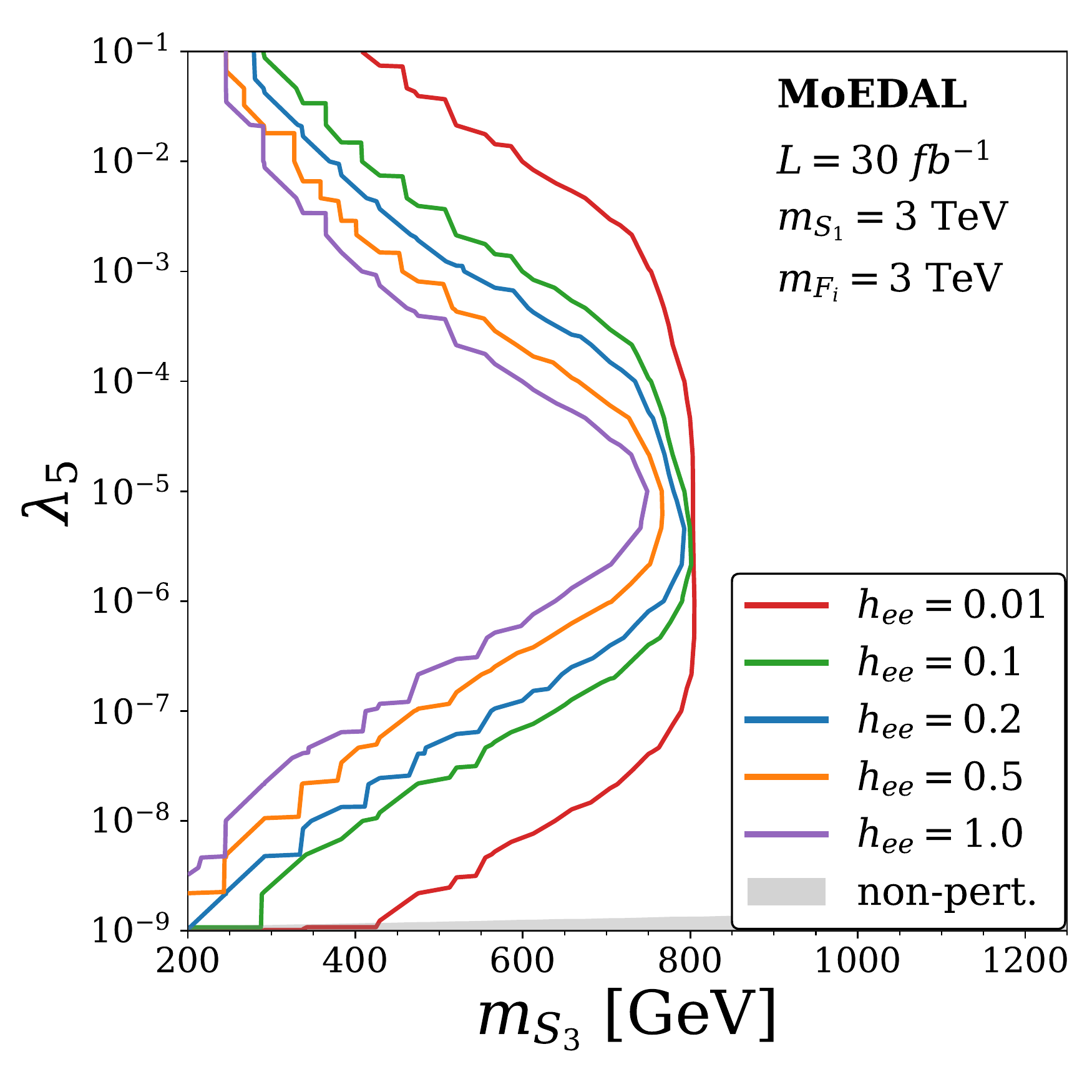} \hspace{5mm}
    \includegraphics[width=0.4\textwidth]{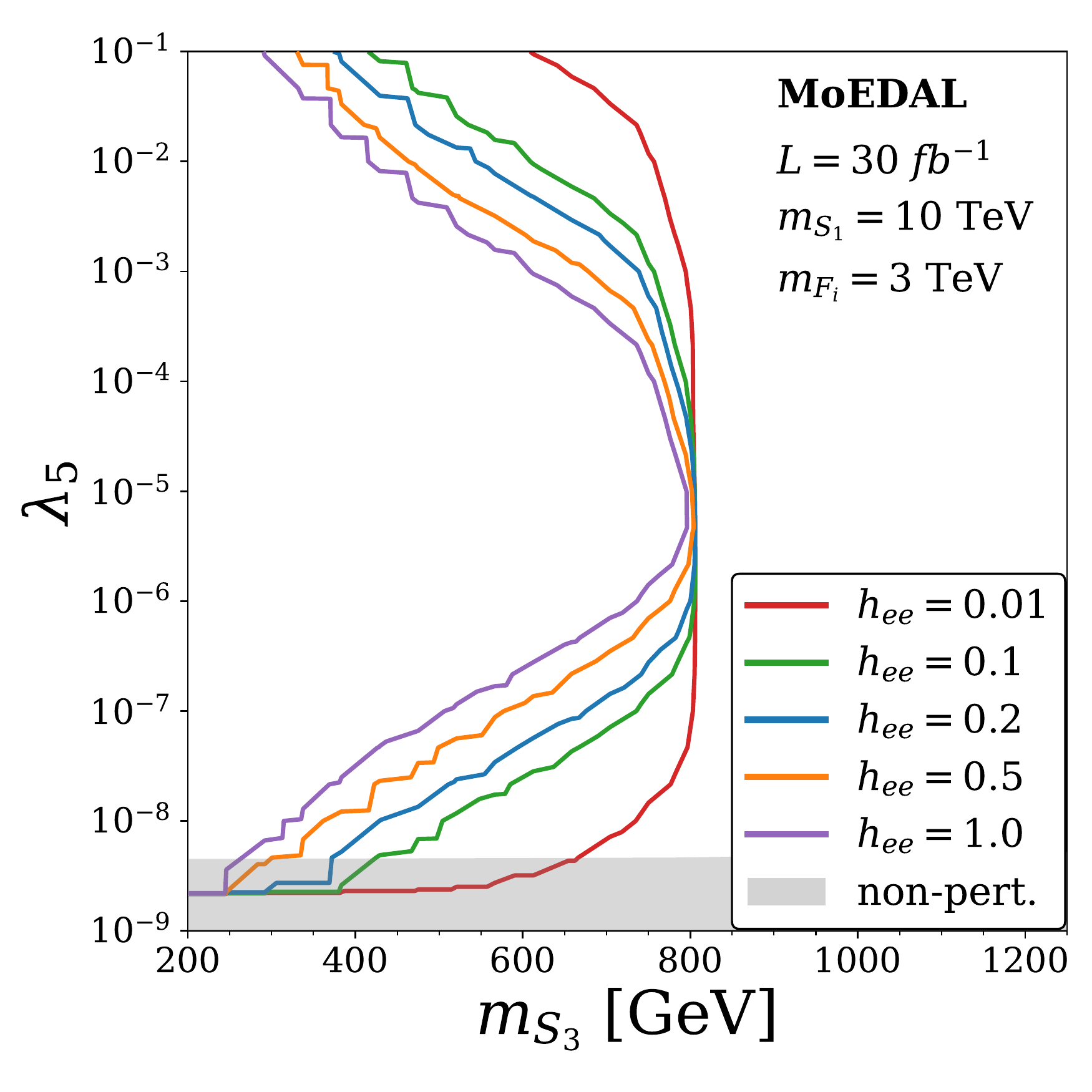}
\caption{$N_{\rm sig} = 1$ contours for different values of $h_{ee}$ in Model-1.
We take $m_{S_1} = 3$ and 10 TeV in the left and right plots, respectively.
We assume $L = 30$ fb$^{-1}$ and $m_{F_i} = 3$ TeV ($i = 1,2,3$). 
}
\label{fig:hee}
\end{figure}

We have so far fixed $h_{ee}$ to be 0.1.  In Fig.~\ref{fig:hee} we
show the impact of $h_{ee}$ on the detectability of the model at
MoEDAL.  In the plots, different curves represent $N_{\rm sig} = 1$
contours with different values of $h_{ee}$.  As previously, we fix
$m_{F_i} = 3$ TeV and take $m_{S_1} = 3$ and 10 TeV in the left and
right plots, respectively.  Both $L$-conserving and $L$-{violating}
decays are proportional to $h_{ee}$ at the amplitude level and the
lifetimes of multi-charged particles in the $S_3$ triplet are
inversely proportional to $|h_{ee}|^2$.  In the left plot with
$m_{F_i} = 3$ TeV, we see that the maximal reach of $m_{S_3}$ becomes
less sensitive to $\lambda_5$ for smaller $h_{ee}$, since the decay
rates are suppressed for smaller $h_{ee}$.  In the right plot with
$m_{S_1} = 10$ TeV this tendency is even stronger since the decay
rates have extra suppression due to the larger value of $m_{F_i}$.


\section{Numerical analysis for coloured models}
\label{sec:model-2}

\subsection{Results for colour-triplet multi-charged particles}

In this section we study the second model (Model-2) of neutrino mass
generation, which we have described in section~\ref{sec:models}.  The
model is obtained from Model-1 by changing the $SU(3)_c$
representation of particles from singlet to (anti-)triplet ($S \to
\tilde S$ and $F \to \tilde F$) as is shown in Eq.~\eqref{eq:replace}.
Another modification is the BSM parity breaking term in the
Lagrangian; $(h_{ee})_{ij} e^c_i e^c_j S_1^\dagger \to (h_{ed})_{ij}
d^c_i e^c_j S_1^\dagger$.  This is necessary for gauge invariance and
making the lightest BSM particle unstable.  With these modifications,
the Lagrangian of Model-2 is the same as that of Model-1 and the
relevant decays are:
\beqn
\begin{array}{lcll}
& & L~{\rm conserving}, & L~{\rm violating} \\
\tilde S^{+10/3} & \to & \ell^+ \ell^+ \ell^+ \bar d, & W^+ W^+ \ell^+ \bar d \\
\tilde S^{+7/3} & \to & \nu \ell^+ \ell^+ \bar d, & W^+ \ell^+ \bar d  \\
\tilde S^{+4/3} & \to & \nu \nu \ell^+ \bar d, & \ell^+ \bar d 
\end{array}
\label{eq:dec_col}
\eeqn
We call the first decay modes in Eq.~\eqref{eq:dec_col} (with
neutrinos or without $W$s) $L$-conserving and the second decay modes
$L$-violating, assigning $L=3$ for $\tilde S_3$.  With the replacement
$\ell_\alpha^+ \to \bar d_\alpha$ and $(h_{ee})_{\alpha \beta} \to
(h_{ed})_{\alpha \beta}$, the decay rate formulae for those decays are
unchanged from those of the corresponding decays in Model-1, which are
found in section~\ref{sec:models}.

There is one possible complication for decays of Model-2 particles.
If, for example, $\tilde S^{10/3}$ forms a hadron together with a
$d$-quark, by moving the $\bar d$ in the final state into the initial
state hadron, one can consider a 3-body decay $(\tilde S^{+10/3}
d)_{\rm had} \to \ell^+ \ell^+ \ell^+$.  However, a dimensional
analysis tells us that the hadronic 3-body decay is subdominant.  For
example for $m_{\tilde S_3} = 600$\,GeV we estimate
\beq
\frac{\Gamma[(\tilde S^{+10/3} d)_{\rm had} \to \ell^+ \ell^+ \ell^+]}{\Gamma[\tilde S^{+10/3} \to \ell^+ \ell^+ \ell^+ \bar d]}
~\sim~ \frac{P_3 f_{\pi}^2}{P_4 m^2_{\tilde S_3}} 
~\sim~ 10^{-5} \,,
\eeq 
where $f_{\pi}$ is the pion decay constant and $P_n \equiv [4 \cdot (4
  \pi)^{2n-3} \cdot (n-1)! \cdot (n-2)!]^{-1}$ is the $n$-body
phase-space factor.  We therefore do not consider this type of decays
in our analysis.

In Model-2, the long-lived multi-charged particles that may be
detected at MoEDAL are the states originated in the $SU(2)_L$ triplet
field $\tilde S_3$, denoted by $\tilde S^{+4/3}$, $\tilde S^{+7/3}$
and $\tilde S^{+10/3}$.  These states correspond to $S^{+2}$, $S^{+3}$
and $S^{+4}$ in Model-1 but there are two main differences.  One is
that their electric charge is smaller by $2/3$ compared to the
corresponding particles in Model-1.  Another difference is that the
multi-charged particles in Model-2 are colour (anti-)triplet states.

Due to the colour charge the particle-anti-particle pair productions,
$pp \to \tilde S^{+10/3} \tilde S^{-10/3}$, $\tilde S^{+7/3} \tilde
S^{-7/3}$ and $\tilde S^{+4/3} \tilde S^{-4/3}$, via QCD interaction
with the $g g$ and $q \bar q$ initial states are the dominant
production mechanism for the long-lived particles in Model-2.  The
contributions from the EW interaction are negligible and the above
three pair production modes have approximately the same cross-section,
which is shown in Fig.~\ref{fig:xsec} by the cyan curve.  As can be
seen in Fig.~\ref{fig:xsec}, the cross-section is almost two orders of
magnitude larger than that of $pp \to S^{+4} S^{-4}$ in Model-1 at
$m_{S_3} \sim 400$ GeV.  The coloured particle production is
relatively more enhanced compared to the non-coloured particle
production at the lower mass region, because the parton distribution
function of gluons gets very large for smaller values of the parton
energy fraction, $x$.

The velocity distribution of the produced coloured particles is shown
in Fig.~\ref{fig:vel} by the cyan curve.  As can be seen, the velocity
of coloured particles are the smallest on average compared to the
non-coloured ones.  This is because unlike the Drell-Yan production of
EW particles, the production rate from the $gg$ initial state does not
suffer from the $p$-wave suppression.  Moreover, as mentioned above,
the parton distribution function of {gluons is strongly enhanced}  
for smaller energy fraction, $x$, preferring near-threshold productions
with low velocities.

Once the multi-charged long-lived particles are produced they
hadronise into colour singlet states before decaying.  After
hadronising, charges of the long-lived particles are shifted by the
constituent quarks.  A precise simulation for predicting the hadronic
final states is complicated and beyond the scope of this study.
Instead, we use the following crude model of hadronisation in order to
see the effect of the charge shift due to hadronisation.  In our
hadronisation model, $\tilde S = (\tilde S^{+10/3}, \, \tilde
S^{+7/3}, \, \tilde S^{+4/3})$ is hadronised into a mesonic state with
the probability $k$ and into a baryonic state with the probability
$1-k$.  We consider the following two mesonic states and assume that
those states appear with the same probability:
\beqn
&& \mbox{Spin-1/2 mesons: probability $k$~~~~~~~} \nonumber \\
&& \tilde S + u_{L/R}  ~~~~ (+ 2/3) \nonumber \\
&& \tilde S + d_{L/R}  ~~~~ (- 1/3) \,,
\eeqn
where the numbers in the brackets represent the charge shift.
For baryonic states, we consider four spin-0 states and
two spin-1 states, assuming the democratic probabilities for appearance:
\beqn
&& \mbox{Spin-0 baryons: probability $\frac{2}{3}(1-k)$} \nonumber \\
&& \tilde S + \bar u_L \bar u_R ~~~~ (- 4/3) \nonumber \\
&& \tilde S + \bar d_L \bar d_R ~~~~  (+ 2/3) \nonumber \\
&& \tilde S + \bar u_L \bar d_R ~~~~ (- 1/3) \nonumber \\
&& \tilde S + \bar d_L \bar u_R ~~~~ (- 1/3) \nonumber \\
&& \mbox{Spin-1 baryons: probability $\frac{1}{3}(1-k)$} \nonumber \\
&& \tilde S + \bar u_L \bar d_L ~~~~ (- 1/3)  \nonumber \\
&& \tilde S + \bar u_R \bar d_R ~~~~ (- 1/3)  \,,
\eeqn 
Finally, we vary the parameter $k$ from 0.3 to 0.7 to roughly
understand the size of the uncertainty from the hadronisation effect.
We emphasis again that we use this naive hadronisation model only for
the purpose of a ballpark estimate of the MoEDAL's detectability of
the model.  More precise treatment of hadronisation may be necessary
to fully understand the MoEDAL's performance.

\begin{figure}[t!]
\centering
       \includegraphics[width=0.45\textwidth]{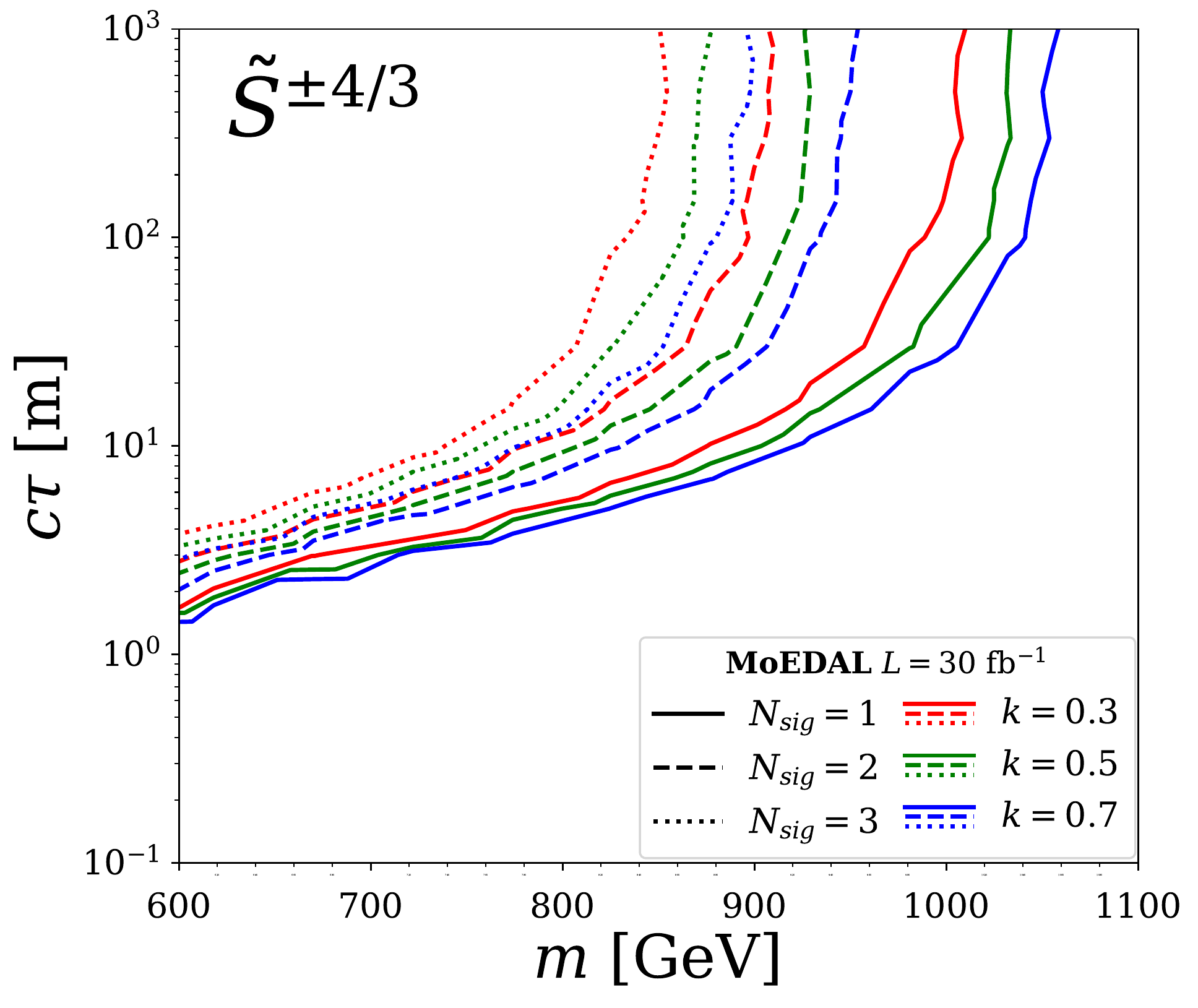} \hspace{4mm}
       \includegraphics[width=0.44\textwidth]{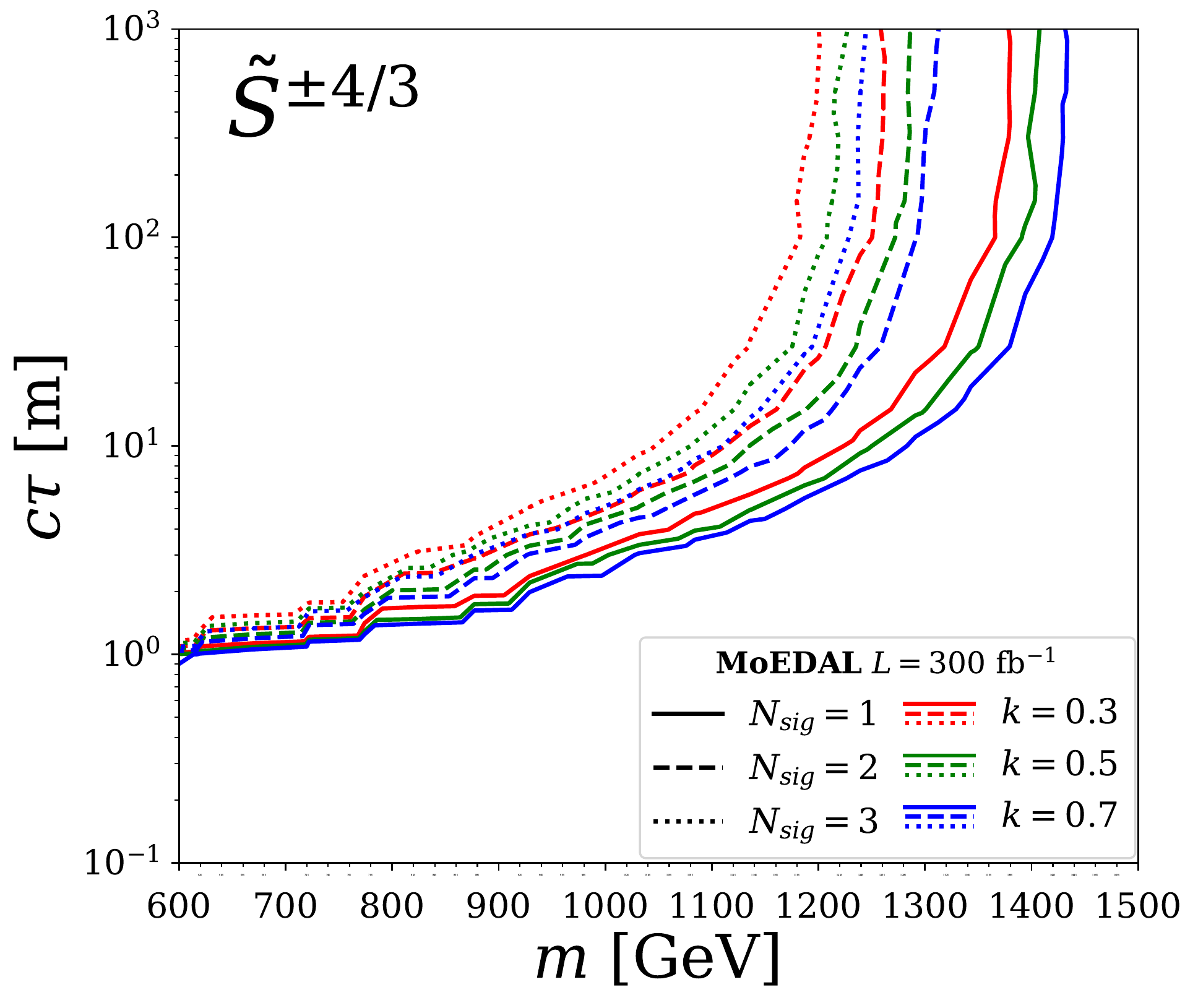}
      \includegraphics[width=0.45\textwidth]{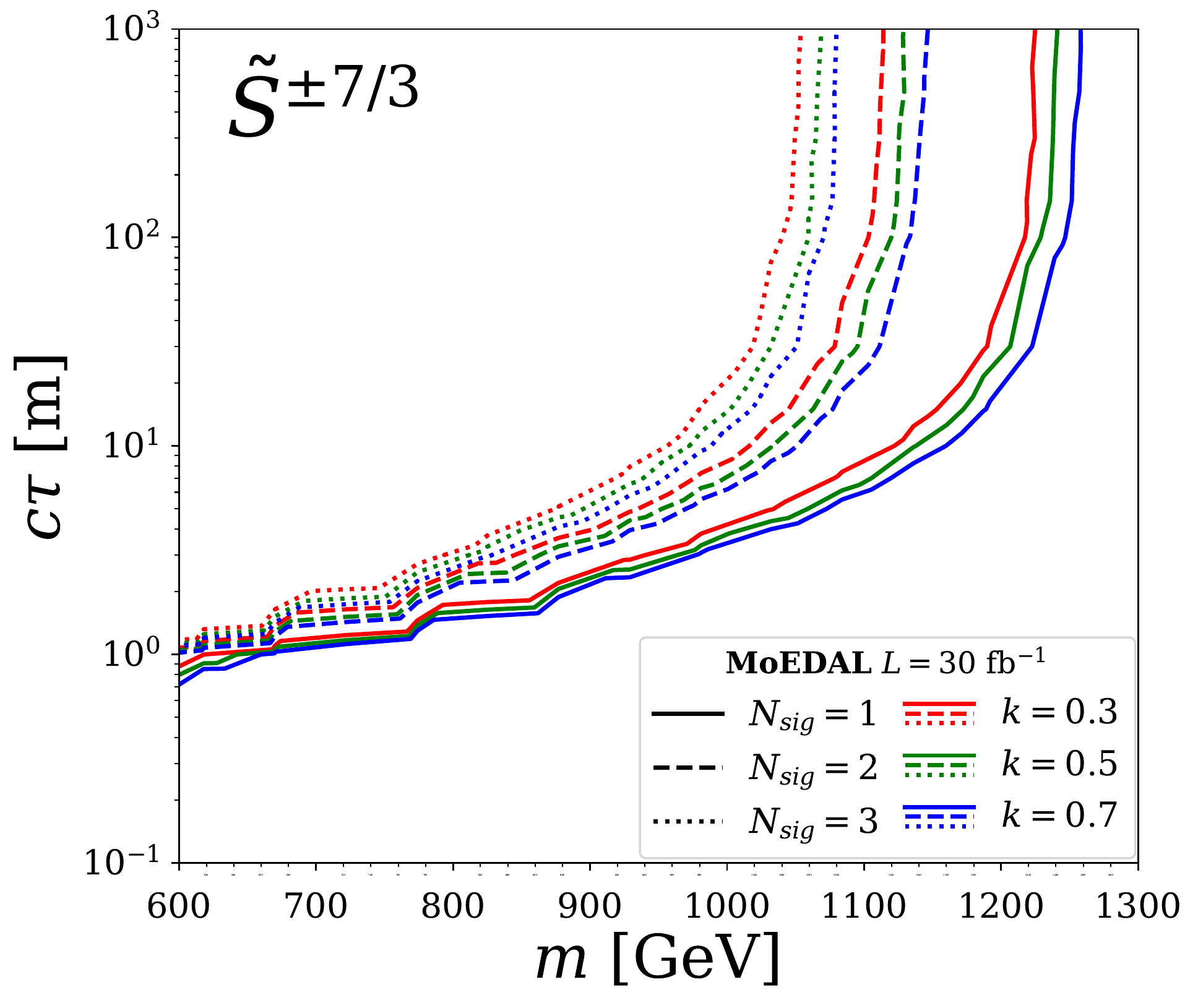} \hspace{4mm}
      \includegraphics[width=0.44\textwidth]{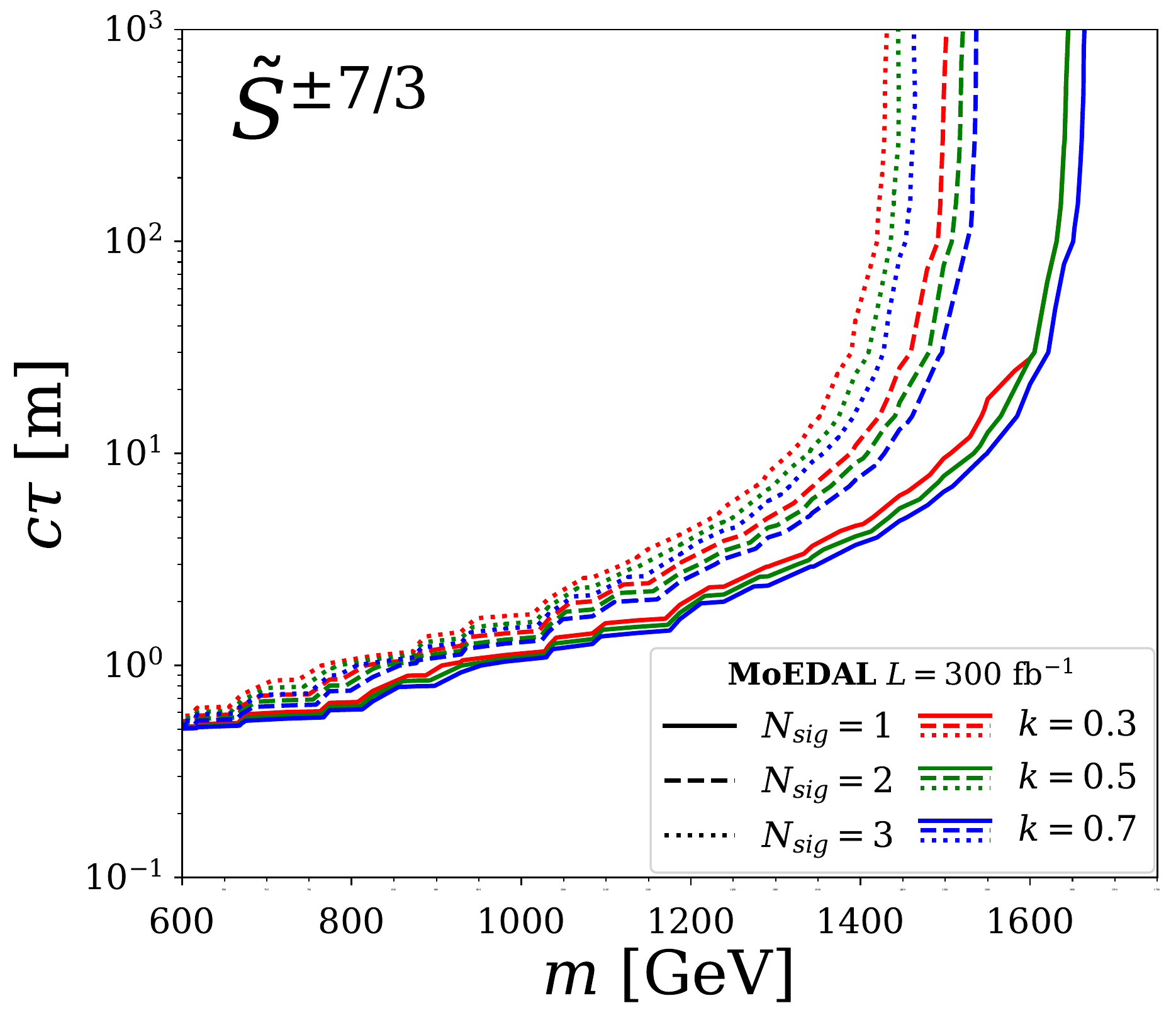}
      \includegraphics[width=0.45\textwidth]{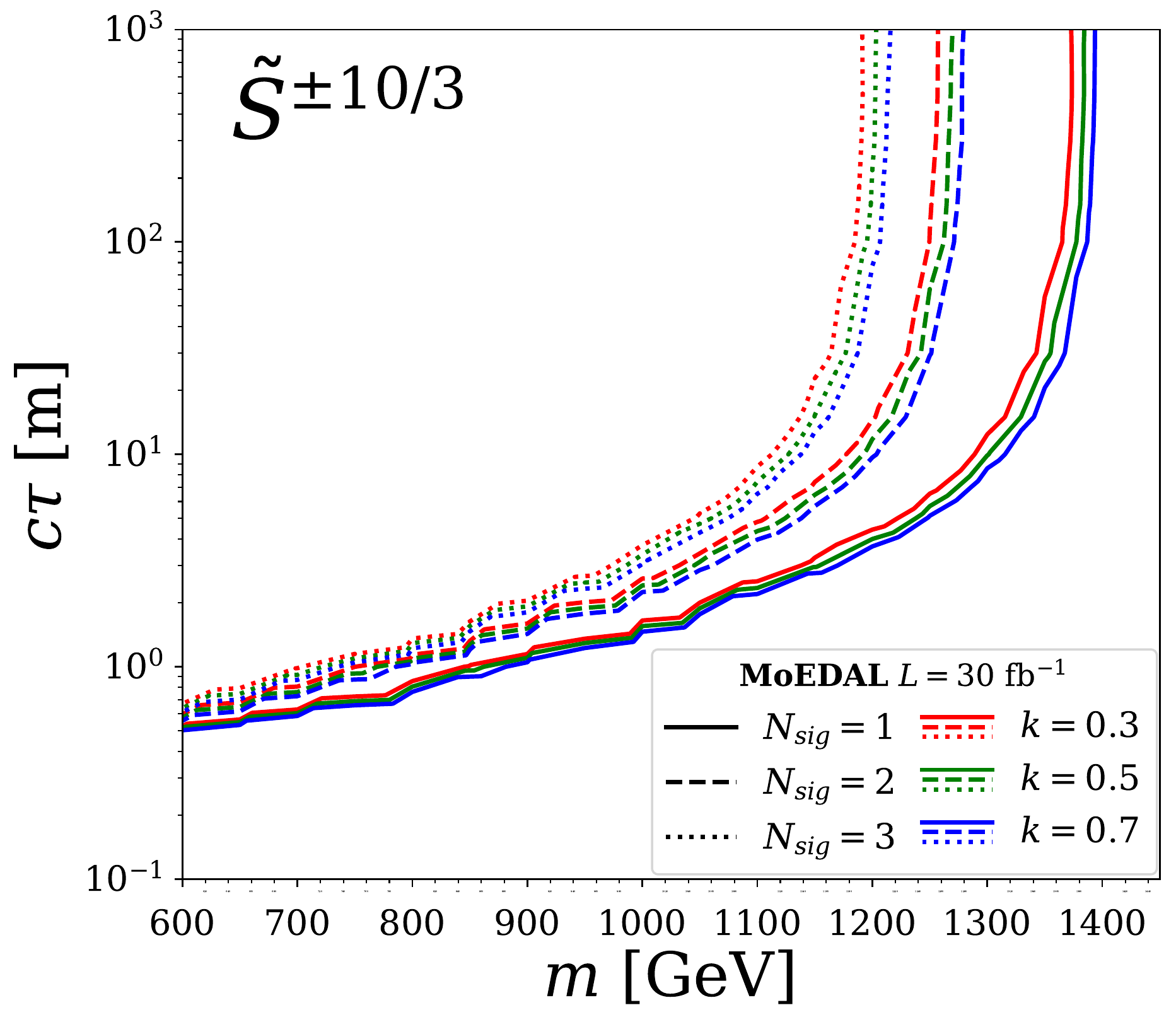} \hspace{4mm}
      \includegraphics[width=0.45\textwidth]{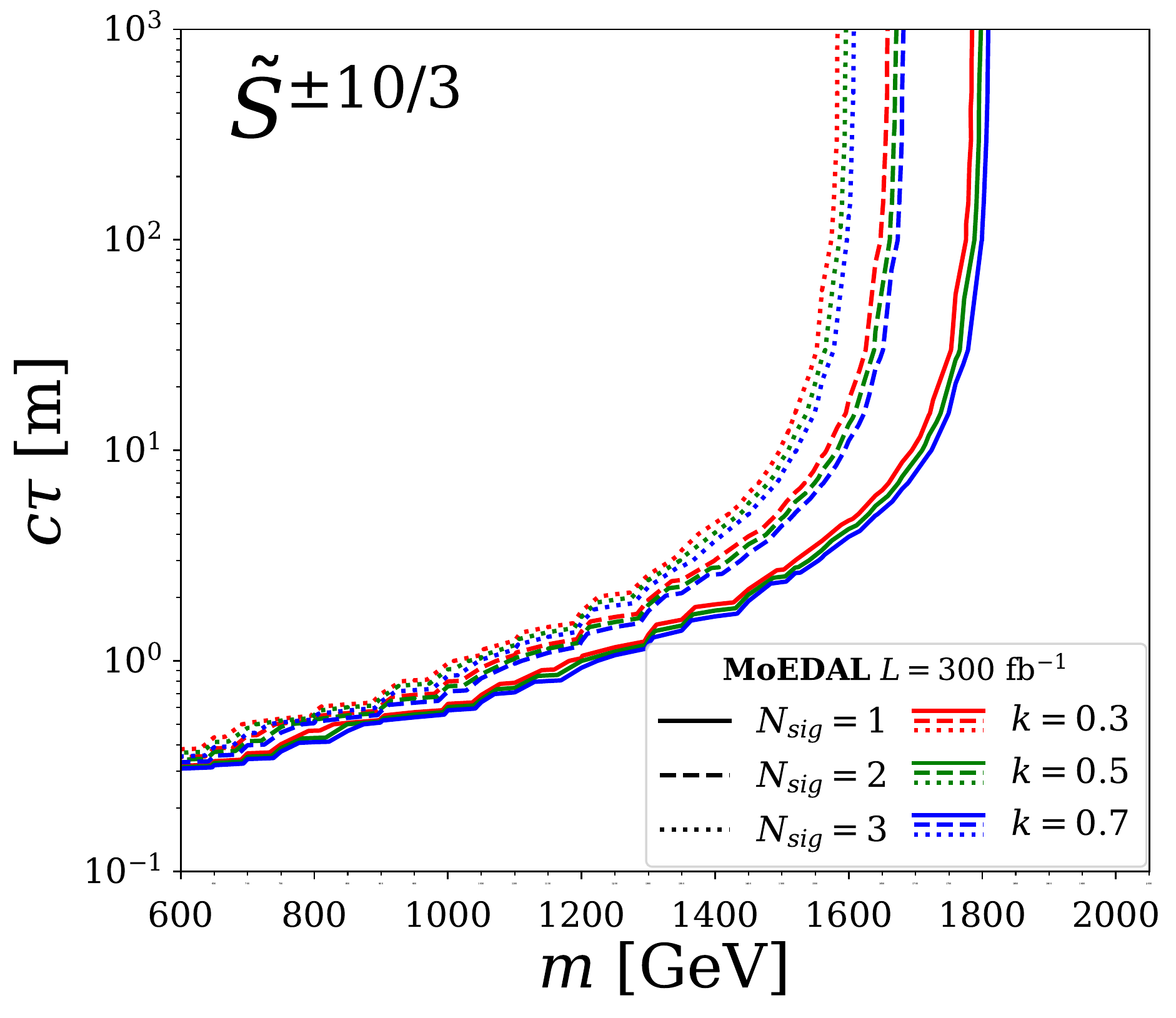}
\caption{
\label{fig:lim_2}
The model-independent detection reach at MoEDAL in the ($m$, $c \tau$)
parameter plane for individual particles of Model-2.  The red, green
and blue contours correspond to the results obtained with the
hadronisation parameter $k = 0.3$, 0.5 and 0.7, respectively, and
solid dashed and dotted contours represent $N_{\rm sig} = 1$, 2 and 3,
respectively.  In the left (right) panel the integrated luminosity of
30 (300) fb$^{-1}$ is assumed.  }
\end{figure}

We show in Fig.~\ref{fig:lim_2} the expected sensitivities for MoEDAL
to detect multi-charged long-lived particles in Model-2.  The results
are regarded as model-independent and presented in the ($m$, $c \tau$)
plane.  The plots on the left (right) column correspond to $L = 30$
(300) fb$^{-1}$, which is achievable in Run-3 LHC (HL-LHC).  Here we
show the sensitivities calculated with the hadronisation parameter $k
= 0.3$, 0.5 and 0.7 with red, green and blue contours, respectively.
One can immediately see that the effect of the hadronisation parameters on
the detection reach is not strong.  From $k = 0.3$ to 0.7, the mass
reach changes about 50 GeV for $\tilde S^{\pm 4/3}$.  For $\tilde
S^{\pm 10/3}$, the effect is even smaller; the impact on the mass
reach is about 30 GeV for varying $k$ from 0.3 to 0.7.  This is
because a change of the signal acceptance in varying the velocity
threshold, $\beta_{\rm th} = 0.15 \cdot Z$, is milder for larger $Z$,
as can be seen in Fig.~\ref{fig:vel}.  We also see that larger $k$
provides a higher mass reach in the plots.  On the other hand, the
solid, dashed and dotted curves correspond to $N_{\rm sig} = 1$, 2 and
3.  The top two plots show the detection reach for $\tilde S^{\pm
  4/3}$.  We see that for $N_{\rm sig} = 1$ (3) MoEDAL can probe
$\tilde S^{\pm 4/3}$ up to $\sim$1050 (880) GeV with $L=30$ fb$^{-1}$.
At HL-LHC with $L = 300$ fb$^{-1}$ the mass reach improves up to $\sim
1400$ (1250) with $N_{\rm sig} = 1$ (3).  The plots in the second line
show MoEDAL's sensitivity for $\tilde S^{\pm 7/3}$.  One can see that
the mass reach with $N_{\rm sig}= 1$ (3) is $\sim$1250 (1080) GeV with
$L = 30$ fb$^{-1}$ (Run-3) and 1650 (150) GeV with $L = 300$ fb$^{-1}$
(HL-LHC).  The detection reaches for $\tilde S^{\pm10/3}$ are shown in
the bottom two plots.  With 30 fb$^{-1}$ (Run-3) the reach is $\sim
1400$ (1200) GeV for $N_{\rm sig} = 1$ (3) and with 300 fb$^{-1}$
(HL-LHC) this is improved up to $\sim 1800$ (1600) GeV.


\begin{table}[t!]
\centering
\renewcommand{\arraystretch}{1.2}
\begin{tabular}{c||c|c||c|c}
           & current HSCP bound & HSCP (Run-3) & MoEDAL (Run-3) & MoEDAL (HL-LHC) \\
           & 36 fb$^{-1}$ \cite{Jager:2018ecz} & 300 fb$^{-1}$ \cite{Jager:2018ecz} & 30 fb$^{-1}$ & 300 fb$^{-1}$ \\        
\hhline{=||=|=||=|=}
$\tilde S^{\pm 4/3}$ & ((1450))  & 1700 & 880 (1050) & 1250 (1400)   \\
\hline
$\tilde S^{\pm 7/3}$ & ((1480))  & 1730 & 1080 (1250) & 1450 (1650)   \\
\hline
$\tilde S^{\pm 10/3}$ & ((1510))  & 1790 & 1200 (1400) & 1600 (1800)   \\
\end{tabular}
\caption{Summary for the model-independent mass reaches (in GeV) of
  the multi-charged particles in Model-2 by MoEDAL (2nd and 3rd
  columns).  In the first column, the numbers in the double-brackets
  show the estimated mass bounds obtained in \cite{Jager:2018ecz} by
  rescaling the 8 TeV CMS result \cite{Chatrchyan:2013oca} to the 13
  TeV LHC with $L = 36$ fb$^{-1}$.  The second column represents the
  projected mass reach for Run-3 (300 fb$^{-1}$) obtained in
  \cite{Jager:2018ecz}.  The numbers outside (inside) the brackets in
  the third and fourth columns represent MoEDAL's mass reaches with
  $N_{\rm sig} \geq 3$ (1) assuming $L = 30$ (Run-3) and 300 (HL-LHC)
  fb$^{-1}$, respectively.}
\label{tab:sum_col}
\end{table}

In Table \ref{tab:sum_col} we summarise the results obtained in this
subsection and compare them with the current bounds and future
projections from HSCP searches (if available).  The mass reaches are
shown in the GeV unit.  In the first column, the numbers in the
double-brackets represent the estimated mass bounds obtained in
\cite{Jager:2018ecz} by rescaling the 8 TeV CMS result
\cite{Chatrchyan:2013oca} to the 13 TeV LHC with $L = 36$ fb$^{-1}$.
The second column represents the projected mass reach for Run-3 (300
fb$^{-1}$) obtained in \cite{Jager:2018ecz}.  The numbers outside
(inside) the brackets in the third and fourth columns represent
MoEDAL's mass reaches with $N_{\rm sig} \geq 3$ (1) assuming $L = 30$
(Run-3) and 300 (HL-LHC) fb$^{-1}$, respectively.  We see in Table
\ref{tab:sum_col} that the estimated current bounds from the HSCP
searches are rather strong and the regions where MoEDAL has
sensitivity for Run-3 are already excluded.  However, MoEDAL can
explore $\tilde S^{\pm 7/3}$ and $\tilde S^{\pm 10/3}$ in some mass
regions that are not excluded at the HL-LHC with $L = 300$ fb$^{-1}$.

\subsection{Interpretation of the results for Model-2}

\begin{figure}[t!]
\centering
      \includegraphics[width=0.46\textwidth]{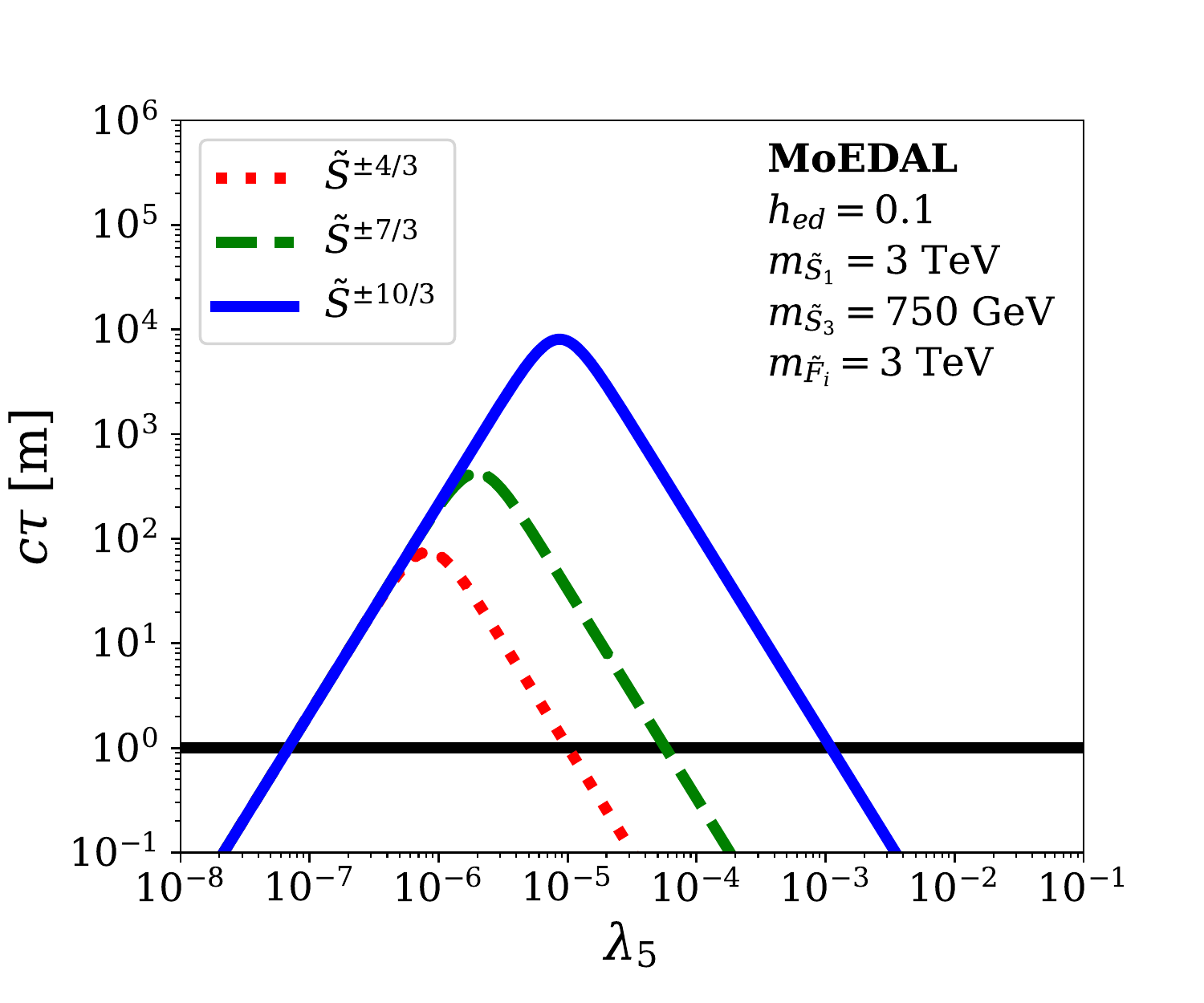}
      \includegraphics[width=0.46\textwidth]{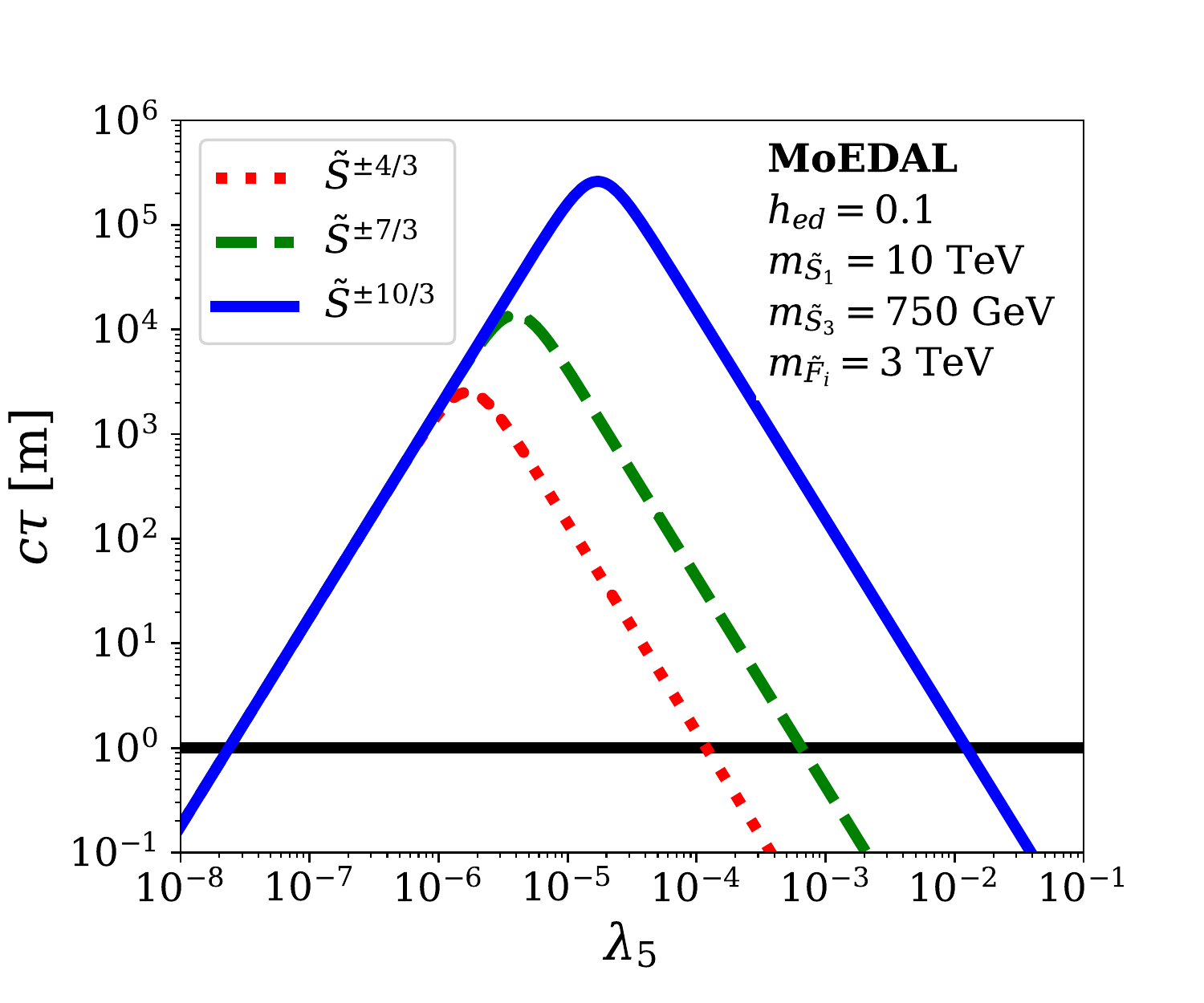}
\caption{Lifetimes of multi-charged particles, $\tilde S^{\pm 10/3}$ (blue solid) $\tilde S^{\pm 7/3}$ (green dashed) and $\tilde S^{\pm 4/3}$ (red dotted), 
in Model-2 as functions of $\lambda_5$, where $h_F$ and $h_{\bar F}$ are fitted to the neutrino data.
The $m_{\tilde S_1}$ is taken to be 3 and 10 TeV in the left and right plots, respectively.
The other parameters are fixed as $h_{ed} = 0.1$, $m_{\tilde S_3} = 750$ GeV and $m_{\tilde F_i} = 3$ TeV ($i = 1,2,3$).
}
\label{fig:lifetime_col}
\end{figure}

We discuss in this section MoEDAL's sensitivity to Model-2 combining
all production modes studied in the previous subsection.  Our goal is
to identify the parameter regions to be explored by MoEDAL at LHC
Run-3 (30 fb$^{-1}$) and HL-LHC (300 fb$^{-1}$) within the subspace of
model parameters in which the experimental values of neutrino mass and
mixing angles are fitted.

We start by showing the lifetimes of multi-charged long-lived
particles in Model-2 as functions of $\lambda_5$ in
Fig.~\ref{fig:lifetime_col}, where $h_F$ and $h_{\bar F}$ are fitted
to the neutrino data.  We fix $m_{\tilde S_3} = 750$ GeV, $m_{\tilde
  F_i} = 3$ TeV and $h_{ed} = 0.1$ and take $m_{\tilde S_1} = 3$ and
10 TeV in the left and right plots, respectively.  As can be seen, the
results are very similar to those for Model-1 shown in
Fig.~\ref{fig:lifetime}, which should be understood since the decay
rate formulae are the same between Model-1 and -2 and the neutrino
mass formula is also the same except for the colour factor, $N_c$.  We
see, however, that the lifetimes for Model-2 particles are slightly
(about a factor of two times) longer than the corresponding particles
in Model-1.  As mentioned in the previous section for
Fig.~\ref{fig:lifetime}, in the large $\lambda_5$ region ($\lambda_5
\gg 10^{-5}$) the $L$-violating decay modes, whose decay rate is
proportional to $|\lambda_5|^2$, dominate over the $L$-conserving
modes.  On the contrary, in the small $\lambda_5$ region ($\lambda_5
\ll 10^{-6}$) the $L$-conserving modes are dominant since their decay
rates are proportional to $h_F h_{\bar F}$, which is inversely related
to $\lambda_5$ through the neutrino masses.  The black horizontal
lines in Fig.~\ref{fig:lifetime_col} represent $c \tau = 1$\,m, which
is the typical lifetime for the MoEDAL detector.  We see that we
expect the multi-charged particles to have long enough lifetime for
MoEDAL when $10^{-7} \lesssim \lambda_5 \lesssim 10^{-3}$ for $m_{S_3}
= 3$ TeV and $5 \cdot 10^{-8} \lesssim \lambda_5 \lesssim 10^{-2}$ for
$m_{S_3} = 10$ TeV.

\begin{figure}[t!]
\centering
      \includegraphics[width=0.4\textwidth]{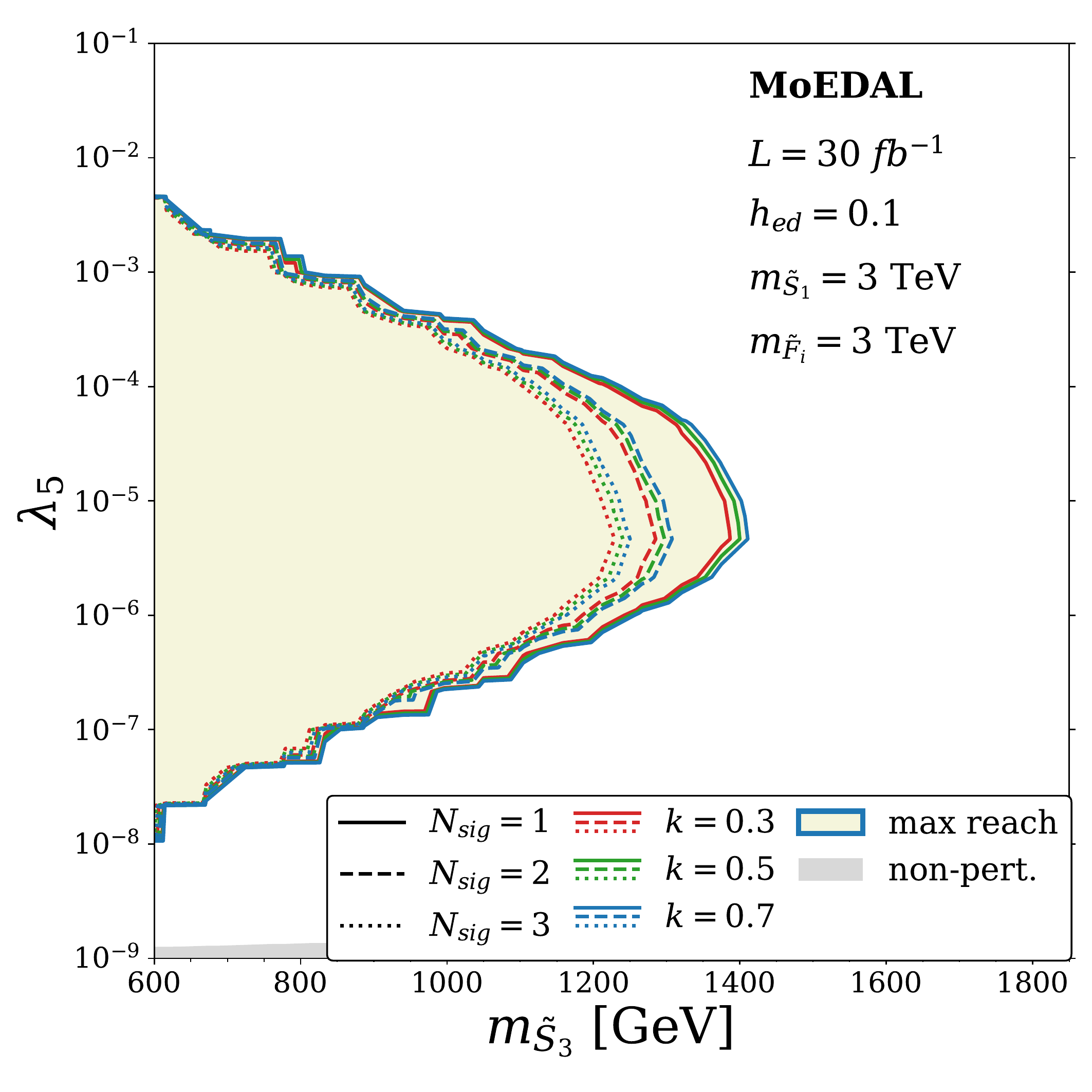} \hspace{5mm}
      \includegraphics[width=0.4\textwidth]{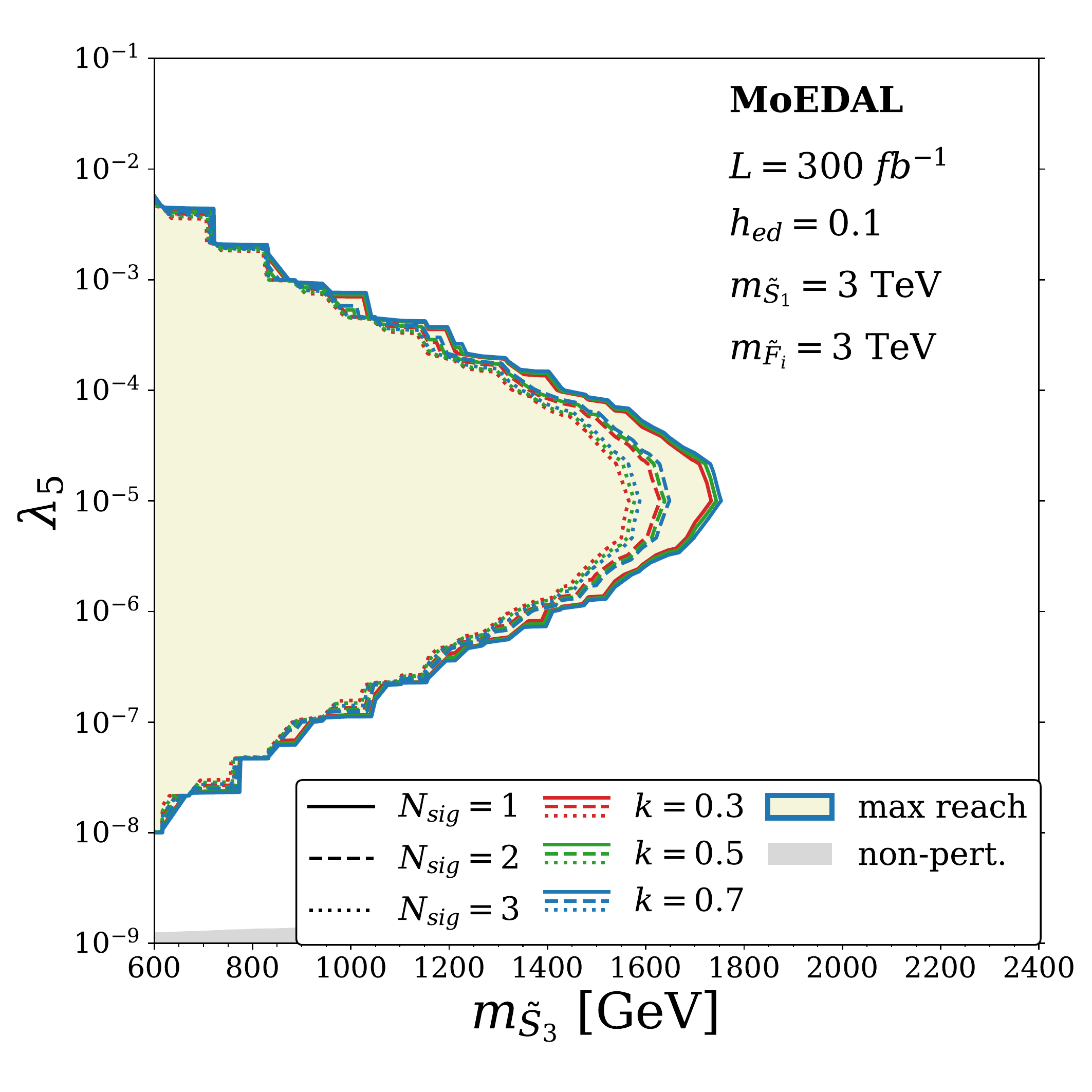}
      \includegraphics[width=0.4\textwidth]{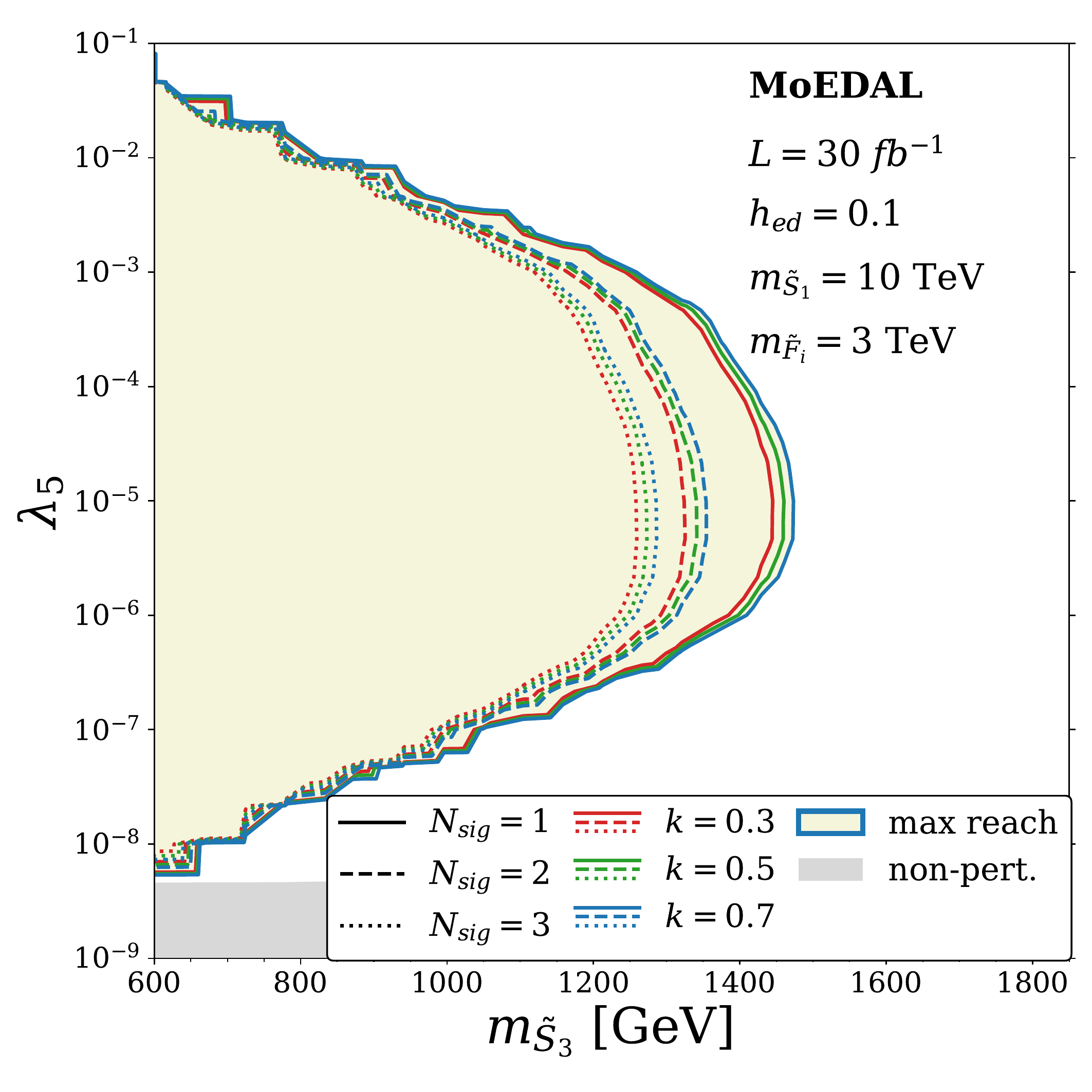} \hspace{5mm}
      \includegraphics[width=0.4\textwidth]{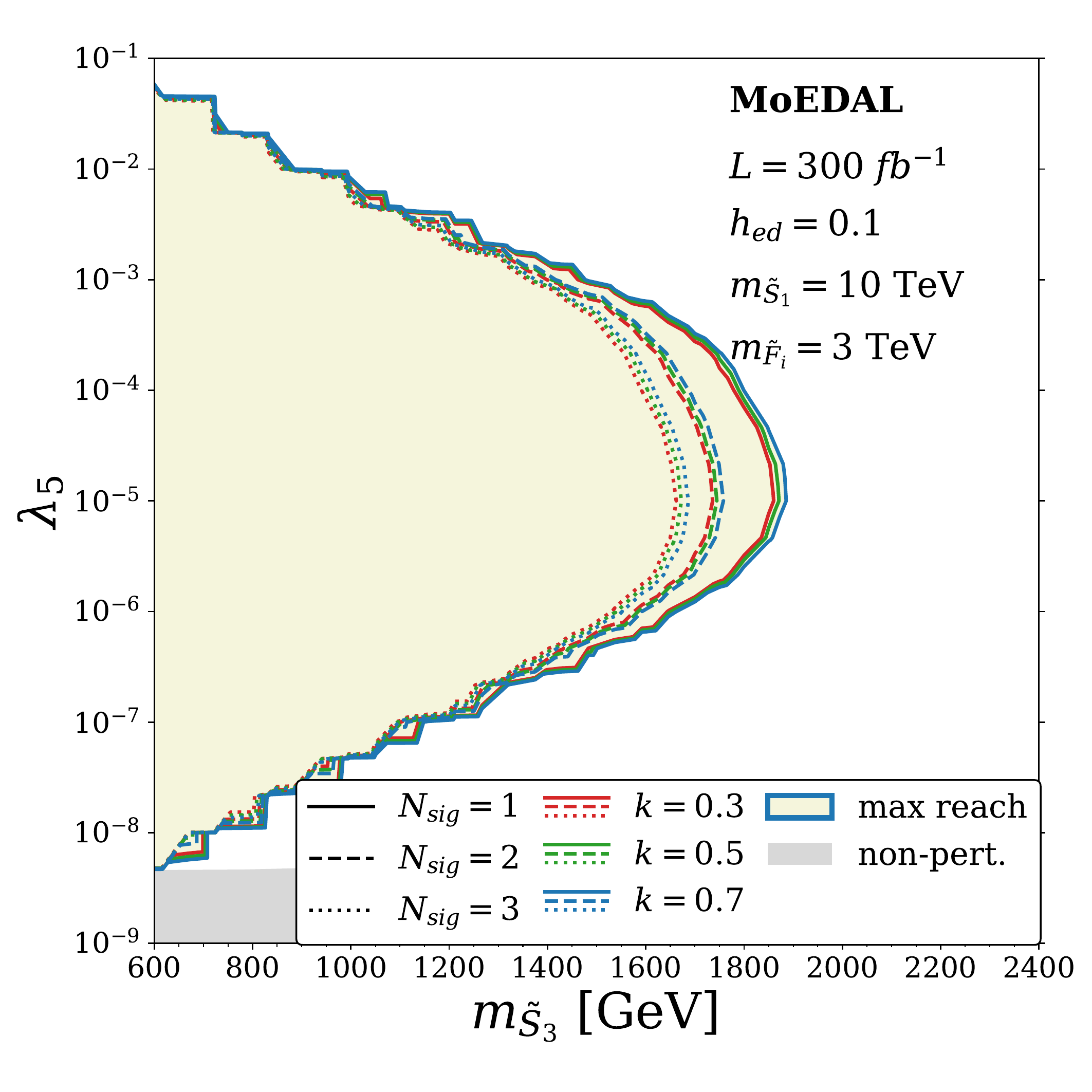}
\caption{MoEDAL's detection sensitivity for Model-2 in the ($m_{S_3}$,
  $\lambda_5$) plane.  The red, green and blue contours correspond to
  the results obtained with the hadronisation parameter $k = 0.3$, 0.5
  and 0.7, respectively.  In the regions inside the solid, dashed and
  dotted contours, MoEDAL expects to observe more than 1, 2 and 3
  events, respectively.  The region with $N_{\rm sig} \geq 3$ will be
  excluded at 95\% CL if MoEDAL does not observe any signal event.
  The left and right panels correspond to $L = 30$ and 300 fb$^{-1}$,
  respectively.  We take $m_{\tilde F_i} = 3$ TeV ($i = 1,2,3$) for
  all plots and $m_{\tilde S_1} = 3$ (10) TeV in the top (bottom)
  panels.  The $h_{F}$ and $h_{\bar F}$ are fitted to the neutrino
  data.  In the grey region one of these couplings tend to be
  non-perturbative; ${\rm max}_{ij} \left( |(h_{F})_{ij}|, |(h_{\bar
    F})_{ij}| \right) \geq 2$.}
\label{fig:nsig_col}
\end{figure}

In Fig.~\ref{fig:nsig_col} we show MoEDAL's sensitivity to Model-2 in
the ($m_{\tilde S_3}$, $\lambda_5$) plane.  The plots on the left
(right) correspond to $L = 30$ (300) fb$^{-1}$ and we take $m_{\tilde
  S_1} = 3$ (10) TeV in the top (bottom) plots.  We fix $m_{\tilde
  F_i} = 3$ TeV and $h_{ed} = 0.1$ for all plots.  As in
Fig.~\ref{fig:lim_2}, we vary the parameter $k$ in our hadronisation
model as $k=0.3$ (red), 0.5 (green) and 0.7 (blue) to see the impact
of the uncertainty of the hadronisation effect.  We present the
contours corresponding to $N_{\rm sig} = 1$, 2 and 3 with the solid,
dashed and dotted curves, respectively.  In the grey region at the
bottom of the plots, we have ${\rm max}_{ij} \left( |(h_{F})_{ij}|,
|(h_{\bar F})_{ij}| \right) \geq 2$ and the result may not be trusted
due to a large coupling.  We first note that varying the hadronisation
parameter $k$ from 0.3 to 0.7, the mass reach changes only $\sim$30
GeV.  We therefore think the conclusion of our analysis is rather
robust despite our crude hadronisation model.  The highest reach for
$m_{\tilde S_3}$ roughly agrees with the model-independent mass reach
of $m_{\tilde S^{\pm 10/3}}$ studied in the previous section, though
the reach for the $m_{\tilde S_3}$ parameter is slightly higher due to
the effect of other particles.

On the top two plots with $m_{\tilde S_1} = 3$ TeV, we see that the
highest reach for $m_{\tilde S_3}$ with $N_{\rm sig} = 1$ is
$\sim$1400 (1800) GeV for $L = 30$ (300) fb$^{-1}$ with $\lambda_5
\sim 10^{-5}$, at which the lifetime of $\tilde S^{\pm 4}$ is
maximized.  Around $\lambda_5 \sim 10^{-3}$ or $10^{-7}$, MoEDAL can
expect to observe more than 1 signal event only up to $m_{\tilde S_3}
\sim 800$ (1000) GeV for $L = 30$ (300) fb$^{-1}$.  Moving to the
bottom plots with $m_{\tilde S_1} = 10$ TeV, we see that the range of
$\lambda_5$ that gives the highest reach for $m_{\tilde S_3}$ becomes
wider compared to the top plots with $m_{\tilde S_1} = 3$ TeV and the
reach of $m_{\tilde S_3}$ becomes slightly higher.  This is because
the lifetimes of the multi-charged particles in the $\tilde S_3$
multiplet get prolonged due to larger $m_{\tilde S_1}$ compared to the
case with $m_{\tilde S_1} = 3$ TeV.  The highest reach for $m_{\tilde
  S_3}$ is obtained around $\lambda_5 \sim 10^{-5}$ and it is $\sim
1500$ (1900) GeV with $L = 30$ (300) fb$^{-1}$ with $N_{\rm sig} = 1$.
The reach is degraded for different values of $\lambda_5$ as can be
seen in Fig.~\ref{fig:nsig_col}.

\begin{figure}[t!]
\centering
  \includegraphics[width=0.4\textwidth]{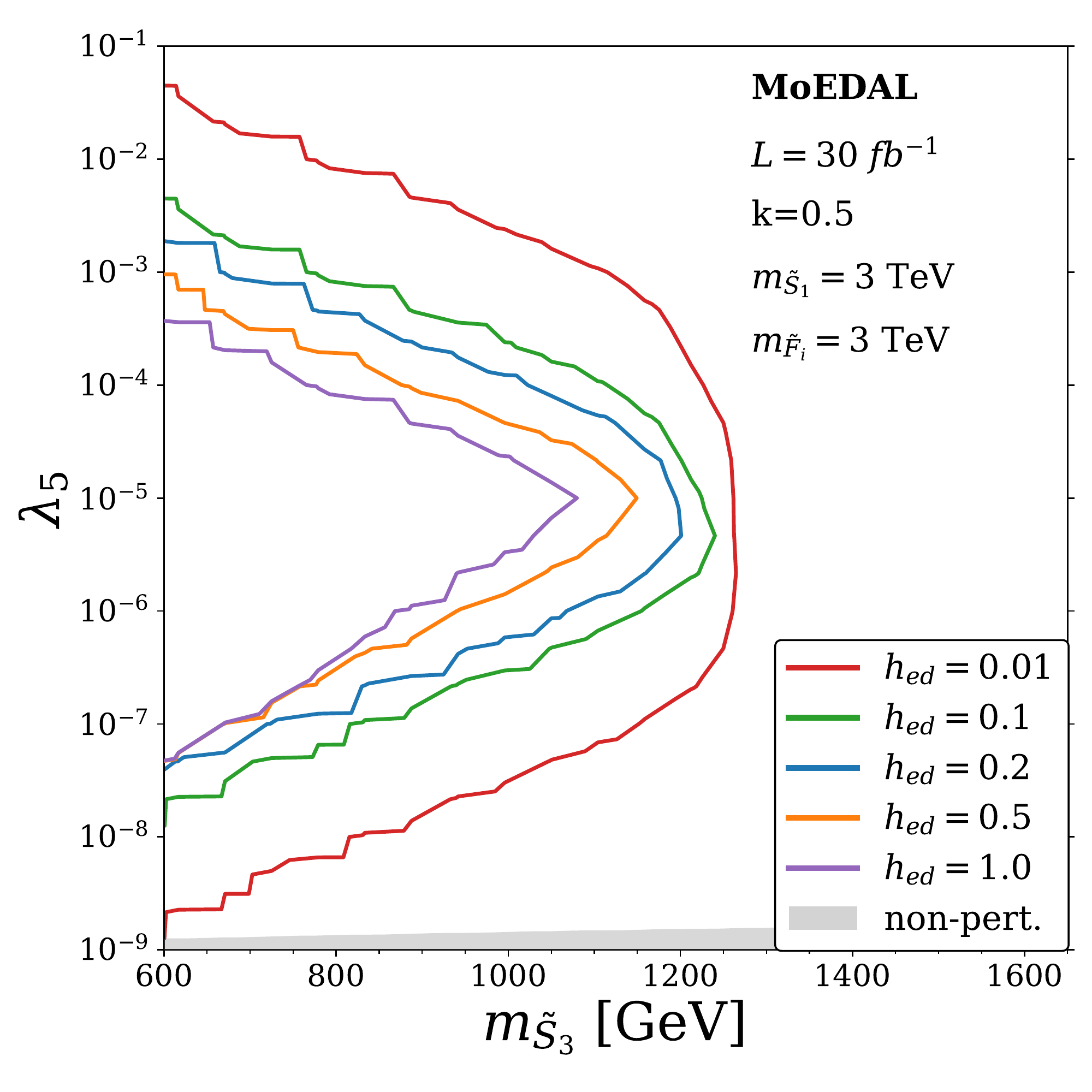}\hspace{5mm}
    \includegraphics[width=0.4\textwidth]{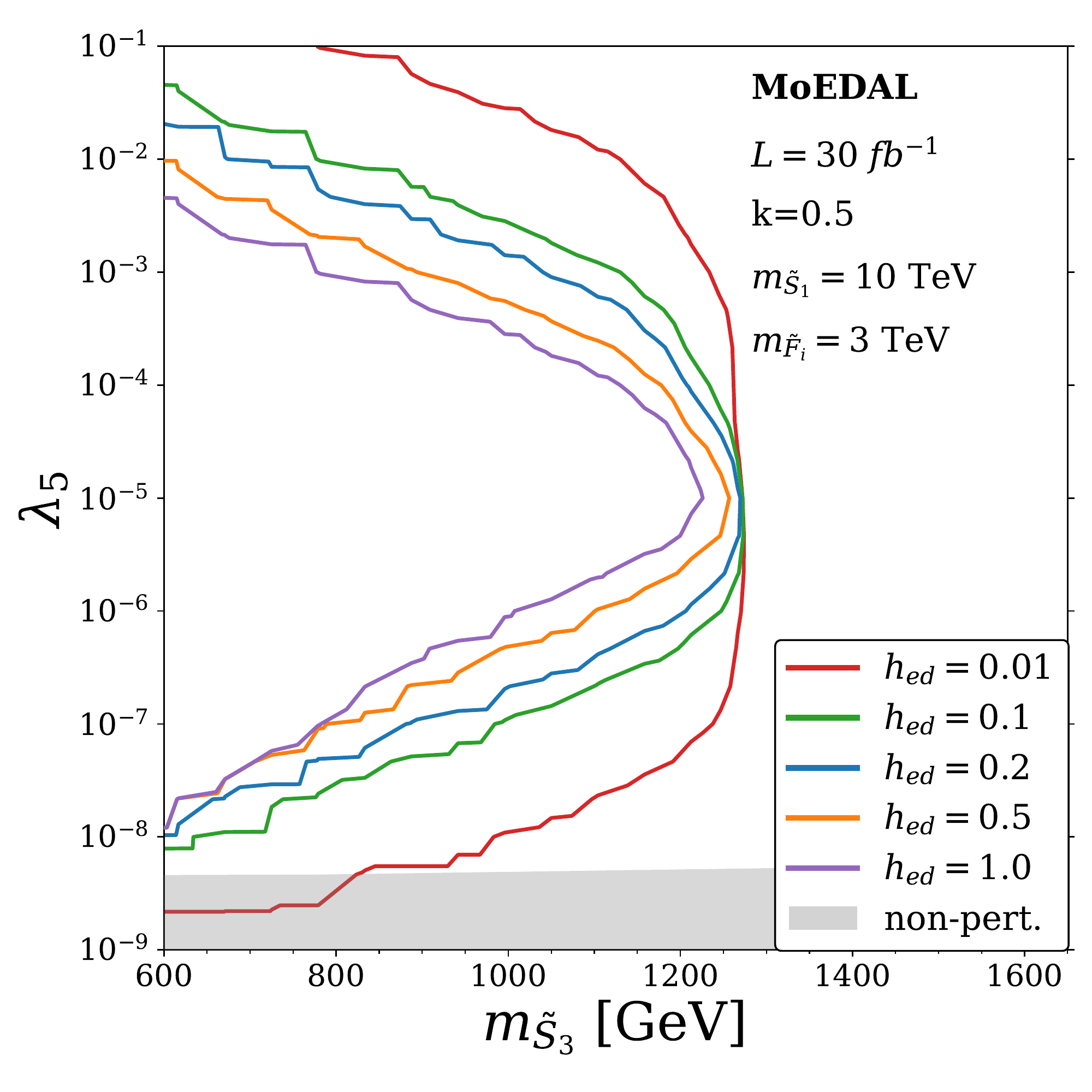}
\caption{$N_{\rm sig} = 1$ contours for different values of $h_{ed}$
  in Model-2.  The hadronisation parameter $k$ is fixed to $0.5$.  We
  take $m_{S_1} = 3$ and 10 TeV in the left and right plots,
  respectively.  We assume $L = 30$ fb$^{-1}$ and $m_{F_i} = 3$ TeV
  ($i = 1,2,3$).
\label{fig:hed}}
\end{figure}

Before closing this section we show the response of MoEDAL's detection
capability to variations of the coupling $h_{ed}$ for Model-2 in the
($m_{\tilde S_3}$, $\lambda_5$) plane in Fig.~\ref{fig:hed}.  Here we
fix $L = 30$ fb$^{-1}$, $k = 0.5$ and $m_{\tilde F_i} = 3$ TeV and
take $m_{\tilde S_1} = 3$ and 10 TeV in the left and right plots,
respectively.  As can be seen, the region in which MoEDAL has a
sensitivity is wider for smaller $h_{ed}$ since the lifetimes of
multi-charged particles are longer.  Comparing these plots with the
corresponding plots in Fig.~\ref{fig:hee} for Model-1, we see that
MoEDAL's is more sensitive to the $h_{ed}$ coupling than the $h_{ee}$
coupling in the Model-1 case.  This is because the electric charges,
$Z$, of the Model-2 particles are smaller than those of the Model-1
particles and the signal efficiency is more sensitive to the lifetimes
in Model-2.

\section{Conclusions}
\label{sec:disc}

In this paper we investigated the possibility of observing long-lived
particles whose electric charges are larger than 1 with the MoEDAL
detector for Run-3 and High-Luminosity phase of the LHC.  In
particular, we took two concrete models of radiative neutrino mass
generation and studied MoEDAL's performance for the long-lived
multi-charged particles found in those models.  In the first model
(Model-1), the BSM sector is non-coloured and has three long-lived
particles ($S^{\pm 2}$, $S^{\pm 3}$, $S^{\pm 4}$) with electric
charges $Z = 2$, 3 and 4.  In the second model (Model-2), the BSM
sector is coloured and the long-lived particles ($\tilde S^{\pm 4/3}$,
$\tilde S^{\pm 7/3}$, $\tilde S^{\pm 10/3}$) have fractional charges
$Z = 4/3$, 7/3 and 10/3.  In both models, the lifetimes of the
multi-charged particles are controlled by the $\lambda_5$, $h_F$,
$h_{\bar F}$ couplings as well as the BSM parity breaking coupling
$h_{ee}$ ($h_{ed}$) for Model-1(2).  In addition $\lambda_5$ and the
product $h_F h_{\bar F}$ are related through the constraints from the
neutrino masses and mixing angles.

We demonstrated that multi-charged scalar particles with a large $Z$
present several advantages for MoEDAL.  First, the ionisation power
increases with larger $Z$ and it relaxes the upper limit on the
production velocity required to be detected by MoEDAL.  Secondly, it
raises the cross-sections due to the large enhancement of the photon
fusion process.  In particular, the Drell-Yan process is $p$-wave
suppressed for scalar particles, meaning that the cross-section
vanishes when the production velocity goes to zero.  The photon fusion
process is free from the $p$-wave suppression and it allows to access
lower production velocities, which helps improve significantly MoEDAL's
detection efficiency.

Our study was comprised of the model-independent and model-specific
parts and in the former we studied MoEDAL's detection capabilities for
individual particles treating their masses and lifetimes as free
parameters.  The results of this part of study are summarised in
Fig.~\ref{fig:lim_1} and Table \ref{tab:sum} for Model-1 and in
Fig.~\ref{fig:lim_2} and Table \ref{tab:sum_col} for Model-2.  By
taking the optimistic criteria $N_{\rm sig} \geq 1$, MoEDAL can
explore the Model-1 particles up to 290, 610 and 960 GeV for $S^{\pm
  2}$, $S^{\pm 3}$ and $S^{\pm 4}$ in Run-3 ($L = 30$ fb$^{-1}$) and
600, 1100 and 1430 GeV at HL-LHC ($L = 300$ fb$^{-1}$).  These numbers
can be compared to the current constraints 650, 780 and 920 GeV from
the HSCP searches.  Although a large part of the region that is
accessible to MoEDAL in Run-3 is already excluded, MoEDAL can explore
some phenomenologically viable regions at HL-LHC.  For Model-2, MoEDAL
can explore the region with $N_{\rm sig} \geq 1$ up to 1050, 1250 and
1400 GeV for $\tilde S^{\pm 4/3}$, $\tilde S^{\pm 7/3}$ and $\tilde
S^{\pm 10/3}$ for Run-3 ($L = 30$ fb$^{-1}$).  At the high-luminosity
phase, these mass reaches are improved up to 1400, 1600 and 1800 GeV.
Comparing these with the estimated current bounds, 1450, 1480 and 1510 GeV
for $\tilde S^{\pm 4/3}$, $\tilde S^{\pm 7/3}$ and $\tilde S^{\pm
  10/3}$, we concluded that although we do not have much hope to see
the exotic particles of Model-2 in Run-3, MoEDAL can explore some part
of the allowed parameter regions at the HL-LHC.

In the model-specific part of the study, we explored the subset of the
parameter space in which the neutrino data are fitted and identify the
regions to which MoEDAL is sensitive.  We find that MoEDAL's
sensitivity is maximised around $\lambda_5 \sim 10^{-5}$, at which the
lifetime of $S^{\pm 4}$ in Model-1 (and $\tilde S^{\pm 10/3}$ in
Model-2) is the longest.  The maximum reach of the $m_{S_3}$
($m_{\tilde S_3}$) parameter was found to be roughly the same as the
corresponding mass reach of $S^{\pm 4}$ ($S^{\pm 10/3}$).

Finally, we comment that 
in this study the NTD response was modelled by only considering the particle velocity and charge. 
In reality, the efficiency may also depend on the incidence angle of the particle trajectory with respect to 
the NTD panel plane.
Although this effect is not expected to be significant,
it may be included in future studies in order to accurately assess the performance of the 
long-lived particle searches at MoEDAL.

\section*{Acknowledgments}

We thank Vasiliki Mitsou for useful comments.
The work of R.M.\ is partially supported by the National Science
Centre, Poland, under research grant 2017/26/E/ST2/00135.  The work of
K.S.\ is partially supported by the National Science Centre, Poland,
under research grant 2017/26/E/ST2/00135 and the Grieg grant
2019/34/H/ST2/00707. M.H. is supported by the Spanish grants
FPA2017-85216-P (MINECO/AEI/FEDER, UE) and PROMETEO/2018/165 grants
(Generalitat Valenciana)


\bibliography{ref}

\providecommand{\href}[2]{#2}\begingroup\raggedright\begin{thebibliography}{10}

\bibitem{Aad:2020nyj}
{\bfseries ATLAS} Collaboration, G.~Aad {\em et~al.}, ``{Search for new
  phenomena in final states with large jet multiplicities and missing
  transverse momentum using $ \sqrt{s} $ = 13 TeV proton-proton collisions
  recorded by ATLAS in Run 2 of the LHC},''
  \href{http://dx.doi.org/10.1007/JHEP10(2020)062}{{\em JHEP} {\bfseries 10}
  (2020) 062}, \href{http://arxiv.org/abs/2008.06032}{{\ttfamily
  arXiv:2008.06032 [hep-ex]}}.

\bibitem{Sirunyan:2018xwt}
{\bfseries CMS} Collaboration, A.~M. Sirunyan {\em et~al.}, ``{Search for black
  holes and sphalerons in high-multiplicity final states in proton-proton
  collisions at $ \sqrt{s}=13 $ TeV},''
  \href{http://dx.doi.org/10.1007/JHEP11(2018)042}{{\em JHEP} {\bfseries 11}
  (2018) 042}, \href{http://arxiv.org/abs/1805.06013}{{\ttfamily
  arXiv:1805.06013 [hep-ex]}}.

\bibitem{Sirunyan:2019bgz}
{\bfseries CMS} Collaboration, A.~M. Sirunyan {\em et~al.}, ``{Search for
  physics beyond the standard model in multilepton final states in
  proton-proton collisions at $\sqrt{s} =$ 13 TeV},''
  \href{http://dx.doi.org/10.1007/JHEP03(2020)051}{{\em JHEP} {\bfseries 03}
  (2020) 051}, \href{http://arxiv.org/abs/1911.04968}{{\ttfamily
  arXiv:1911.04968 [hep-ex]}}.

\bibitem{Aad:2021egl}
{\bfseries ATLAS} Collaboration, G.~Aad {\em et~al.}, ``{Search for new
  phenomena in events with an energetic jet and missing transverse momentum in
  $pp$ collisions at $\sqrt{s} = 13$ TeV with the ATLAS detector},''
  \href{http://arxiv.org/abs/2102.10874}{{\ttfamily arXiv:2102.10874
  [hep-ex]}}.

\bibitem{Sirunyan:2017hci}
{\bfseries CMS} Collaboration, A.~M. Sirunyan {\em et~al.}, ``{Search for dark
  matter produced with an energetic jet or a hadronically decaying W or Z boson
  at $ \sqrt{s}=13 $ TeV},''
  \href{http://dx.doi.org/10.1007/JHEP07(2017)014}{{\em JHEP} {\bfseries 07}
  (2017) 014}, \href{http://arxiv.org/abs/1703.01651}{{\ttfamily
  arXiv:1703.01651 [hep-ex]}}.

\bibitem{Aad:2019xav}
{\bfseries ATLAS} Collaboration, G.~Aad {\em et~al.}, ``{Search for long-lived
  neutral particles produced in $pp$ collisions at $\sqrt{s} = 13$ TeV decaying
  into displaced hadronic jets in the ATLAS inner detector and muon
  spectrometer},'' \href{http://dx.doi.org/10.1103/PhysRevD.101.052013}{{\em
  Phys. Rev. D} {\bfseries 101} no.~5, (2020) 052013},
  \href{http://arxiv.org/abs/1911.12575}{{\ttfamily arXiv:1911.12575
  [hep-ex]}}.

\bibitem{Sirunyan:2020cao}
{\bfseries CMS} Collaboration, A.~M. Sirunyan {\em et~al.}, ``{Search for
  long-lived particles using displaced jets in proton-proton collisions at
  $\sqrt{s} = $ 13 TeV},'' \href{http://arxiv.org/abs/2012.01581}{{\ttfamily
  arXiv:2012.01581 [hep-ex]}}.

\bibitem{Aaboud:2017mpt}
{\bfseries ATLAS} Collaboration, M.~Aaboud {\em et~al.}, ``{Search for
  long-lived charginos based on a disappearing-track signature in pp collisions
  at $ \sqrt{s}=13 $ TeV with the ATLAS detector},''
  \href{http://dx.doi.org/10.1007/JHEP06(2018)022}{{\em JHEP} {\bfseries 06}
  (2018) 022}, \href{http://arxiv.org/abs/1712.02118}{{\ttfamily
  arXiv:1712.02118 [hep-ex]}}.

\bibitem{Sirunyan:2020pjd}
{\bfseries CMS} Collaboration, A.~M. Sirunyan {\em et~al.}, ``{Search for
  disappearing tracks in proton-proton collisions at $\sqrt{s} =$ 13 TeV},''
  \href{http://dx.doi.org/10.1016/j.physletb.2020.135502}{{\em Phys. Lett. B}
  {\bfseries 806} (2020) 135502},
  \href{http://arxiv.org/abs/2004.05153}{{\ttfamily arXiv:2004.05153
  [hep-ex]}}.

\bibitem{Alimena:2019zri}
J.~Alimena {\em et~al.}, ``{Searching for long-lived particles beyond the
  Standard Model at the Large Hadron Collider},''
  \href{http://dx.doi.org/10.1088/1361-6471/ab4574}{{\em J. Phys. G} {\bfseries
  47} no.~9, (2020) 090501}, \href{http://arxiv.org/abs/1903.04497}{{\ttfamily
  arXiv:1903.04497 [hep-ex]}}.

\bibitem{Acharya:2020uwc}
B.~Acharya, A.~De~Roeck, J.~Ellis, D.~Ghosh, R.~Mase\l{}ek, G.~Panizzo,
  J.~Pinfold, K.~Sakurai, A.~Shaa, and A.~Wall, ``{Prospects of searches for
  long-lived charged particles with MoEDAL},''
  \href{http://dx.doi.org/10.1140/epjc/s10052-020-8093-5}{{\em Eur. Phys. J. C}
  {\bfseries 80} no.~6, (2020) 572},
  \href{http://arxiv.org/abs/2004.11305}{{\ttfamily arXiv:2004.11305
  [hep-ph]}}.

\bibitem{Felea:2020cvf}
D.~Felea, J.~Mamuzic, R.~Mase\l{}ek, N.~Mavromatos, V.~Mitsou, J.~Pinfold,
  R.~Ruiz~de Austri, K.~Sakurai, A.~Santra, and O.~Vives, ``{Prospects for
  discovering supersymmetric long-lived particles with MoEDAL},''
  \href{http://dx.doi.org/10.1140/epjc/s10052-020-7994-7}{{\em Eur. Phys. J. C}
  {\bfseries 80} no.~5, (2020) 431},
  \href{http://arxiv.org/abs/2001.05980}{{\ttfamily arXiv:2001.05980
  [hep-ph]}}.

\bibitem{Acharya:2014nyr}
{\bfseries MoEDAL} Collaboration, B.~Acharya {\em et~al.}, ``{The Physics
  Programme Of The MoEDAL Experiment At The LHC},''
  \href{http://dx.doi.org/10.1142/S0217751X14300506}{{\em Int. J. Mod. Phys. A}
  {\bfseries 29} (2014) 1430050},
  \href{http://arxiv.org/abs/1405.7662}{{\ttfamily arXiv:1405.7662 [hep-ph]}}.

\bibitem{Pinfold:2009oia}
{\bfseries MoEDAL} Collaboration, J.~Pinfold {\em et~al.}, ``{Technical Design
  Report of the MoEDAL Experiment},''.

\bibitem{Pinfold:2019nqj}
J.~L. Pinfold, ``{The MoEDAL Experiment at the LHC\textemdash{}A Progress
  Report},'' \href{http://dx.doi.org/10.3390/universe5020047}{{\em Universe}
  {\bfseries 5} no.~2, (2019) 47}.

\bibitem{Acharya:2019vtb}
{\bfseries MoEDAL} Collaboration, B.~Acharya {\em et~al.}, ``{Magnetic Monopole
  Search with the Full MoEDAL Trapping Detector in 13 TeV pp Collisions
  Interpreted in Photon-Fusion and Drell-Yan Production},''
  \href{http://dx.doi.org/10.1103/PhysRevLett.123.021802}{{\em Phys. Rev.
  Lett.} {\bfseries 123} no.~2, (2019) 021802},
  \href{http://arxiv.org/abs/1903.08491}{{\ttfamily arXiv:1903.08491
  [hep-ex]}}.

\bibitem{Zee:1980ai}
A.~Zee, ``{A theory of lepton number violation, neutrino Majorana mass, and
  oscillation},''
\href{http://dx.doi.org/10.1016/0370-2693(80)90349-4,
  10.1016/0370-2693(80)90349-4}{{\em Phys.Lett.} {\bfseries B93} (1980) 389}.

\bibitem{Cheng:1980qt}
T.~P. Cheng and L.-F. Li, ``{Neutrino masses, mixings and oscillations in
  $SU(2) x U(1)$ models of electroweak interactions},''
\href{http://dx.doi.org/10.1103/PhysRevD.22.2860}{{\em Phys. Rev.} {\bfseries
  D22} (1980) 2860}.

\bibitem{Zee:1985id}
A.~Zee, ``{Quantum numbers of Majorana neutrino masses},''
\href{http://dx.doi.org/10.1016/0550-3213(86)90475-X}{{\em Nucl. Phys.}
  {\bfseries B264} (1986) 99--110}.

\bibitem{Babu:1988ki}
K.~S. Babu, ``{Model of ``calculable'' Majorana neutrino masses},''
\href{http://dx.doi.org/10.1016/0370-2693(88)91584-5}{{\em Phys. Lett.}
  {\bfseries B203} (1988) 132--136}.

\bibitem{Cai:2017jrq}
Y.~Cai, J.~Herrero-García, M.~A. Schmidt, A.~Vicente, and R.~R. Volkas,
  ``{From the trees to the forest: a review of radiative neutrino mass
  models},'' \href{http://dx.doi.org/10.3389/fphy.2017.00063}{{\em Front.in
  Phys.} {\bfseries 5} (2017) 63},
\href{http://arxiv.org/abs/1706.08524}{{\ttfamily arXiv:1706.08524 [hep-ph]}}.

\bibitem{Bonnet:2012kz}
F.~Bonnet, M.~Hirsch, T.~Ota, and W.~Winter, ``{Systematic study of the d=5
  Weinberg operator at one-loop order},''
  \href{http://dx.doi.org/10.1007/JHEP07(2012)153}{{\em JHEP} {\bfseries 07}
  (2012) 153},
\href{http://arxiv.org/abs/1204.5862}{{\ttfamily arXiv:1204.5862 [hep-ph]}}.

\bibitem{Sierra:2014rxa}
D.~Aristizabal~Sierra, A.~Degee, L.~Dorame, and M.~Hirsch, ``{Systematic
  classification of two-loop realizations of the Weinberg operator},''
  \href{http://dx.doi.org/10.1007/JHEP03(2015)040}{{\em JHEP} {\bfseries 03}
  (2015) 040},
\href{http://arxiv.org/abs/1411.7038}{{\ttfamily arXiv:1411.7038 [hep-ph]}}.

\bibitem{Cepedello:2018rfh}
R.~Cepedello, R.~M. Fonseca, and M.~Hirsch, ``{Systematic classification of
  three-loop realizations of the Weinberg operator},''
  \href{http://dx.doi.org/10.1007/JHEP10(2018)197,
  10.1007/JHEP06(2019)034}{{\em JHEP} {\bfseries 10} (2018) 197},
  \href{http://arxiv.org/abs/1807.00629}{{\ttfamily arXiv:1807.00629
  [hep-ph]}}.
[erratum: JHEP06,034(2019)].

\bibitem{Arbelaez:2020xcg}
C.~Arbel\'aez, G.~Cottin, J.~C. Helo, and M.~Hirsch, ``{Long-lived charged
  particles and multi-lepton signatures from neutrino mass models},''
  \href{http://dx.doi.org/10.1103/PhysRevD.101.095033}{{\em Phys. Rev. D}
  {\bfseries 101} no.~9, (2020) 095033},
  \href{http://arxiv.org/abs/2003.11494}{{\ttfamily arXiv:2003.11494
  [hep-ph]}}.

\bibitem{Ma:2006km}
E.~Ma, ``{Verifiable radiative seesaw mechanism of neutrino mass and dark
  matter},'' \href{http://dx.doi.org/10.1103/PhysRevD.73.077301}{{\em Phys.
  Rev. D} {\bfseries 73} (2006) 077301},
  \href{http://arxiv.org/abs/hep-ph/0601225}{{\ttfamily arXiv:hep-ph/0601225}}.

\bibitem{Cordero-Carrion:2019qtu}
I.~Cordero-Carri\'on, M.~Hirsch, and A.~Vicente, ``{General parametrization of
  Majorana neutrino mass models},''
  \href{http://dx.doi.org/10.1103/PhysRevD.101.075032}{{\em Phys. Rev. D}
  {\bfseries 101} no.~7, (2020) 075032},
  \href{http://arxiv.org/abs/1912.08858}{{\ttfamily arXiv:1912.08858
  [hep-ph]}}.

\bibitem{Alwall:2007st}
J.~Alwall, P.~Demin, S.~de~Visscher, R.~Frederix, M.~Herquet, F.~Maltoni,
  T.~Plehn, D.~L. Rainwater, and T.~Stelzer, ``{MadGraph/MadEvent v4: The New
  Web Generation},''
  \href{http://dx.doi.org/10.1088/1126-6708/2007/09/028}{{\em JHEP} {\bfseries
  09} (2007) 028}, \href{http://arxiv.org/abs/0706.2334}{{\ttfamily
  arXiv:0706.2334 [hep-ph]}}.

\bibitem{Alwall:2011uj}
J.~Alwall, M.~Herquet, F.~Maltoni, O.~Mattelaer, and T.~Stelzer, ``{MadGraph 5
  : Going Beyond},'' \href{http://dx.doi.org/10.1007/JHEP06(2011)128}{{\em
  JHEP} {\bfseries 06} (2011) 128},
  \href{http://arxiv.org/abs/1106.0522}{{\ttfamily arXiv:1106.0522 [hep-ph]}}.

\bibitem{Alwall:2014hca}
J.~Alwall, R.~Frederix, S.~Frixione, V.~Hirschi, F.~Maltoni, O.~Mattelaer,
  H.~S. Shao, T.~Stelzer, P.~Torrielli, and M.~Zaro, ``{The automated
  computation of tree-level and next-to-leading order differential cross
  sections, and their matching to parton shower simulations},''
  \href{http://dx.doi.org/10.1007/JHEP07(2014)079}{{\em JHEP} {\bfseries 07}
  (2014) 079}, \href{http://arxiv.org/abs/1405.0301}{{\ttfamily arXiv:1405.0301
  [hep-ph]}}.

\bibitem{Staub:2013tta}
F.~Staub, ``{SARAH 4 : A tool for (not only SUSY) model builders},''
  \href{http://dx.doi.org/10.1016/j.cpc.2014.02.018}{{\em Comput. Phys.
  Commun.} {\bfseries 185} (2014) 1773--1790},
  \href{http://arxiv.org/abs/1309.7223}{{\ttfamily arXiv:1309.7223 [hep-ph]}}.

\bibitem{Staub:2012pb}
F.~Staub, ``{SARAH 3.2: Dirac Gauginos, UFO output, and more},''
  \href{http://dx.doi.org/10.1016/j.cpc.2013.02.019}{{\em Comput. Phys.
  Commun.} {\bfseries 184} (2013) 1792--1809},
  \href{http://arxiv.org/abs/1207.0906}{{\ttfamily arXiv:1207.0906 [hep-ph]}}.

\bibitem{Porod:2003um}
W.~Porod, ``{SPheno, a program for calculating supersymmetric spectra, SUSY
  particle decays and SUSY particle production at e+ e- colliders},''
  \href{http://dx.doi.org/10.1016/S0010-4655(03)00222-4}{{\em Comput. Phys.
  Commun.} {\bfseries 153} (2003) 275--315},
  \href{http://arxiv.org/abs/hep-ph/0301101}{{\ttfamily arXiv:hep-ph/0301101}}.

\bibitem{Porod:2011nf}
W.~Porod and F.~Staub, ``{SPheno 3.1: Extensions including flavour, CP-phases
  and models beyond the MSSM},''
  \href{http://dx.doi.org/10.1016/j.cpc.2012.05.021}{{\em Comput. Phys.
  Commun.} {\bfseries 183} (2012) 2458--2469},
  \href{http://arxiv.org/abs/1104.1573}{{\ttfamily arXiv:1104.1573 [hep-ph]}}.

\bibitem{Manohar:2016nzj}
A.~Manohar, P.~Nason, G.~P. Salam, and G.~Zanderighi, ``{How bright is the
  proton? A precise determination of the photon parton distribution
  function},'' \href{http://dx.doi.org/10.1103/PhysRevLett.117.242002}{{\em
  Phys. Rev. Lett.} {\bfseries 117} no.~24, (2016) 242002},
\href{http://arxiv.org/abs/1607.04266}{{\ttfamily arXiv:1607.04266 [hep-ph]}}.

\bibitem{Manohar:2017eqh}
A.~V. Manohar, P.~Nason, G.~P. Salam, and G.~Zanderighi, ``{The Photon Content
  of the Proton},'' \href{http://dx.doi.org/10.1007/JHEP12(2017)046}{{\em JHEP}
  {\bfseries 12} (2017) 046},
\href{http://arxiv.org/abs/1708.01256}{{\ttfamily arXiv:1708.01256 [hep-ph]}}.

\bibitem{Butterworth:2015oua}
J.~Butterworth {\em et~al.}, ``{PDF4LHC recommendations for LHC Run II},''
  \href{http://dx.doi.org/10.1088/0954-3899/43/2/023001}{{\em J. Phys.}
  {\bfseries G43} (2016) 023001},
\href{http://arxiv.org/abs/1510.03865}{{\ttfamily arXiv:1510.03865 [hep-ph]}}.

\bibitem{Aaboud:2018kbe}
{\bfseries ATLAS} Collaboration, M.~Aaboud {\em et~al.}, ``{Search for heavy
  long-lived multicharged particles in proton-proton collisions at $\sqrt{s}$ =
  13 TeV using the ATLAS detector},''
  \href{http://dx.doi.org/10.1103/PhysRevD.99.052003}{{\em Phys. Rev. D}
  {\bfseries 99} no.~5, (2019) 052003},
  \href{http://arxiv.org/abs/1812.03673}{{\ttfamily arXiv:1812.03673
  [hep-ex]}}.

\bibitem{Jager:2018ecz}
S.~J\"ager, S.~Kvedarait\.{e}, G.~Perez, and I.~Savoray, ``{Bounds and
  prospects for stable multiply charged particles at the LHC},''
  \href{http://dx.doi.org/10.1007/JHEP04(2019)041}{{\em JHEP} {\bfseries 04}
  (2019) 041}, \href{http://arxiv.org/abs/1812.03182}{{\ttfamily
  arXiv:1812.03182 [hep-ph]}}.

\bibitem{Chatrchyan:2013oca}
{\bfseries CMS} Collaboration, S.~Chatrchyan {\em et~al.}, ``{Searches for
  Long-Lived Charged Particles in $pp$ Collisions at $\sqrt{s}$=7 and 8 TeV},''
  \href{http://dx.doi.org/10.1007/JHEP07(2013)122}{{\em JHEP} {\bfseries 07}
  (2013) 122}, \href{http://arxiv.org/abs/1305.0491}{{\ttfamily arXiv:1305.0491
  [hep-ex]}}.

\end{thebibliography}\endgroup
\bibliographystyle{utphys}

\end{document}